\documentclass{aa}  

\usepackage{graphicx}
\usepackage{amssymb}
\usepackage{lipsum}
\usepackage{subcaption}
\usepackage{txfonts}

\begin{document}

   \title{VLBA observations of a sample of low-power compact symmetric objects}

   \author{M. Orienti
          \inst{1}\
          \and
          F. D'Ammando\inst{1}
          \and
          D. Dallacasa\inst{2,1}
          \and
          G. Migliori\inst{1}
          \and
          P. Rossi\inst{3}
          \and
          G. Bodo\inst{3}
}
   \institute{INAF - Istituto di Radioastronomia, Via P. Gobetti 101, I-40129 Bologna, Italy\\
              \email{orienti@ira.inaf.it}
         \and
             Dipartimento di Fisica e Astronomia, Universit\`a di Bologna, Via Gobetti 93/2, I-40129 Bologna, Italy
             \and
             INAF - Osservatorio Astrofisico di Torino, Strada Osservatorio 20, I-10025 Pino Torinese, Italy
             }

   \date{Received <date> / Accepted <date>}

 
  \abstract
      {Compact symmetric objects (CSOs) are intrinsically compact extragalactic radio sources that are thought to be the progenitors of classical radio galaxies. To date, evolutionary models have mainly focused on the formation and growth of high-power radio sources, leaving unanswered many questions related to low-power objects, whose relativistic jets are likely more prone to instabilities. We present a new sample of candidate low-power CSOs selected from the Faint Images of the Radio Sky at Twenty-cm (FIRST) survey. The main selection criteria are (i)  a parsec-scale double radio morphology from archival Very Long Baseline Array (VLBA) images and (ii) a VLBA total flux density consistent with that from the FIRST survey, which rules out the presence of significant radio emission extending beyond the parsec scale. The final sample consists of 60 sources with radio luminosities between 10$^{24}$ and 10$^{27}$ W Hz$^{-1}$ at 1.4 GHz and projected linear sizes between 45 and 430 pc, which fill a region in the radio power-size plane that is currently underpopulated. We carried out VLBA observations at 4.98 GHz of a sub-sample of 20 sources among the brightest candidate CSOs with the aim of confirming their classification. We classify 12 sources as CSOs on the basis of their radio structure and spectral index distribution. In two out of the four CSOs with core identification, the asymmetry in the flux density of the outer components is in agreement with light travel time effects, and there is no evidence of jet-cloud interaction. If we assume a simplistic parametric model, most of the sources in the total sample have a jet power of $\sim 10^{44} - 10^{45}$ erg s$^{-1}$, making their evolutionary paths sensitive to the individual conditions of the jet and its environment.}

\titlerunning{VLBA observations of low-power CSOs}
\authorrunning{M. Orienti et al.}
   \maketitle

%

\section{Introduction}

In the framework of the self-similar deployment of a radio source, it is commonly accepted that the linear size (LS) of the radio
emission, which ranges from (sub)parsec to megaparsec scales in the largest Fanaroff-Riley (FR) I/II radio galaxies, is indicative of the evolutionary
stage of the radio emission. Compact symmetric objects (CSOs), with LSs $< 1$ kpc and typical ages of 10$^{2 - 4}$ yr,  
are intrinsically compact radio sources in an early evolutionary stage \citep{wilkinson94}. CSOs should evolve into medium-sized symmetric objects (MSOs) with LSs between 1 and 20 kpc and ages of $\sim 10^{5-6}$ yr as the relativistic jet propagates through the interstellar medium (ISM) of the host galaxy, and ultimately into classical large-scale radio sources with LSs $>$ 20 kpc \citep{fanti95,readhead96}. However, there is growing evidence that not all CSOs become FRI/FRII sources, and a fraction of them may be short-lived objects \citep[e.g.,][]{kiehlmann24b,readhead24}. High-power CSOs  ($L > 10^{26}$ W/Hz) represent a significant fraction of radio sources in flux-density-limited samples, the percentage varying depending on the frequency and flux density threshold  \citep[e.g.,][]{fanti90,odea98,callingham17,ballieux24,kiehlmann24a}, suggesting an excess of young radio sources that cannot be reconciled with source counts even if luminosity evolution is taken into account. 

Several evolutionary models have been proposed to describe how the physical parameters (e.g., luminosity, expansion velocity, and magnetic field) evolve as the
relativistic jet propagates within the host galaxy and interacts with the ISM \citep[e.g.,][]{kaiser97,alexander00,snellen00,perucho02}. 
Studies of ionized and atomic gas have noted the presence of an inhomogeneous medium enshrouding CSOs, which may slow the source growth \citep[e.g.,][]{morganti04,labiano06,mo08,struve12,morganti13}. Evidence that jet-ISM interactions do occur in this class of sources comes from the asymmetric radio structure observed in a significant fraction of CSOs in which the brightest hotspot is that closest to the core, which is inconceivable with light travel time effects \citep[e.g.,][]{jeyakumar00,mo07a,dd13,mo14}. These asymmetries become less prominent as the radio source grows, becoming almost negligible once the relativistic jets leave the ISM and enter the intracluster medium \citep{mo16}.

Relativistic magnetohydrodynamic (RMHD) simulations enable the characterization of the morphology and dynamics of relativistic jets in terms of their physical parameters, such as the jet-to-ambient density ratio, Mach number, and magnetic field strength \citep[e.g.,][]{rossi17}. 
Although most studies have focused on powerful radio galaxies (L$_{\rm 1.4 GHz} > 10^{26}$ W/Hz), interest in low-power radio galaxies with low Mach number jets is growing  \citep{massaglia19,rossi20,rossi24}. Several theoretical and numerical works have dealt with the dynamical development of relativistic jets \citep[e.g.,][]{scheuer74,kaiser97,carvalho02a,carvalho02b,krause03}.
Based on 3D RMHD simulations, \citet{massaglia22} find that all jets produce FRII-like radio morphology with a strong termination shock during the early evolutionary phase, independent of the kinetic power. Only at a later stage the jet evolution diverges depending on the jet-to-ambient density ratio and the magnetization parameter.
Weak jets are more prone to instabilities and seem to have more individualized  evolution schemes than their high-power counterparts \citep[e.g.,][]{mukherjee20}. The interaction between the relativistic plasma and clouds of gas may be able to decelerate and/or disrupt the jet even in the presence of a continuous supply of relativistic particles \citep[e.g.,][]{kaiser97,alexander00,perucho14,perucho20}.
A large fraction of the energy of low-power jets is
then deposited into the host galaxy, which potentially impacts the distribution and kinematics of the ISM for a longer time than for high-power jets \citep[e.g.,][]{massaglia16,mukherjee18,murthy19,mukherjee20,webster21}.  

So far, only a few samples of candidate low-power CSOs have been compiled \citep[e.g.,][]{snellen98,cstan09}. Increasing the number of confirmed low-power CSOs is thus of great importance  for improving our knowledge of the evolutionary path of radio emission and the interplay between relativistic jets and the host galaxy. \\

In this paper we present a new sample of candidate low-power CSOs with the aim of constraining the physical properties of these sources, and we investigate how jet-ISM interaction may influence both the jet evolution and the ISM properties. Very Long Baseline Array (VLBA) observations at 4.98 GHz of a subsample of 20 low-power CSO candidates were performed in order to confirm or disprove their CSO classification, by determining their milliarcsecond-scale radio structure and spectral index distribution, and to investigate jet-ISM interaction by detecting flux-density and/or arm-length asymmetries between the lobes and/or hotspots of the sources. \\

The paper is organized as follows: Sect. \ref{sec-sample} presents the
sample, and Sect. \ref{sec-data} describes the radio observations.
Results are presented in Sect. \ref{sec-result} and discussed in Sect. \ref{discussion}. In
Sect. \ref{summary} we draw our conclusions.

Throughout this paper, we assume the following cosmology:
$H_{0} =70\; {\rm km\,s^{-1}\, Mpc^{-1}}$, $\Omega_{\rm M} = 0.27,$ and $\Omega_{\rm \Lambda} = 0.73$ in a flat Universe. 
The spectral index, $\alpha,$ is defined as S($\nu$) $\propto \nu^{- \alpha}$.

\section{The sample}
\label{sec-sample}

The best way to select a CSO is by its parsec-scale two-sided structure, and this requires milliarcsecond-resolution very long baseline interferometry  (VLBI) images. Only a few samples targeting high-power CSOs have been compiled on the basis of VLBI information \citep[e.g.,][]{peck00,sokolovsky11,tremblay16}. On the other hand, samples of low-power CSOs have been selected mainly by their spectral properties \citep[e.g.,][]{snellen98,cstan09}, and follow-up VLBI observations were performed in a second time to discriminate between genuine CSOs and beamed core-jet sources 
\citep[e.g.,][]{snellen00b,mo12,mo20}.

We used the data from the mJy imaging of VLBA exploration project \citep[mJIVE-20;][]{deller14} to construct a sample of low-power CSOs selected by their double radio morphology. The mJIVE-20 is a VLBA large project that made use of VLBA filler time to systematically inspect a large sample of millijansky radio sources, preselected from the Faint Images of the Radio Sky at Twenty-cm (FIRST) survey \citep{becker95}, with the aim of identifying any compact emission that may be present \citep{deller14}.

Our selection criteria are:

\begin{itemize}

\item The source is unresolved in the FIRST survey but shows a well-resolved two-sided structure (according to visual inspection) in mJIVE-20 VLBA images.\\

\item The VLBI total flux density is consistent, within uncertainties, with the FIRST flux density, indicating that no radio emission is present on scales between those sampled by VLBI and the resolution of the FIRST.\\

\item No significant flux density variability ($<$10\%), within uncertainties, is found when comparing VLA Sky Survey (VLASS) data taken at three different epochs and spanning a time interval of about 5 years \citep{lacy20}.

\end{itemize}

We ended up with a sample of 60 objects. In Table \ref{sample-table} we present the selected sources as well as the peak flux density at 1.38 GHz from mJIVE-20 data, flux densities at 1.4 and 3.0 GHz from the FIRST and VLASS surveys, respectively, the spectral index $\alpha_{1.4}^{3.0}$ between FIRST and VLASS data, and the angular size measured on mJIVE-20 images. When available, we report optical and infrared information of the host galaxy from the Dark Energy Spectroscopic Instrument (DESI) Legacy Imaging Surveys \citep{dey19}, Sloan Digital Sky Survey \citep[SDSS;][]{abdurrouf22}, and Wide-field Infrared Survey Explorer \citep[WISE;][]{wright10}. 

The identification of compact doubles in snapshot VLBA images is not an easy task due to the high sidelobe level, and more CSOs may be unveiled when additional flagging, cleaning, and possibly self-calibration are performed on the a priori calibrated data \citep{deller14}. Although the sample is not statistically complete, it can be considered representative of the compact double population in the mJIVE-20 survey (see Sect. \ref{discussion}).\\

The selection criteria should reduce the contamination from beamed blazar sources. However,
since only a single frequency is available, we were not able to unambiguously discriminate between a
double-lobed object, as expected for genuine de-beamed CSOs, and a core-jet structure, typical of blazars. 

We performed VLBA observations at 4.98 GHz of a subsample of 20 sources selected from those with a peak flux density at 1.38 GHz S$_{\rm p, 1.38 GHz} \geq 9$ mJy/beam and with right ascension and declination that minimize slewing time and overheads. The goal is to determine the nature of each source component by studying the spectral index distribution.  
The CSO nature of the radio source J1209+4115 was already confirmed by \citet{devries09} on the basis of multifrequency VLBI observations, and no further VLBA observations were requested.

\section{Radio data}
\label{sec-data}

\subsection{VLBA observations}

VLBA observations in the C band (central frequency 4.98 GHz) were carried out between August and December 2021 in dual polarization mode and with an aggregate bit rate of 2 Gbps (project code BO064). Observations were divided into eight short scheduling blocks (from 4 to 5.5 h each) depending on the right ascension and declination of target sources, in order to reduce slewing time and overheads (Table \ref{log-table}). No data from the antenna in Pie Town or Fort Davis were acquired in runs BO064B and BO064G, respectively. Each source was observed for about 1 h, in 12 5-minute scans. Observations were performed in phase-referencing mode, since our sources were not expected to be bright enough for fringe-fitting. Fringe finders, bandpass, and phase calibrators are reported in Table \ref{log-table}.\\

Calibration and data reduction were performed following the standard procedures described in the Astronomical Image Processing System (AIPS) cookbook. Antenna system temperatures and antenna gains were used to calibrate the amplitudes. The uncertainty in the amplitude calibration is about 5\%. The AIPS task {\tt IMAGR} was used to create the images.
When the peak flux density of the source was S$_{\rm p}$ $>$ 5 mJy beam$^{-1}$ at 4.98 GHz, we performed a few iterations of phase-only self-calibration before producing the final images. In the case of a peak flux density S$_{\rm p} >$ 10 mJy beam$^{-1}$, phase-only self-calibration iterations were followed by a single amplitude self-calibration with a solution interval of the scan length. 

The rms measured on the image plane is close to the expected thermal noise, and is about 25 $\mu$Jy beam$^{-1}$. 

\subsection{mJIVE-20 data}

For each source, we downloaded the calibrated datasets from the mJIVE-20 database\footnote{https://safe.nrao.edu/vlba/mjivs/home.html}. The data were then imported into AIPS, inspected for bad data, and then flagged. We produced the final images after a few phase-only self-calibration iterations. As we did for the VLBA data at 4.98 GHz, a final step of amplitude self-calibration with a solution interval of the scan length was attempted for the brightest sources only. 

The rms measured on the image plane is between 60 and 200 $\mu$Jy beam$^{-1}$.

\section{Results}
\label{sec-result}

The final images at 1.38 and 4.98 GHz of the subsample of 20 candidate CSOs are presented in Fig. \ref{vlba-fig}. The flux density of the extended components was measured using {\tt TVSTAT}, which extracts the flux density on a selected polygonal area on the image plane. For unresolved or marginally resolved components, we measured the flux density and the deconvolved angular size using the AIPS task {\tt JMFIT}, which performs a 2D elliptical Gaussian fit on the image plane. Flux density errors are estimated by $\sigma = \sqrt{\sigma_{\rm cal}^2 + \sigma_{\rm rms}^2}$, where $\sigma_{\rm cal}$ and $\sigma_{\rm rms}$ are the uncertainty in the amplitude calibration and the rms on the image plane, respectively. We conservatively estimate the amplitude uncertainty at 1.38 GHz to be 10\%. The $\sigma_{\rm rms}$ contribution is $\sigma_{\rm rms} = {\rm rms} \times \sqrt{\theta_{\rm ext}/ \theta_{\rm beam}}$, where $\theta_{\rm ext}$ and $\theta_{\rm beam}$ are the area of the source component and the beam area, respectively. \\

The spectral index computed from the flux densities measured on the full-resolution images may be artificially steeper than the real value owing to the mismatch between the visibility ranges at the two frequencies. Sources J0958$+$3208 and J1127$+$2540 are two examples in which the 4.98-GHz data cannot recover the flux density from the extended structure. To tackle this issue,
in addition to the full-resolution images, we created a set of images at both 1.38 and 4.98 GHz with the same shortest and longest baseline in wavelengths (approximately $9 - 40.5$ M$\lambda$), beam size, and pixel size common to both frequencies. 
For completeness, the spectral indices computed from the flux densities measured on full-resolution images, $\alpha_{\rm F}$, and those computed using the flux densities measured on low-resolution images, $\alpha_{\rm L}$, are reported in Table \ref{flux-table}. Spectral index errors were computed following the error propagation theory. For the discussion about the spectral index, we refer to $\alpha_{\rm L}$, unless otherwise stated.

\subsection{Source classification}

Our VLBA observations at 4.98 GHz could detect and well resolve all target sources. The availability of dual-frequency VLBA data enabled us to determine the spectral index of each component. 
Sources with two-sided steep-spectrum components are considered CSOs, while those showing a steep-spectrum one-sided jet emerging from a (usually) flat-spectrum compact component are considered blazars. 

Following the aforementioned criteria, we classified 12 objects as CSOs, 4 as blazars showing a core-jet structure, and 4 as blazar candidates (Table \ref{flux-table}). Of the CSOs, 4 objects (about 33\%) have a triple structure at 4.98 GHz. \\

We investigated source asymmetries for the sources classified as CSO, by computing the flux density ratio, R$_{\rm S}$, between the outer components. When the core is detected we also computed the arm-length ratio, R$_{\rm L}$, between the outer components and the core.
A significant flux density asymmetry between the lobes (flux density ratio, R$_{\rm S} >$ 2 at at least one frequency) is found in 5 CSOs (Table \ref{asym}). 
No significant arm-length asymmetry (arm-length ratio, R$_{\rm L} > 2$) is found in the four CSOs with triple structure. Among them, flux density asymmetry is found in two objects, with indication that the brighter component is on the approaching side, in line with light travel time effects. 

A short description of each source with the motivation for its classification, either a CSO or blazar (or blazar candidate), is reported in Sect. \ref{sec-notes}.

\subsection{Notes on individual objects}
\label{sec-notes}

Here we present a brief description of the sources studied in this paper.
VLBA images of the sources discussed below are presented in Fig. \ref{vlba-fig}. Flux density, deconvolved component size, spectral index between 1.38 and 4.98 GHz, and source classification are reported in Table \ref{flux-table}, while asymmetry parameters, R$_{\rm S}$ and R$_{\rm L}$, for asymmetric CSOs can be found in Table \ref{asym}. The projected largest angular size of the entire radio source is reported in Table \ref{sample-table}, and is measured either from the lowest reliable contour in Fig. \ref{vlba-fig} (if the component structure is resolved, e.g., J1107+1619 and J1230+3728) or (for most sources) from the distance between the peaks if the outer components are unresolved (e.g., J1008+1328 and J1554+1134).

\subsubsection{J0834+4111 $-$ CSO}

This source has an asymmetric double structure elongated in the N-S direction, and an angular size of $\sim 22$ mas, which corresponds to a projected LS of about 184 pc at the estimated photometric redshift of the source ($z$ = 1.137). Both components have steep spectra ($\alpha_{\rm L} \sim 0.9$). The flux density of component S is roughly double that of component N at both frequencies (Table \ref{asym}). 

\subsubsection{J0855+2030 $-$ CSO}

This source has a roughly symmetric double structure elongated in the E-W direction, and an angular size of $\sim 16$ mas, which corresponds to a projected LS of about 63 pc at the estimated photometric redshift of the source ($z$ = 0.25). The spectra of neither components are particularly steep ($\alpha_{\rm L} = 0.4 - 0.6$), likely indicating a dominant contribution from the hotspots rather than the lobes.

\subsubsection{J0956+2529 $-$ blazar}

This source has a one-sided structure elongated in the E-W direction and an angular size of $\sim 26$ mas, which corresponds to a projected LS of about 220 pc at the estimated photometric redshift of the source ($z$ = 1.248). At 4.98 GHz the source is resolved into several subcomponents. The flat spectrum core is hosted in component E1, from which a jet emerges. We remove this source from our final sample of genuine CSOs.

\subsubsection{J0958+3208 $-$ blazar?}

This source has a very asymmetric structure elongated in the SE-NW direction and an angular size of $\sim$ 25 mas, which corresponds to a projected LS of about 207 pc at the estimated photometric redshift of the source ($z$ = 1.09). Component W has a very steep spectrum ($\alpha_{\rm L} \sim 1.2$) and is slightly misaligned (about 20$^{\circ}$) with respect to the elongation of component E. The spectrum of component E is not as flat as one would expect for a blazar core. This may be related to the low resolution of the data at 1.38 GHz that is not adequate to separate the contribution from E1 and E2, which may have a different spectral index. We interpret this source as a one-sided core-jet blazar, where the jet emerges from the core hosted in E1. The source might also be a very asymmetric CSO, with R$_{\rm S} >$ 13 between the outer components E1 and W and R$_{\rm L} > 5$, where the brighter component is closer to the core, similar to the CSO J1335+5844 \citep{mo14}. However, the misalignment observed between the elongation of component E and the position of component W would imply a different orientation of the jet and counter-jet on parsec scales; this is quite uncommon even in the most asymmetric CSOs, where the hotspot-core-counter-hotspot are usually well aligned and only lobes or tails of extended emission may have a different orientation \citep[e.g.,][]{peck00,sokolovsky11,dd13,mo14,tremblay16,dd21,cstan25}. On the other hand, jet bending at such scales is quite common in blazars \citep[e.g.,][]{lister13,jorstad17}. The source might also be a restarted object, with component E being the new CSO and component W the relic of a previous activity phase. Deep VLBI observations at higher frequencies are needed to unambiguously classify this source.

\subsubsection{J1008+1328 $-$ CSO}

This source has a triple structure elongated in the SE-NW direction and an angular size of $\sim 22$ mas. The central component, with an inverted spectrum ($\alpha_{\rm L} \sim -0.2$), is the core and is roughly midway between the outer steep-spectrum components, R$_{\rm L} \sim 1.1$. We interpret the elongation of component C toward component N as the inner part of the approaching jet. The flux density of component N is roughly twice that of component S.

\subsubsection{J1107+1619 $-$ CSO}

This source has a roughly symmetric double structure elongated in the E-W direction and an angular size of $\sim 19$ mas. At 4.98 GHz, both components, which both have a steep spectrum ($\alpha_{\rm L} \sim 1.1$), are resolved into two subcomponents.

\subsubsection{J1120-0607 $-$ blazar}

The source structure is elongated in the E-W direction with an angular size of $\sim 14$ mas. Component E, which is the brightest component at 1.38 GHz, becomes the faintest at 4.98 GHz, showing a very steep spectrum ($\alpha_{\rm L} \sim 1.3$). Component W is characterized by a flat spectrum ($\alpha_{\rm L} \sim 0.2$) and hosts the source core, though its location, either in subcomponents W1 or W2, cannot be securely determined. We suggest that W1 is the flat-spectrum core, whereas W2 is a jet knot with a steeper spectrum. Deep VLBI observations at higher frequencies are needed to infer the position of the core. We remove this source from our final sample of genuine CSOs.

\subsubsection{J1127+2540 $-$ blazar?}

The source structure is elongated in the N-S direction and is characterized by a compact component from which arises a diffuse emission extending for about 50 mas in length. At 4.98 GHz only the northern component is detected, while the southern component is resolved out with no sign of compact regions. The steep spectrum of component N ($\alpha_{\rm L} \sim 1.1$) makes the classification of this source nontrivial. It may be a blazar, with the core and jet emissions blended together owing to insufficient resolution, or a fading CSO. However, the morphology at low frequency is reminiscent of a one-sided core-jet structure, and for this reason we conservatively classify this source as a candidate blazar. Deeper VLBI observations are needed to unambiguously determine its nature. 

\subsubsection{J1206+2821 $-$ CSO}

This source has a double structure elongated in the SE-NW direction and an angular size of $\sim$ 30 mas. The flux density at 4.98 GHz of component N is roughly twice that of component S, while their ratio is close to unity at 1.38 GHz. Both components have a steep spectrum ($\alpha_{\rm L} \geq$ 1.0).

\subsubsection{J1214+3438 $-$ CSO}

This source has a triple structure elongated in the NE-SW direction and an angular size of $\sim$ 18 mas. The central component is roughly midway between the outer steep-spectrum components, R$_{\rm L} \sim 1.2$. We considered component C as the source core, though the low resolution at 1.38 GHz prevents us from estimating its spectral index. At 1.38 GHz the flux density ratio between the outer components is R$_{\rm S} \sim 2.8$, while it is close to unity at 4.98 GHz.

\subsubsection{J1219+5845 $-$ blazar}

This is a very asymmetric double source elongated in the N-S direction with an angular size of $\sim$ 20 mas. The northern component, the faintest at 1.38 GHz, is the brightest at 4.98 GHz. Its spectrum turns out to be flat/inverted ($\alpha_{\rm L} = -0.1$), whereas component S, likely a jet component, has a steep spectrum ($\alpha_{\rm L} = 1.5$). We remove this source from our final sample of genuine CSOs.

\subsubsection{J1230+3728 $-$ CSO}

This source has a symmetric double structure elongated in the NE-SW direction and an angular size of $\sim$ 21 mas, which corresponds to a projected LS of about 126 pc at the estimated photometric redshift of the source ($z$ = 0.479). In the 4.98-GHz image, both components are resolved in a direction roughly perpendicular to the source elongation, making the radio structure reminiscent of Z/S-shaped radio sources, a morphology that has been observed in other CSOs, such as 0108+388 and 1031+567 \citep{cstan25}. Both components E and W have steep spectra ($\alpha_{\rm L} \sim 1.1$) with a flux density ratio close to unity at both frequencies.

\subsubsection{J1239+3713: A CSO}

This source has a triple structure elongated in the NE-SW direction and an angular size of $\sim$ 25 mas, which corresponds to a projected LS of about 202 pc at the estimated photometric redshift of the source ($z$ = 0.984). Component C is roughly midway between the outer steep-spectrum components with R$_{\rm L} \sim 1.1$. We considered component C as the source core, though the low resolution at 1.38 GHz prevents us from estimating its spectral index. The flux density ratio between components E and W is close to unity at both observing frequencies.

\subsubsection{J1431+5427 $-$ blazar?}

This source has a double structure elongated in the E-W direction and an angular size of $\sim$ 22 mas. The flux density ratio between component E and W is roughly unity at low frequency, then reaching R$_{\rm S} \sim 4$ at 4.98 GHz. Both components have very steep spectra ($\alpha_{\rm L} > 1.0$). These spectral indices might be artificially steeper than the real value owing to the presence of significant diffuse emission, mainly from the western component, which is not recovered at high frequency owing to a combination of sensitivity and sparse (uv) coverage. The morphology in the 1.38-GHz image, an extended structure emerging from a compact component, is reminiscent of a one-sided core-jet object, and for this reason we conservatively classify this source as a candidate blazar, similar to the case of J1127+2540. VLBI data at higher frequencies are needed to unambiguously determine its nature.

\subsubsection{J1500+4507 $-$ blazar?}

The source has a radio structure elongated in the N-S direction and an angular size $\sim$ 30 mas. At 4.98 GHz,  
component N is slightly misaligned with respect to the elongation of component S, which is resolved into three subcomponents. Components N and S have steep spectra ($\alpha_{\rm L} \geq 1.0$). However, as in the case of J1431+5427, there may be an artificial steepening of the spectra due to the diffuse emission, about 30 mas in size, which is not recovered in the data at 4.98 GHz. The morphology can be interpreted in terms of either an asymmetric CSO, where S3 might be the core and S1 a hotspot, or a restarted CSO with component N being the relic of a past activity, or a one-sided core-jet blazar, where S1 is the core, while S2 and S3 are jet knots. We conservatively classify this source as a candidate blazar, similar to the case of J0958+3208. VLBI data at higher frequencies are needed to unambiguously determine its nature.

\subsubsection{J1554+1134 $-$ CSO}

This source has a triple structure elongated in the NE-SW direction and an angular size of $\sim$ 26 mas. The central component is roughly midway between the outer steep-spectrum components, with R$_{\rm L} \sim 1.1$. At 4.98 GHz, component C, hosting the core, is resolved into two subcomponents, which are likely the core and the jet that extends toward component N. The combination of core and jet emission may explain the steep spectrum ($\alpha_{\rm L} \sim 0.9$) of component C.
The flux density ratio between components N and S is close to unity at both observing frequencies.

\subsubsection{J1600+3304 $-$ CSO}

This is a roughly symmetric double source elongated in the NE-SW direction with an angular size of $\sim$ 28 mas. In the 4.98 GHz image, both components are slightly resolved. Although they have a steep spectrum, the flatter spectral index ($\alpha_{\rm L} \sim 0.6$) of component E suggests a dominant contribution from the hotspot region. Their flux density ratio is close to unity at both observing frequencies.

\subsubsection{J1603+2944 $-$ CSO}

This is an asymmetric double source elongated in the E-W direction with an angular size of $\sim$ 23 mas, which corresponds to a projected LS of about 190 pc at the estimated photometric redshift of the source ($z$ = 1.058). Both components have a relatively steep spectrum ($\alpha_{\rm L} \geq 0.8$). The flux density of component W roughly doubles that of component E at both frequencies. 

\subsubsection{J1603+3317 $-$ CSO}

This is a double source elongated in the NE-SW direction with an angular size of $\sim$ 19 mas, which corresponds to a projected LS of about 152 pc at the estimated photometric redshift of the source ($z$ = 0.939). The flux density at 4.98 GHz of component S is roughly twice that of component N, while their ratio is close to unity at 1.38 GHz. Both components have steep spectra ($\alpha_{\rm L} \geq$ 1.0).

\subsubsection{J1604+1758 $-$ blazar}

This is an asymmetric double source elongated in the E-W direction with an angular size of $\sim$ 19 mas, which corresponds to a projected LS of about 146 pc at the estimated photometric redshift of the source ($z$ = 0.821). Component E, the faintest at 1.38 GHz, is the brightest at 4.98 GHz. Its spectrum turns out to be relatively flat ($\alpha_{\rm L} \sim 0.4$). In the 4.98 GHz image, it is resolved into a compact bright component, likely the core, and an extended structure, likely the jet. Component W has a steep spectrum ($\alpha_{\rm L} \sim 0.9$) and may be a jet knot. We remove this source from our final sample of genuine CSOs.

\begin{table}
\caption{CSOs with flux density asymmetry.}
\begin{center}
\begin{tabular}{cccc}
\hline
Source & R$_{\rm S, 1.38}$ & R$_{\rm S, 4.98}$ & R$_{\rm L}$ \\
\hline
&&&\\
J0834+4111 & 2.3 & 2.1 & - \\
J1008+1328& 1.7 & 2.1 & 1.1 \\
J1214+3438 & 2.8 & 1.1 & 1.2 \\
J1603+2944 & 2.0 & 2.4 & - \\
J1603+3317 & 1.2 & 4.0 & - \\
\hline
\end{tabular}
\end{center}
\label{asym}
\end{table}

\section{Discussion}
\label{discussion}

Determining the physical properties of low-power CSOs, the interplay between their relativistic jet and the surrounding medium, and their evolutionary paths are key factors in building a complete picture of the life cycle of extragalactic radio sources. 
Most of the CSOs studied so far have radio power $P_{\rm 1.4 GHz} > 10^{26}$ W/Hz \citep[e.g.,][]{kb10,an12a,readhead24}, while the population of low-power CSOs is largely unexplored. \\

\begin{figure}
\begin{center}
\includegraphics[width=1.0\columnwidth]{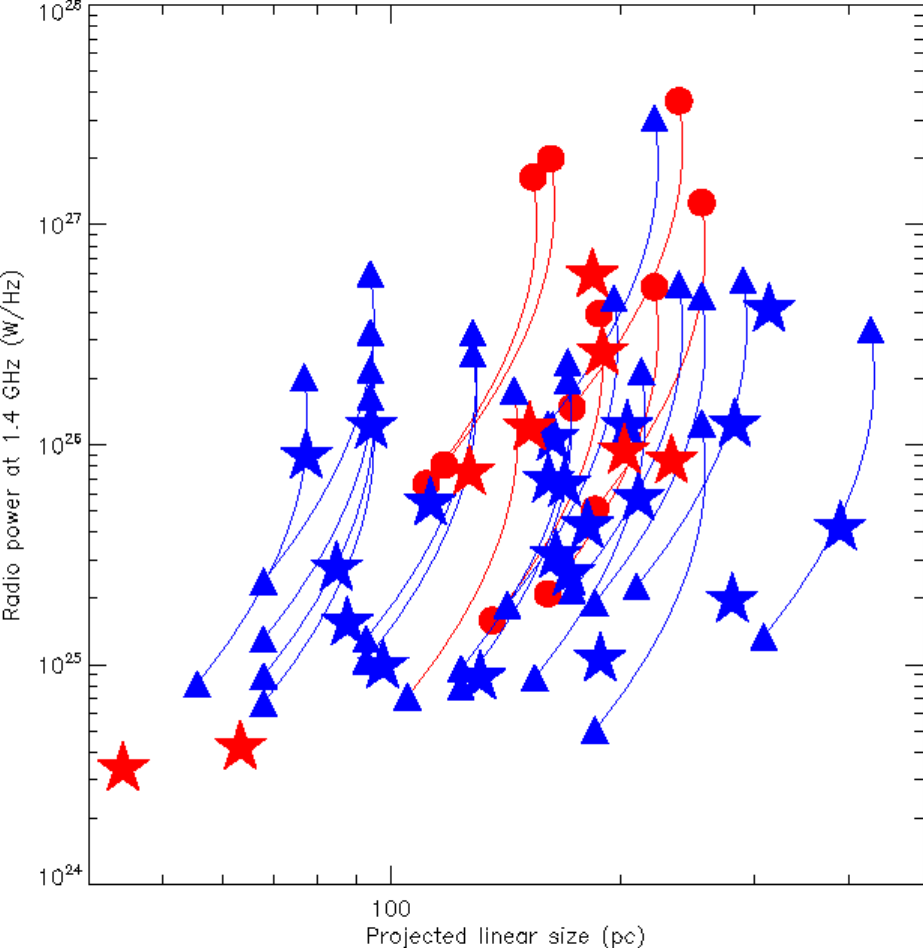}
\caption{Power versus size for the CSOs and CSO candidates of the sample. Red
symbols are the sources already observed at 4.98 GHz with the VLBA (core-jet blazars have been removed), and the blue
symbols are the remaining objects. Stars indicate objects with known redshifts. When the redshift is unknown, the radio power and LS have been computed for $z$ between 0.5 and 2. The solid lines represent how the power and size vary as a function of redshift.}
\label{size-lum}
\end{center}
\end{figure}

\begin{figure}
\begin{center}
\includegraphics[width=1.0\columnwidth]{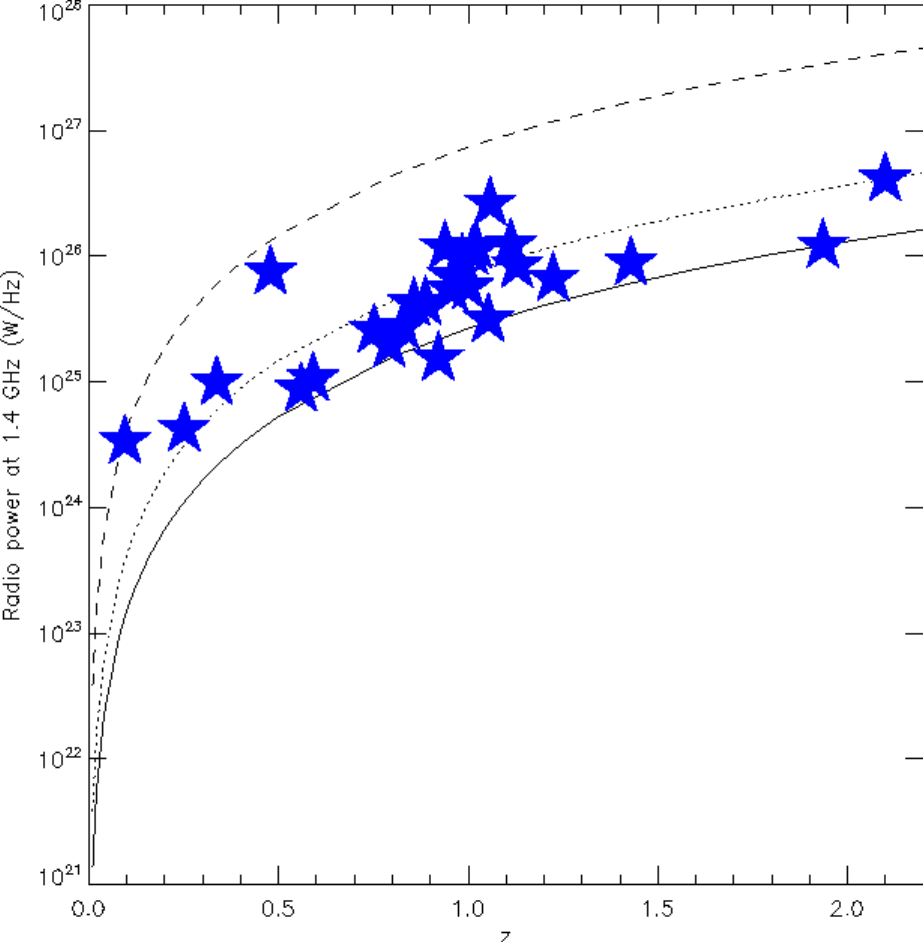}
\caption{Luminosity as a function of redshift for sources with flux density $S_{\rm 1.4 GHz} =$ 6.1 mJy (solid line), 17.1 mJy (dotted line), and 169.7 mJy (dashed line).\ These correspond to the faintest source with an unknown redshift in the sample, the faintest source with an unknown redshift studied in this paper, and the brightest source of the sample with an unknown redshift, respectively. Stars are sources with known redshifts.}
\label{lum-z}
\end{center}
\end{figure}

Figure \ref{size-lum} shows the radio power at 1.4 GHz versus (projected) LS for the candidate CSOs from our sample. The radio power is computed as

\begin{equation}
P = 4 \pi D_{\rm L}^2 \frac{S_{\rm 1.4}}{(1+z)^{1 - \alpha}}
\label{lum-eq}
,\end{equation}

\noindent where $D_{\rm L}$ is the luminosity distance, $S_{\rm 1.4}$ is the flux density at 1.4 GHz from the FIRST survey, and $\alpha$ is the spectral index. 
For sources without an optical counterpart, we computed the physical
parameters assuming the redshift range $z = 0.5 - 2.0$. 
This is the
redshift range spanned by most of the CSOs in our sample with optical identification (Fig. \ref{lum-z}). 
This redshift range is also in agreement with what is expected for massive early-type galaxies, similar to the hosts of gigahertz-peaked spectrum galaxies \citep[e.g.,][]{snellen96,snellen02}, with a mid-infrared 3.4 $\mu$m W1 WISE magnitude comparable to those of our sources \citep[see, e.g.,][]{yan13}. We notice that some sources do not have a counterpart in DESI and WISE images. Their hosts may be massive early-type galaxies at high redshift ($z > 2$) or intrinsically faint, less massive galaxies. Dedicated infrared/optical observations are needed to characterize the properties of the host galaxies.  \\ 

The radio power at 1.4 GHz of the sources in our sample is (0.32 – 59.1)$\times$10$^{25}$ W/Hz, or up to 3.5$\times$10$^{27}$ W/Hz at $z = 2$ (Fig. \ref{lum-z}). The (projected) LS is between 45 and 390 pc, or up to 430 pc at $z=2$.

The lack of very compact sources in our sample is a bias related to the selection criteria. Sources smaller than $9-10$ mas appear unresolved with the angular resolution of the mJIVE-20 images ($\sim$ 16$\times$6.5 mas$^{2}$), implying a lower limit to the LS of the sources of $\sim$ 55 pc at $z=0.5$. The only exception is J1209+4115 (LS $\sim$ 45 pc), which is the closest object of the sample, at $z=0.095$. A similar bias applies to the luminosity range of the selected sources, since the flux density of the sources targeted by the mJIVE-20 survey ranges between 1 and 150 mJy, with a small fraction of sources with $S_{\rm 1.4} > 200$ mJy \citep{deller14}.

\begin{figure}
\begin{center}
\includegraphics[width=1.0\columnwidth]{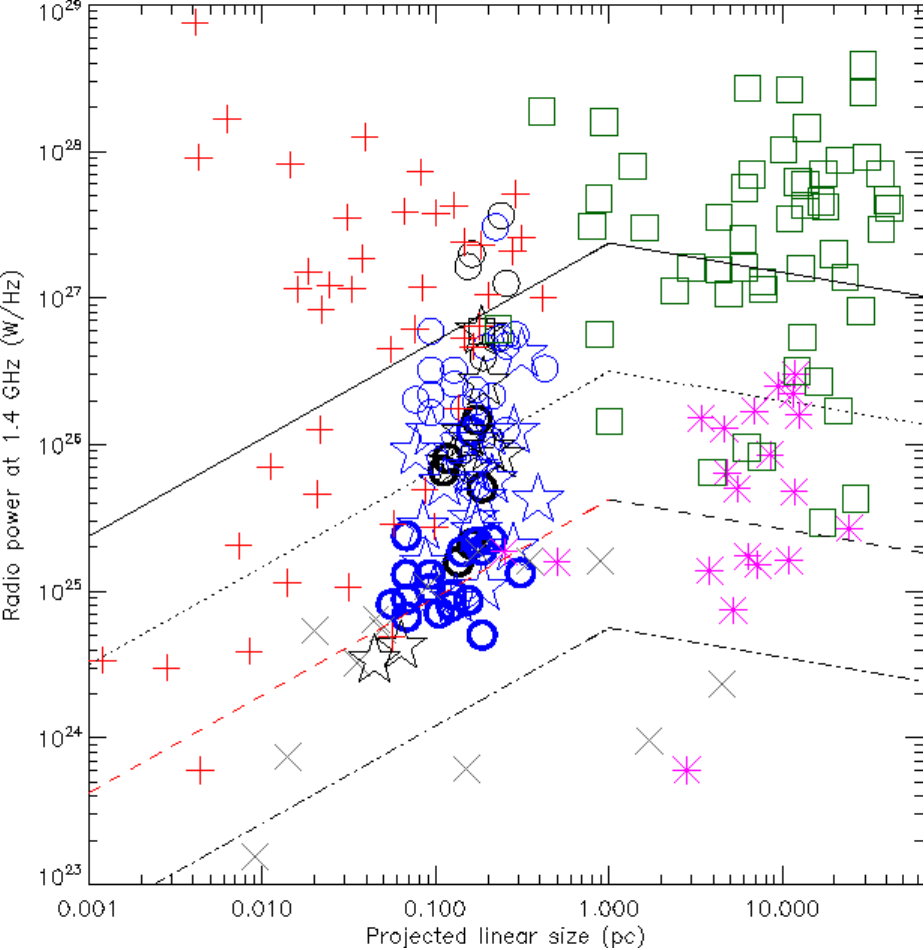}
\caption{Power versus size for the CSOs and MSOs from various samples: red plus signs are CSOs from \citet{an12a}, gray crosses are low-$z$ CSOs from \citet{snellen04}, magenta asterisks are MSOs from \citet{kb10}, and green squares are CSOs and MSOs from \citet{fanti01}.
Circles, triangles, and stars are the CSOs and CSO candidates from our sample: black symbols are CSOs already observed with the VLBA at 4.98 GHz (core-jet blazars have been removed), and the blue symbols are the remaining objects. Thick and thin symbols refer to values computed assuming redshifts of 0.5 and 2, respectively. Stars indicate objects with known redshifts. Illustrative evolutionary tracks are depicted for jet powers of 10$^{46}$ erg s$^{-1}$ (solid line), 10$^{45}$ erg s$^{-1}$ (dotted line), 10$^{44}$ erg s$^{-1}$ (dashed line), and 10$^{43}$ erg s$^{-1}$ (dash-dot line).}
\label{diagram}
\end{center}
\end{figure}

Figure \ref{diagram} shows the radio power-size diagram in which, in addition to the sources of our sample, CSOs and MSOs from other samples have been included \citep{fanti01,snellen04,kb10,an12a}. For the objects from these samples we scaled the luminosity and LS assuming the cosmology considered in this paper. We checked the literature and revised or updated the redshifts for some sources. From the sample compiled by \citet{an12a} we removed some sources that turned out to be beamed objects. 

The sources of our sample are located in a region of the power-size diagram that is currently underpopulated. Most of them are fainter (and less powerful assuming the same redshift range) than the CSOs in \citet{an12a}, which have been selected from classical radio samples such as \citet{pw81}, \citet{pr88}, \citet{polatidis95}, \citet{beasley02}, and \citet{helmboldt07}. Although the majority of the sources in our sample have $z > 0.5$, their luminosity is comparable to the luminosity of the closest ($z < 0.2$) sources from \citet{fanti01} and the high-power tail of the CORALZ sample \citep[$z<0.16$;][]{snellen04}. \\

In Fig. \ref{diagram} we show the evolutionary tracks of radio power as a function of the LS for different values of jet power $P_{j}$. These are illustrative evolutionary tracks that are based on a simple parametric model that assumes $P \propto P_{j}^{7/8} LS^{2/3}$ during the CSO phase, and $P \propto P_{j}^{7/8} LS^{-0.2}$ in the MSO phase \citep[see, e.g.,][]{snellen00,an12a}.
In this model, the transition between the CSO and MSO phases takes place when the jets reach the size of $\sim$1 kpc, exiting the densest region of the host galaxy.

The bulk of the CSOs in our sample is consistent with $P_{j} \sim 10^{44 - 45}$ erg s$^{-1}$, while the high-redshift tail ($z \sim 2$) may have $P_{j} \sim 10^{46}$ erg s$^{-1}$. This is the range of jet power considered in the simulations presented in \citet{mukherjee20}. In these simulations, the low-power CSOs with $P_{j} \sim 10^{44}$ erg s$^{-1}$ are affected by instabilities causing strong jet deceleration. These low-power CSOs might be the progenitors of low-power MSOs discussed in \citet{kb10}. On the other hand, the high-power (high-$z$) tail, with $P_{j} \geq 10^{46}$ erg s$^{-1}$, is less prone to instabilities and should evolve into classical high-power MSOs and then FRII sources. We notice that individual source evolution is influenced by many conditions, either intrinsic to the jet or related to the surrounding medium, and different initial jet parameters may result in different outcomes. For example, if low-power CSOs reside in galaxies without a rich environment, their jets may remain un-disrupted as they expand through the ISM and then the intracluster medium. These low-power CSOs might be the progenitors of the population of low-power FRII found in the LOFAR Two-Metre Sky Survey \citep{mingo19}. 

We do not see any evidence of jet-ISM interaction in the low-power CSOs observed so far, suggesting that the jet is expanding in a comparatively low density and homogeneous medium. However, assessing the incidence of jet-ISM interaction requires information from a larger sample. VLBA observations at 5 GHz of the remaining sources are necessary to constrain the physical properties of the low-power CSOs and the interplay between the radio source and the ambient medium.
Moreover, new multifrequency high-sensitivity VLBI observations of the blazar candidates will be performed to confirm or reject their tentative  classification.

Finally, dedicated photometric and spectroscopic IR and optical observations of the sources that turn out to be genuine CSOs will be performed to characterize the host galaxy and determine (or set tight constraint on) the redshift, which is a key factor in computing the physical parameters of the radio sources.\\

\section{Conclusions}
\label{summary}

In this paper we present a new sample of candidate low-power CSOs selected from the FIRST survey that show a compact two-sided structure on snapshot VLBA images at 1.38 GHz from the mJIVE-20 survey. For 20 of the brightest sources we carried out additional VLBA observations at 4.98 GHz to confirm, or not, the CSO nature of the sources, by determining their radio morphology and spectral index distribution.\\
The conclusions that we can draw from this study are as follows:

\begin{itemize}

\item We classified 12 sources ($\sim$ 60\%  of the subsample with 4.98-GHz VLBA observations) as CSOs. In 4 CSOs we detected the core component, which is roughly midway between the outer components.\\

\item Asymmetry in the flux density between the lobes is found in five CSOs ($\sim$ 41\%). Of the sources with core detection, two objects have significant flux density asymmetry that can be interpreted in terms of light travel time effects. No arm-length asymmetry is found in any CSOs with core detection, suggesting that the jets are expanding in a rather homogeneous medium. \\

\item Although our source selection criteria should prevent the inclusion of beamed sources, of the brightest sources of the sample, about 40\% were blazars or blazar candidates. This fraction is comparable to the percentage of blazars in the COINS sample \citep{peck00}, which was selected on the basis of parsec-scale morphology, while it is significantly smaller than in samples selected on the basis of the spectral shape in the gigahertz regime \citep[e.g.,][]{torniainen05,mo07,mingaliev12}. \\

\item The low-power CSOs from the sample presented in this paper fill a region of the power-size plane that is underpopulated. Their radio luminosity at 1.4 GHz is on average below 10$^{26}$ W/Hz, with the high-$z$ tail exceeding this value. Assuming a simplistic parametric model, we found that the majority of the sources have a jet power in the range 10$^{44 - 45}$ erg s$^{-1}$, making their evolutionary paths sensitive to jet parameters and the conditions of the surrounding environment.

\end{itemize}

Once the characterization of the sample is completed, the properties of low-power CSOs will be compared with those of low-power MSOs and FRI/FRII radio galaxies in order to investigate whether the former are the progenitors of the latter, and investigate the role of the environment in the source evolution. Moreover, by
analogy with what has been done for high-power radio sources,
observational results will be compared with the outcome of
dedicated RMHD simulations of low-power decelerating jets. This will allow us to study various evolutionary paths for low-power CSOs
and investigate the importance of jet deceleration and jet-ISM interaction.
Low-power radio sources represent the vast majority of the radio sources detected in recent radio surveys and will
represent the bulk of radio sources detected by forthcoming facilities, like the Square Kilometre Array and the next generation-VLA (e.g., Best et al.
2005). Therefore, understanding the physical properties of these objects, their interplay with the ambient medium, and
their evolutionary path is crucial for building a complete picture of the radio source phenomenon.


\begin{acknowledgements}

We thank the anonymous referee for reading the manuscript carefully and making valuable suggestions.
The VLBA is operated by the US 
National Radio Astronomy Observatory which is a facility of the National Science Foundation operated under cooperative agreement by Associated Universities, Inc. AIPS is produced and maintained by the National Radio Astronomy Observatory, a facility of the National Science Foundation operated under cooperative agreement by Associated Universities, Inc. CIRADA is funded by a grant from the Canada Foundation for Innovation 2017 Innovation Fund (Project 35999), as well as by the Provinces of Ontario, British Columbia, Alberta, Manitoba and Quebec. This work has made use of the NASA/IPAC Extragalactic Database (NED) which is operated by the JPL, California Institute of Technology, under contract with the National Aeronautics and Space Administration. 
M.O. and F.D. acknowledge financial support from INAF 2022 fundamental research programme ob. fun. 1.05.12.05.15.
This publication makes use of data products from the Wide-field Infrared Survey Explorer, which is a joint project of the University of California, Los Angeles, and the Jet Propulsion Laboratory/California Institute of Technology, funded by the National Aeronautics and Space Administration.
The Legacy Surveys consist of three individual and complementary projects: the Dark Energy Camera Legacy Survey (DECaLS; Proposal ID \#2014B-0404; PIs: David Schlegel and Arjun Dey), the Beijing-Arizona Sky Survey (BASS; NOAO Prop. ID \#2015A-0801; PIs: Zhou Xu and Xiaohui Fan), and the Mayall z-band Legacy Survey (MzLS; Prop. ID \#2016A-0453; PI: Arjun Dey). DECaLS, BASS and MzLS together include data obtained, respectively, at the Blanco telescope, Cerro Tololo Inter-American Observatory, NSF’s NOIRLab; the Bok telescope, Steward Observatory, University of Arizona; and the Mayall telescope, Kitt Peak National Observatory, NOIRLab. Pipeline processing and analyses of the data were supported by NOIRLab and the Lawrence Berkeley National Laboratory (LBNL). The Legacy Surveys project is honored to be permitted to conduct astronomical research on Iolkam Du’ag (Kitt Peak), a mountain with particular significance to the Tohono O’odham Nation.

NOIRLab is operated by the Association of Universities for Research in Astronomy (AURA) under a cooperative agreement with the National Science Foundation. LBNL is managed by the Regents of the University of California under contract to the U.S. Department of Energy.

This project used data obtained with the Dark Energy Camera (DECam), which was constructed by the Dark Energy Survey (DES) collaboration. Funding for the DES Projects has been provided by the U.S. Department of Energy, the U.S. National Science Foundation, the Ministry of Science and Education of Spain, the Science and Technology Facilities Council of the United Kingdom, the Higher Education Funding Council for England, the National Center for Supercomputing Applications at the University of Illinois at Urbana-Champaign, the Kavli Institute of Cosmological Physics at the University of Chicago, Center for Cosmology and Astro-Particle Physics at the Ohio State University, the Mitchell Institute for Fundamental Physics and Astronomy at Texas A\&M University, Financiadora de Estudos e Projetos, Fundacao Carlos Chagas Filho de Amparo, Financiadora de Estudos e Projetos, Fundacao Carlos Chagas Filho de Amparo a Pesquisa do Estado do Rio de Janeiro, Conselho Nacional de Desenvolvimento Cientifico e Tecnologico and the Ministerio da Ciencia, Tecnologia e Inovacao, the Deutsche Forschungsgemeinschaft and the Collaborating Institutions in the Dark Energy Survey. The Collaborating Institutions are Argonne National Laboratory, the University of California at Santa Cruz, the University of Cambridge, Centro de Investigaciones Energeticas, Medioambientales y Tecnologicas-Madrid, the University of Chicago, University College London, the DES-Brazil Consortium, the University of Edinburgh, the Eidgenossische Technische Hochschule (ETH) Zurich, Fermi National Accelerator Laboratory, the University of Illinois at Urbana-Champaign, the Institut de Ciencies de l’Espai (IEEC/CSIC), the Institut de Fisica d’Altes Energies, Lawrence Berkeley National Laboratory, the Ludwig Maximilians Universitat Munchen and the associated Excellence Cluster Universe, the University of Michigan, NSF’s NOIRLab, the University of Nottingham, the Ohio State University, the University of Pennsylvania, the University of Portsmouth, SLAC National Accelerator Laboratory, Stanford University, the University of Sussex, and Texas A\&M University.

BASS is a key project of the Telescope Access Program (TAP), which has been funded by the National Astronomical Observatories of China, the Chinese Academy of Sciences (the Strategic Priority Research Program “The Emergence of Cosmological Structures” Grant \# XDB09000000), and the Special Fund for Astronomy from the Ministry of Finance. The BASS is also supported by the External Cooperation Program of Chinese Academy of Sciences (Grant \# 114A11KYSB20160057), and Chinese National Natural Science Foundation (Grant \# 12120101003, \# 11433005).

The Legacy Survey team makes use of data products from the Near-Earth Object Wide-field Infrared Survey Explorer (NEOWISE), which is a project of the Jet Propulsion Laboratory/California Institute of Technology. NEOWISE is funded by the National Aeronautics and Space Administration.

The Legacy Surveys imaging of the DESI footprint is supported by the Director, Office of Science, Office of High Energy Physics of the U.S. Department of Energy under Contract No. DE-AC02-05CH1123, by the National Energy Research Scientific Computing Center, a DOE Office of Science User Facility under the same contract; and by the U.S. National Science Foundation, Division of Astronomical Sciences under Contract No. AST-0950945 to NOAO. The Photometric Redshifts for the Legacy Surveys (PRLS) catalog used in this paper was produced thanks to funding from the U.S. Department of Energy Office of Science, Office of High Energy Physics via grant DE-SC0007914. Funding for the Sloan Digital Sky Survey IV has been provided by the Alfred P. Sloan Foundation, the U.S. 
Department of Energy Office of Science, and the Participating Institutions. SDSS-IV acknowledges support and resources from the Center for High Performance Computing at the University of Utah. The SDSS website is www.sdss4.org. SDSS-IV is managed by the Astrophysical Research Consortium for the Participating Institutions of the SDSS Collaboration including the Brazilian Participation Group, the Carnegie Institution for Science, Carnegie Mellon University, Center for Astrophysics | Harvard \& Smithsonian, the Chilean Participation Group, the French Participation Group, Instituto de Astrof\'isica de Canarias, The Johns Hopkins University, Kavli Institute for the Physics and Mathematics of the 
Universe (IPMU) / University of Tokyo, the Korean Participation Group, Lawrence Berkeley National Laboratory, Leibniz Institut f\"ur Astrophysik Potsdam (AIP),  Max-Planck-Institut f\"ur Astronomie (MPIA Heidelberg), Max-Planck-Institut f\"ur Astrophysik (MPA Garching), Max-Planck-Institut f\"ur Extraterrestrische Physik (MPE), National Astronomical Observatories of China, New Mexico State University, 
New York University, University of Notre Dame, Observat\'ario Nacional / MCTI, The Ohio State University, Pennsylvania State University, Shanghai Astronomical Observatory, United Kingdom Participation Group, 
Universidad Nacional Aut\'onoma de M\'exico, University of Arizona, University of Colorado Boulder, 
University of Oxford, University of Portsmouth, University of Utah, University of Virginia, University 
of Washington, University of Wisconsin, Vanderbilt University, and Yale University.

\end{acknowledgements}

\begin{appendix}

\onecolumn
\section{Tables}

\begin{table*}[h!]
\caption{Sample of candidate low-power CSOs.}
\begin{center}
\begin{tabular}{ccccccccccc}
 \hline
Name & mJIVE-20 name& Id. & m$_{\rm r}$ & z & m$_{\rm W1}$ &S$_{\rm p, 1.38}$ & S$_{\rm FIRST}$ & S$_{\rm VLASS}$ & $\alpha^{3.0}_{1.38}$ &LAS\\
   & MJV &  & mag & & mag &mJy/b& mJy & mJy& & mas \\
\hline
&&&&&&&\\
J0802+6125 & 00141 & s & 24.37 & 1.129$\pm$0.151 & 17.52 & 5.7 & 14.6 & 7.8 & 0.8 & 28 \\
J0810+5011 & 01651 & G & 22.13 & 1.053$\pm$0.434 & 16.20 & 3.8 & 8.1 & 11.4 & $-$0.5 & 20 \\ 
J0810+4117 & 09498 & G & 23.38& 0.957$\pm$0.158 & 16.21 & 4.9 & 13.6 & 9.3 & 0.5 & 14 \\
J0816+3616 & 04053 & E & & & & 4.7 & 8.8 & 5.5 & 0.6 & 17 \\
{\bf J0834+4111}& 19180 & G & 24.28 & 1.137$\pm$0.142 & 16.22 & 39.1 & 100.6 & 56.0& 0.7 & 22 \\
{\bf J0855+2030}& 01713 & G &18.16& 0.251$\pm$0.011 & 14.83 & 11.8 & 25.1 & 14.0 & 0.7 & 16 \\ 
J0857+2128 & 22448 & G & 21.33 & 0.857$\pm$0.055 & 16.03 & 3.2 & 12.1 & 6.2 & 0.9 & 50 \\
J0901+4256 & 17295 & Q & 19.31 & 1.011sp & 15.19 & 10.9 & 24.5 & 15.5 & 0.6 & 20 \\
J0902+4703 & 09733 & s & 23.85 & 1.020$\pm$0.155& & 14.8 & 23.6 & 19.4& 0.3 & 25 \\
J0917+3351 & 16793 & E &  & & 17.30 & 8.0 & 25.8 & 14.2 & 0.8 & 34 \\
J0924+4035 & 14046 & G & 24.19& 1.112$\pm$0.239& 17.29 & 3.4 & 22.7 & 10.0 & 1.1 & 34 \\
J0941+2600 & 04207 & G & 24.09 & 1.224$\pm$0.331 & 16.97 & 2.3 & 9.9 & 4.8 & 0.9 & 20 \\
{\bf J0956+2529}& 11834 & s & 24.88& 1.248$\pm$0.375 & 17.09 & 16.9 & 47.1 & 23.0 & 0.9 & 26\\
J0957+2541 & 11874 & E & & & & 99.0 & 142.5 & 96.4 & 0.5 & 26 \\
{\bf J0958+3208}& 04323 & G & 23.65& 1.090$\pm$0.465 & 17.33 & 37.8 & 60.5 & 34.8 & 0.7 & 25 \\
{\bf J1008+1328}& 17434 & E & &  &  & 9.1 & 17.1 & 11 & 0.6 & 22 \\
J1017+3532 & 08253  & G & 19.68 & 0.337sp & 15.47 & 19.0 & 28.4 & 17.0 & 0.7 & 20 \\
J1022+3459 & 15244 & Q & 19.33 & 1.935sp & 15.17 & 3.0 & 6.6 & 4.1 & 0.6 & 11 \\
J1037+0514 & 12154 & Q & 20.59 & 2.0995sp & 16.38 & 5.0 & 16.5 & 8.0 & 0.9 &  37 \\
J1041+5344 & 13694 & G & 22.40 & 1.000$\pm$0.130 & 16.36 & 4.8 & 16.4 & 9.9 & 0.6 & 26 \\  
J1102+3815 & 06949 & E & & &  17.40 & 3.1 & 6.1 & 2.9 & 1.0 & 11 \\
{\bf J1107+1619}& 19609 & E & & & & 39.1 & 96.4 & 45.0 & 1.0 & 19 \\
J1107+4346 & 05578 & E & & & & 6.2 & 10.5 & 6.5 & 0.6 & 11 \\
J1112+1445 & 08326 & E & & & & 2.8 & 6.4 & 4.5 & 0.5 & 30 \\
{\bf J1120$-$0607}& 10165 & E & & & & 62.3 & 122.4 & 89.0 & 0.4 & 14 \\
{\bf J1127+2540}& 07057 & E & & & 17.23 &19.7 & 65.5 & 31.0 & 1.0 & 55 \\
J1201$-$0503 & 12448 & G & 22.23& 0.752$\pm$0.054 & 16.24 & 3.8 & 10.8 & 5.2 & 0.9 & 23 \\
{\bf J1206+2821}& 13722 & E & & & &  17.6 & 56.2 & 27.0 & 1.0 & 30 \\
J1209+4115 & 00873 & G &18.1 & 0.095sp & 13.18 & 68.6 & 146.8 & 120.1 & 0.3 & 25 \\
{\bf J1214+3438} & 15828 & E & &  & 17.61 & 48.3 & 78.4 & 30.0 & 1.3 & 18 \\
J1217$-$0049 & 20203 & G & 23.38 & 0.887$\pm$0.069 & &  8.0 & 13.7 & 7.3 & 0.8 & 23 \\
J1218+3813 & 20308 & E & & & &  6.0 & 12.1 & 7.1 & 0.7 & 20 \\
{\bf J1219+5845} & 12583 & E & & & & 11.1 & 17.5 & 14.0 & 0.3 & 20 \\
{\bf J1230+3728}& 07332 & G & 21.89& 0.479$\pm$0.093 & 17.26 & 42.6 & 91.0 & 46.0 & 0.9 & 21 \\
{\bf J1239+3713}& 20358 & G & 24.83 & 0.984$\pm$0.219& 16.08 & 10.0 & 21.4 & 11.0 & 0.9 & 25 \\
J1242+3709 & 20362 & E & & & & 4.6 & 8.4 & 5.3 & 0.6 & 9 \\
J1254+1156 & 07387 & G & 20.47 & 0.591$\pm$0.067 & 15.59 & 2.8 & 7.7 & 3.6 & 1.0 & 28 \\ 
J1257+3244 & 04612 & G & 22.38 & 0.792$\pm$0.073 & 16.67 & 2.6 & 8.1 & 5.0 & 0.6 & 37 \\
J1258+3252 & 04637 & G & 19.68 & 0.559$\pm$0.031 & 15.10 & 3.8 & 7.4 & 5.5 & 0.4 & 20 \\ 
J1310+1204 & 15395 & E & & & & 7.0 & 12.8 & 6.0 & 1.0 & 25 \\    
J1311+3221 & 02847 & G & 23.49 & 0.960$\pm$0.105 & 17.01 & 5.6 & 16.3 & 11.3 & 0.5 & 20 \\
J1334+1623 & 20701 & G & 23.91& & 16.78 & 16.0 & 26.2 & 15.2 & 0.7 & 28 \\
J1337+4603 & 06117 & E & & & 17.77 & 12.7 & 27.6 & 14.4 & 0.8 & 11 \\
J1343+2730 & 14513 & E & & & & 8.0 & 15.5 & 6.8 & 1.1 & 11\\
J1357+1922 & 10688 & E & & & & 7.0 & 11.5 & 8.7 & 0.4 & 15 \\
&&&&&&\\
\hline
\end{tabular}
\end{center}
\label{sample-table}
\tablefoot{Column 1: Source name; Column 2: mJIVE-20 identifier; Column 3: optical identification: G = galaxy; Q = quasar; E = empty field in optical images; s = stellar object. Column 4: optical $r$-band ab magnitude from the DESI Legacy Surveys. A * indicates z-band magnitude; Column 5: photometric redshift from DESI Legacy Surveys Data Release 9 \citep{zhou23}. An sp denotes a spectroscopic redshift from SDSS data; Column 6: W1-band (3.4$\mu$m) VEGA magnitude from WISE;
Column 7: VLBI peak flux density (mJy beam$^{-1}$); Columns 8 and 9: total flux density (mJy), at 1.4 GHz from FIRST and 3.0 GHz from VLASS data, respectively; Column 10: spectral index between 1.4 GHz (from FIRST data) and 3.0 GHz (from VLASS); Column 11: largest angular scale. The sources discussed in this paper are marked in bold.}
\end{table*}

\addtocounter{table}{-1}
\begin{table*}
\caption{continued.}
\begin{center}
\begin{tabular}{ccccccccccc}
 \hline
Name & mJIVE-20 name & Id. & m$_{\rm r}$ & $z$ & m$_{\rm W1}$ &S$_{\rm p, 1.38}$ & S$_{\rm FIRST}$ & S$_{\rm VLASS}$ & $\alpha^{3.0}_{1.38}$ &LAS\\
  &MJV    &  & mag & & mag &mJy/b& mJy & mJy& & mas \\
\hline
&&&&&&&\\
{\bf J1431+5427}& 10812 & E & & & 17.85 & 12.3 & 24.9 & 14.0 & 0.7 & 22 \\
{\bf J1500+4507}& 21436 & E & & & & 18.2 & 38.5 & 17.8 & 1.0 & 30 \\
J1514+2318 & 21003 & G & 21.70 & 0.921$\pm$0.257 & 14.92 & 2.0 & 4.3 & 2.7 & 0.6 & 11 \\
{\bf J155441134}& 24364 & E & & & & 9.0 & 30.2 & 13.3 & 1.1 & 26 \\
{\bf J1600+3304}& 07918 & E & & & & 71.8 &  169.7 & 107.0 & 0.6 & 28 \\
{\bf J1603+2944}& 16889 & G & 23.54 & 1.058$\pm$0.111 & 16.67 & 32.0 & 47.5 & 27.0 & 0.7 & 23 \\
{\bf J1603+3317}& 07954 & G & 22.81$^{*}$& 0.939$\pm$0.318 & & 13.5 & 30.1 & 13.0 & 1.1 & 19 \\
{\bf J1604+1758}& 21587 & G & 22.15 & 0.821$\pm$0.103 & & 15.4 & 34.9 & 23.0 & 0.5 & 19 \\
J1607+3128 & 11252 & E & & & & 16.0 & 23.4 & 16.5 & 0.4 & 30 \\
J1607+3011 & 16956 & G & 22.28 & 0.828$\pm$0.205 & & 3.9 & 10.7 & 7.4 & 0.5 & 11 \\
J1610+1907 & 04895 & E & & & & 5.5 & 10.1 & 3.7 & 1.3 & 20 \\
J1621+5717 & 13091 & E & & & & 6.0 & 14.2 & 8.0 & 0.7 & 15 \\
J1654+3952 & 18493 & E & & & & 5.7 & 21.9 & 11.9 & 0.8 & 23 \\
J2352$-$0004 & 03886 & G & 24.87 & & 17.52 & 5.0 & 13.5 & 8.1 & 0.7 & 50 \\
J2353$-$0022 & 03850 & G & 24.52 & 1.429$\pm$0.290 & & 3.5 & 8.5 & 4.7 & 0.8 & 9 \\
&&&&&&&\\
\hline
\end{tabular}
\end{center}
\end{table*}

\begin{table}
\caption{Log of VLBA observations.}
\begin{center}
\begin{tabular}{cccc}
\hline
\hline
Code & Obs. date & BP \& FR cal. & Phase cal. \\
\hline
&&&\\
BO064 A & 2021-09-04 & J0555+3948    & J0854+2006 \\
       &            & J0927+3902 & J0836+4125 \\
BO064 B & 2021-10-10 & J0555+3948    & J0958+3224 \\
       &            & J0927+3902 & J0956+2515 \\
       &            &            & J1007+1356 \\
BO064 C & 2021-08-31 & J0927+3902 & J1125+2610 \\
       &            &            & J1121-0553 \\
       &            &            & J1107+1628 \\
BO064 D & 2021-08-29 & J0927+3902 & J1159+2914 \\
       &            &            & J1215+3448 \\
BO064 E & 2021-12-09 & J0927+3902 & J1228+3706 \\
       &            &            & J1217+5835 \\
BO064 F & 2021-08-20 & J0927+3902 & J1429+5406 \\
       &            & J1642+3948 & J1455+4431 \\
BO064 G & 2021-09-21 & J1642+3848 & J1558+3323 \\
       &            &            & J1555+1111 \\
BO064 H & 2021-09-20 & J1642+3948 & J1606+3124 \\
       &            & J1800+3848 & J1606+1814 \\
&&&\\
\hline
\end{tabular}
\end{center}
\label{log-table}
\end{table}

\begin{table*}
\caption{Observational parameters of the subsample of candidate CSOs.}
\begin{center}
\begin{tabular}{ccccccccccc}
\hline
\hline
Source & Code& Morph. & comp. & S$_{\rm 1.38}$ &  S$_{\rm 4.98}$ & $\alpha_{\rm F}$ & $\alpha_{\rm L}$ & $\theta_{\rm maj}$ & $\theta_{\rm min}$ & PA\\
     & & & & mJy & mJy & &  & mas & mas & deg\\
\hline
J0834+4111 & A & CSO & N & 27.3$\pm$2.7 & 8.8$\pm$0.5 & 0.9$\pm$0.1 &0.9$\pm$0.1 & 3.77$\pm$0.05 & 2.28$\pm$0.04 & 177$\pm$1 \\
           &     &  & S & 63.6$\pm$6.4 & 18.9$\pm$1.0& 0.9$\pm$0.1 &0.9$\pm$0.1 & 5.38$\pm$0.03 & 3.34$\pm$0.02 & 147$\pm$1 \\ 
J0855+2030 & A & CSO & E &  8.6$\pm$0.9 &  4.4$\pm$0.3& 0.5$\pm$0.1 &0.4$\pm$0.1 & 3.35$\pm$0.05 & 2.86$\pm$0.13 & 70$\pm$10\\
           &     &  & W & 14.1$\pm$1.4 &  5.4$\pm$0.3& 0.7$\pm$0.1 &0.6$\pm$0.1 & 3.41$\pm$0.05 & 2.41$\pm$0.11 & 88$\pm$4 \\
J0956+2529 & B & CJ  & W &  37.8$\pm$4.0& 10.8$\pm$0.6& 1.0$\pm$0.1 &0.3$\pm$0.1 & 8.63$\pm$0.18 & 3.12$\pm$0.12 & 77$\pm$1\\
           &      & & E &   9.5$\pm$1.0&            & 0.7$\pm$0.1 &0.1$\pm$0.1 & & & \\   
           &      & & E1&              & 1.7$\pm$0.1 &   & & 0.92$\pm$0.13 & $<$0.3 & 155$\pm$7  \\
           &      & & E2&              & 0.8$\pm$0.1 &   & & 1.08$\pm$0.50 & $<$0.3 & - \\
           &      & & E3 &             & 0.7$\pm$0.1&   &  & 2.0$\pm$0.23 & $<$0.3 & 170$\pm$4\\
J0958+3208 & B & CJ?   & W &   8.2$\pm$0.8& 0.8$\pm$0.1 & 1.8$\pm$0.1 &1.2$\pm$0.1 & 2.35$\pm$0.23 & $<$0.2 & 140$\pm$6\\ 
           &      & & E &  52.7$\pm$5.3&             &  0.8$\pm$0.1 &0.6$\pm$0.1 & & \\
           &      & & E1&              & 10.9$\pm$0.6& & & 1.97$\pm$0.02 & 1.75$\pm$0.02 & 21$\pm$5\\
           &      & & E2&              & 6.6$\pm$0.4 & & & 1.16$\pm$0.02 & 0.65$\pm$0.05 & 95$\pm$2\\
J1008+1328 & B & CSO & N &  8.9$\pm$1.0 & 3.2$\pm$0.3 & 0.8$\pm$0.1 & 0.7$\pm$0.1 & 2.18$\pm$0.15 & 1.45$\pm$0.15 & 140$\pm$9\\  
           &      & & C &  2.0$\pm$0.2 & 0.9$\pm$0.1 & 0.6$\pm$0.1 & $-$0.2$\pm$0.1 & 3.04$\pm$0.34 & $<$0.5 & 150$\pm$6\\
           &      & & S &  5.3$\pm$0.5 & 1.5$\pm$0.2 & 1.0$\pm$0.1 & 0.7$\pm$0.1 & 1.83$\pm$0.25 & 0.80$^{+0.20}_{-0.6}$ & 138$\pm$12\\
J1107+1619 & C & CSO  & E & 37.6$\pm$4.0 & 8.7$\pm$0.9 & 1.1$\pm$0.1 &1.1$\pm$0.1 & 2.96$\pm$0.02 & 0.98$\pm$0.05 & 108$\pm$1\\  
           &      & & W & 51.4$\pm$5.2 &12.3$\pm$1.3 & 1.1$\pm$0.1 &1.1$\pm$0.1 & 1.72$\pm$0.01 & 0.72$\pm$0.04 & 63$\pm$1\\
J1120$-$0607& C & CJ  & E & 67.3$\pm$7.0 & 12.0$\pm$1.2 & 1.3$\pm$0.1 &1.3$\pm$0.1 & 1.00$\pm$0.02 & 0.50$\pm$0.02 & 35$\pm$1 \\
           &      & & W & 49.8$\pm$5.0 & 38.9$\pm$4.1& 0.2$\pm$0.1 &0.2$\pm$0.1 & 0.80$\pm$0.02 & 0.43$\pm$0.01 & 73$\pm$1\\
J1127+2540$^{a}$ & C & CJ?   & N & 29.4$\pm$3.0 &  7.4$\pm$0.8& 1.1$\pm$0.1 &1.1$\pm$0.1 & 1.39$\pm$0.02 & 0.93$\pm$0.03 & 126$\pm$2\\  
           &      & & S & 34.2$\pm$3.4 &             &     & & 14.32$\pm$0.50 & 10.82$\pm$0.42 & 162$\pm$6      \\ 
J1206+2821 & D & CSO  & N & 25.2$\pm$2.7 &  7.3$\pm$0.4 & 1.0$\pm$0.1 &1.0$\pm$0.1 & 1.88$\pm$0.02 & 0.68$\pm$0.03 & 143$\pm$1\\
           &      & & S & 24.2$\pm$2.5 &  3.8$\pm$0.2 & 1.4$\pm$0.1 &1.4$\pm$0.1 & 2.08$\pm$0.05 & 1.27$\pm$0.09 & 118$\pm$3\\
J1214+3438 & D & CSO  & E & 54.3$\pm$5.5 &  5.9$\pm$0.3 & 1.7$\pm$0.1 &1.6$\pm$0.1 & 1.82$\pm$0.02 & 1.28$\pm$0.06 & 64$\pm$3\\
           &      & & C &              &  2.0$\pm$0.1 &   & & $<$0.6 & $<$0.3 & 130$\pm$12          \\
           &      & & W & 19.0$\pm$2.0 &  5.2$\pm$0.3 & 1.0$\pm$0.1 & 1.0$\pm$0.1 & 0.96$\pm$0.04 & 0.64$\pm$0.04 & 23$\pm$6\\
J1219+5845 & E & CJ  & N &  5.8$\pm$0.6 &  6.8$\pm$0.4 & $-$0.1$\pm$0.1 & $-$0.1$\pm$0.1 & 0.82$\pm$0.06 & 0.22$\pm$0.02 & 178$\pm$2\\ 
           &      & & S & 12.3$\pm$1.3 &  1.7$\pm$0.2 & 1.5$\pm$0.1 & 1.5$\pm$0.1 & 1.54$\pm$0.15 & 0.54$^{+0.14}_{-0.09}$ & 15$\pm$6\\
J1230+3728 & E & CSO  & E & 45.1$\pm$4.5 & 11.3$\pm$0.6 & 1.1$\pm$0.1 & 1.1$\pm$0.1 & 3.50$\pm$0.02 & 1.41$\pm$0.04 & 65$\pm$1\\
           &      & & W & 50.8$\pm$5.1 & 12.5$\pm$0.7 & 1.1$\pm$0.1 & 1.1$\pm$0.1 & 2.86$\pm$0.02 & 2.16$\pm$0.03 & 67$\pm$1\\
J1239+3713 & E & CSO  & E & 12.1$\pm$1.2 &  4.4$\pm$0.3 & 0.8$\pm$0.1 & 0.8$\pm$0.1 & 4.30$\pm$0.10 & 2.56$\pm$0.06 & 172$\pm$2\\
           &      & & C &              &  0.6$\pm$0.1 & & & 2.28$\pm$0.46 & $<$0.7 & 125$\pm$15\\
           &      & & W & 10.1$\pm$1.0 &  3.6$\pm$0.2 & 0.8$\pm$0.1 & 0.8$\pm$0.1 & 2.37$\pm$0.09 & 1.93$\pm$0.11 & 21$\pm$8\\
J1431+5427 & F & CJ? & E & 12.8$\pm$1.3 &  3.3$\pm$0.2 & 1.1$\pm$0.1 & 1.1$\pm$0.1 & 3.16$\pm$0.11 & 1.31$\pm$0.25 & 77$\pm$3\\ 
           &      & & W & 10.7$\pm$1.1 &  0.8$\pm$0.1 & 2.0$\pm$0.1 & 2.0$\pm$0.1 & 1.86$\pm$0.44 & 1.13$^{+0.19}_{-0.43}$ & 177$\pm$20 \\
J1500+4507 & F & CJ?   & N & 22.4$\pm$2.5 &  1.0$\pm$0.1 &2.4$\pm$0.1 & 2.3$\pm$0.1 & 1.50$\pm$0.25 & 0.88$\pm$0.17 & 166$\pm$15\\
           &      & & S & 18.0$\pm$1.8 &                & 1.0$\pm$0.1 &1.0$\pm$0.1 \\
           &      & & S1 &             &  3.4$\pm$ 0.2 &   & & 1.36$\pm$0.06 & 0.76$\pm$0.06 & 157$\pm$3\\
           &      & & S2 &             &  1.2$\pm$0.1  &   &  & 2.80$\pm$0.17 & 0.82$\pm$0.06 & 4$\pm$2\\
           &      & & S3 &             &  0.5$\pm$0.1  &  &  & 2.90$\pm$0.40 & $<$0.5 & 11$\pm$5 \\
J1554+1134 & G & CSO  & N & 12.0$\pm$1.2 &  1.7$\pm$0.1 & 1.5$\pm$0.1 &1.4$\pm$0.1 & 0.54$\pm$0.13 & $<$0.1 & 143$\pm$13\\
           &      & & C &  3.8$\pm$0.4 &  1.1$\pm$0.1 & 1.0$\pm$0.1 &0.9$\pm$0.1 & 1.95$\pm$0.11 & 0.75$\pm$0.12 & 16$\pm$4\\
           &      & & S &  9.2$\pm$1.0 &  1.8$\pm$0.1 & 1.3$\pm$0.1 & 1.2$\pm$0.1 & $<$0.5 & $<$0.4 & - \\
J1600+3304 & G & CSO  & E & 86.0$\pm$9.0 & 36.9$\pm$1.9 & 0.7$\pm$0.1 & 0.6$\pm$0.1 & 2.51$\pm$0.02 & 1.16$\pm$0.06 & 44$\pm$1 \\ 
           &      & & W & 86.2$\pm$9.0 & 27.4$\pm$1.4 & 0.9$\pm$0.1 & 0.9$\pm$0.1 & 2.88$\pm$0.02 & 1.44$\pm$0.02 & 51$\pm$1\\
J1603+2944 & H & CSO  & E & 17.4$\pm$1.5 &  5.1$\pm$0.3 & 1.0$\pm$0.1 & 1.0$\pm$0.1 & 4.24$\pm$0.05 & 1.45$\pm$0.03 & 27$\pm$1\\
           &      & & W & 35.9$\pm$3.6 & 12.1$\pm$0.6 & 0.8$\pm$0.1 & 0.8$\pm$0.1 & 2.74$\pm$0.03 & 2.07$\pm$0.03 & 68$\pm$2\\
J1603+3317 & G & CSO  & N & 13.3$\pm$1.4 &  0.8$\pm$0.1 & 2.2$\pm$0.1 & 1.8$\pm$0.1 & 1.79$\pm$0.14 & 1.23$\pm$0.23 & 55$\pm$12\\
           &      & & S & 15.8$\pm$1.6 &  3.7$\pm$0.2 & 1.1$\pm$0.1 & 1.0$\pm$0.1 & 3.06$\pm$0.06 & 1.17$\pm$0.04 & 174$\pm$1\\
J1604+1758 & H & CJ   & E & 15.3$\pm$1.6 &  9.6$\pm$0.5 & 0.4$\pm$0.1 & 0.4$\pm$0.1 & 1.49$\pm$0.02 & 1.19$\pm$0.01 & 4$\pm$1\\
           &      & & W & 18.9$\pm$2.0 &  6.0$\pm$0.3 & 0.9$\pm$0.1 & 0.9$\pm$0.1 & 2.94$\pm$0.02 & 1.24$\pm$0.02 & 31$\pm$1\\
\hline           
\end{tabular}
\end{center}
\label{flux-table}
\tablefoot{Column 1: Source name. Column 2: Code of observing run, from Table \ref{log-table}. Column 3: Radio morphology. CSO = two-sided compact symmetric object; CJ= one-sided core-jet structure. A ? indicates that the classification of the radio structure is still tentative (see Sect. \ref{sec-notes}). Column 4: source component. Columns 5 and 6: flux density at 1.38 and 4.98 GHz, respectively. Columns 7 and 8: spectral index between 1.38 and 4.98 GHz computed using full-resolution images, and low-resolution images produced with visibility range, pixel size, and beam size common to both frequencies, respectively. Columns 9 and 10: deconvolved major and minor axis, respectively, measured from the full-resolution image at 4.98 GHz. Column 11: position angle of the major axis. $^a$: the deconvolved major and minor axes, and the position angle are measured from the full-resolution image at 1.38 GHz.} 
\end{table*}

\onecolumn
\section{VLBA images of the target sources}

Full-resolution VLBA images at 1.38 and 4.98 GHz of the target sources are presented in Fig. \ref{vlba-fig}. Source name, observing frequency, peak brightness, and the first contour (f.c.) are reported on each image. The first contour corresponds to three times the off-source noise level measured on the image plane. Contours increase by a factor of 2. The restoring beam is plotted in the bottom left-hand corner of each image. Images were produced with the AIPS task {\tt KNTR}. \\

\twocolumn

\begin{figure*}[h!]
\begin{center}
\includegraphics[width=0.60\columnwidth]{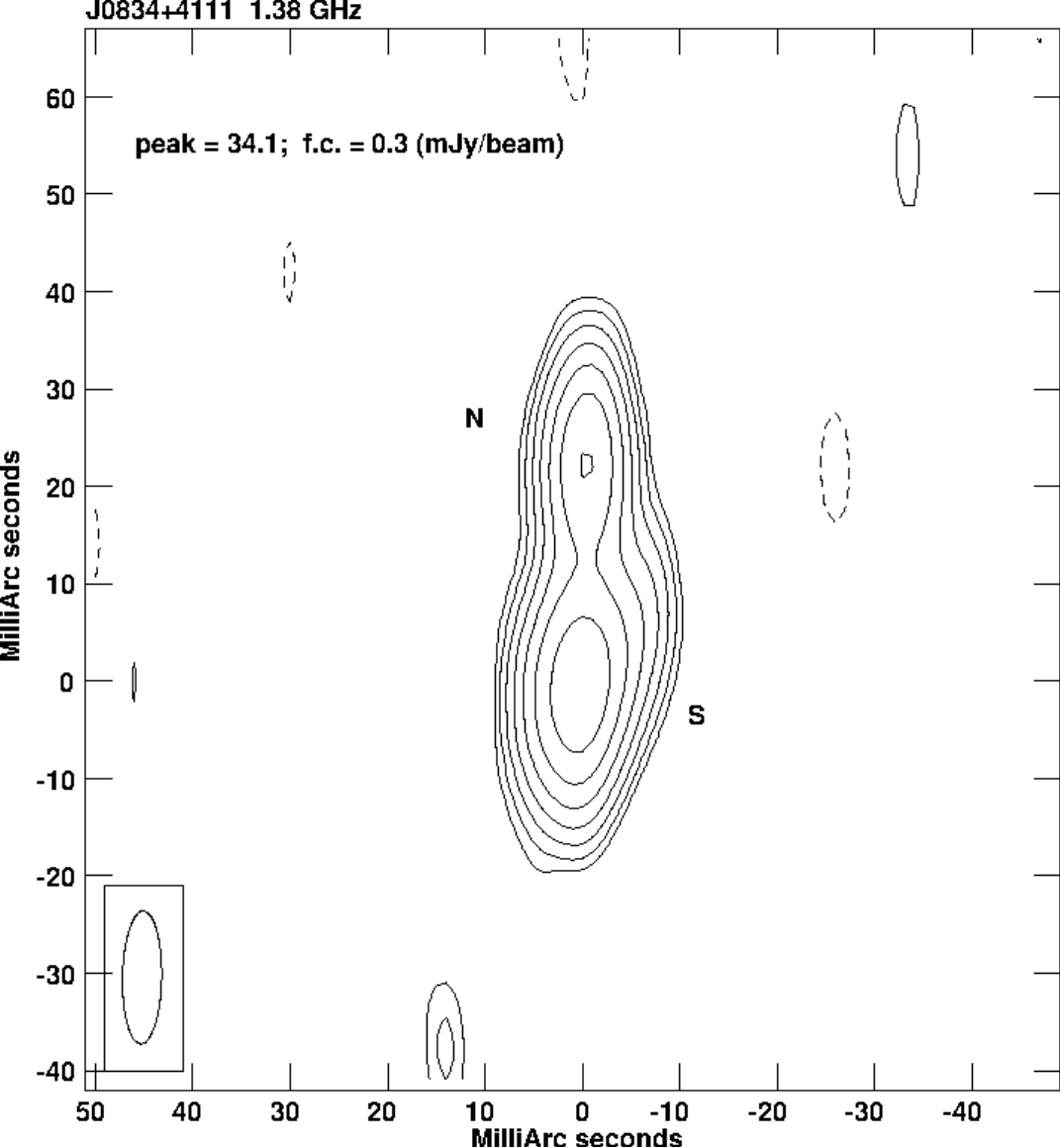}
\includegraphics[width=0.60\columnwidth]{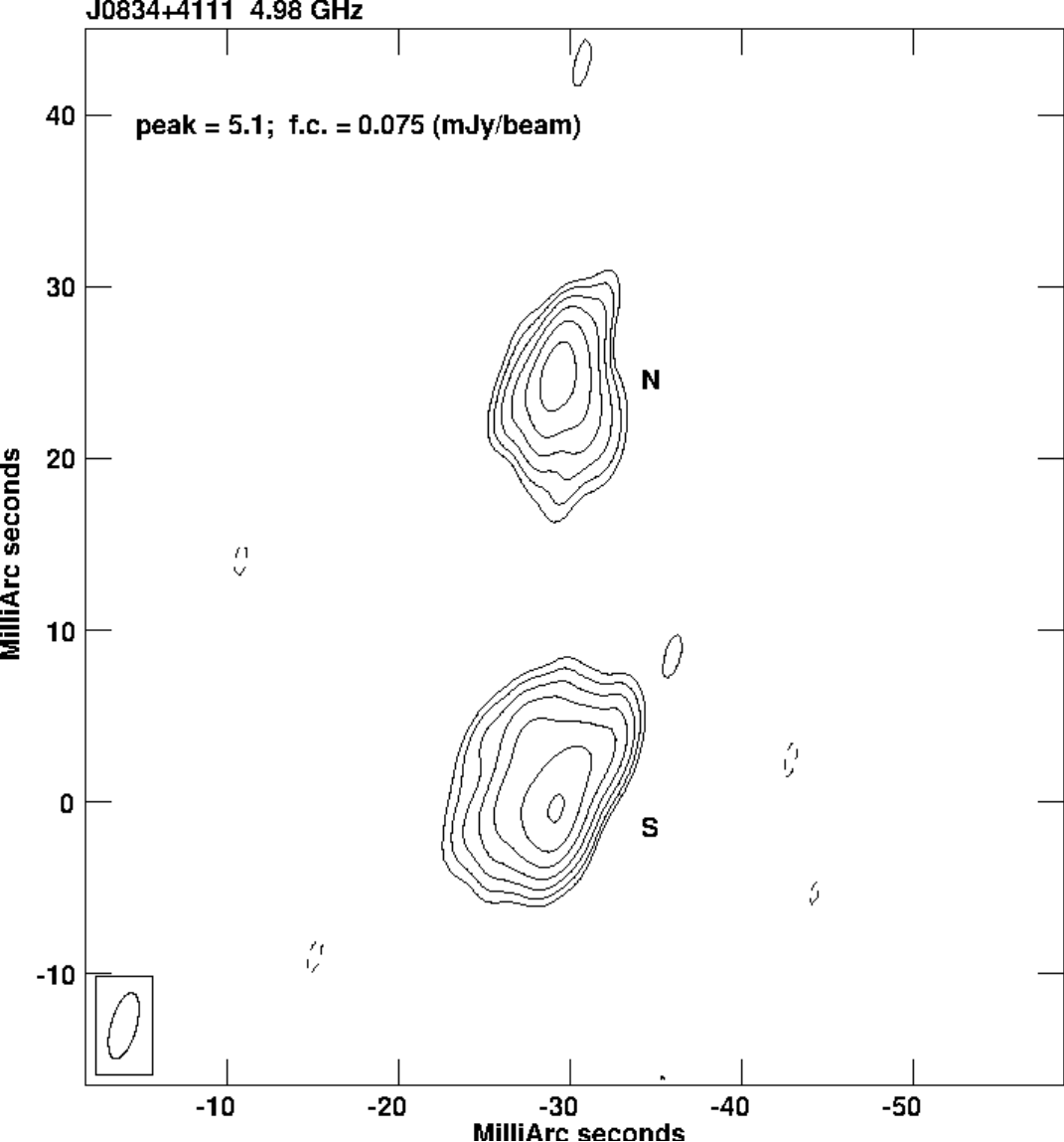}\\
\includegraphics[width=0.60\columnwidth]{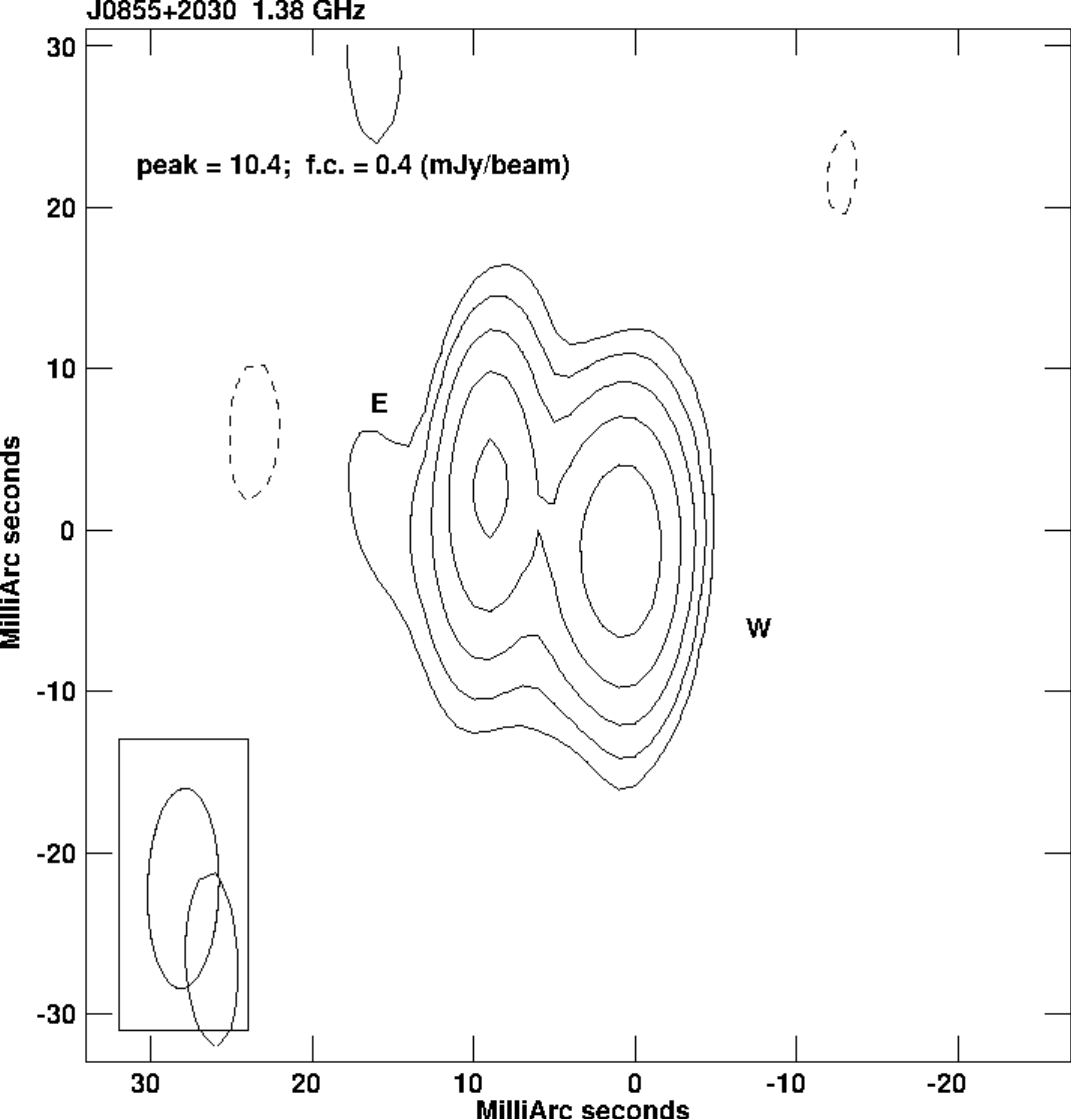}
\includegraphics[width=0.60\columnwidth]{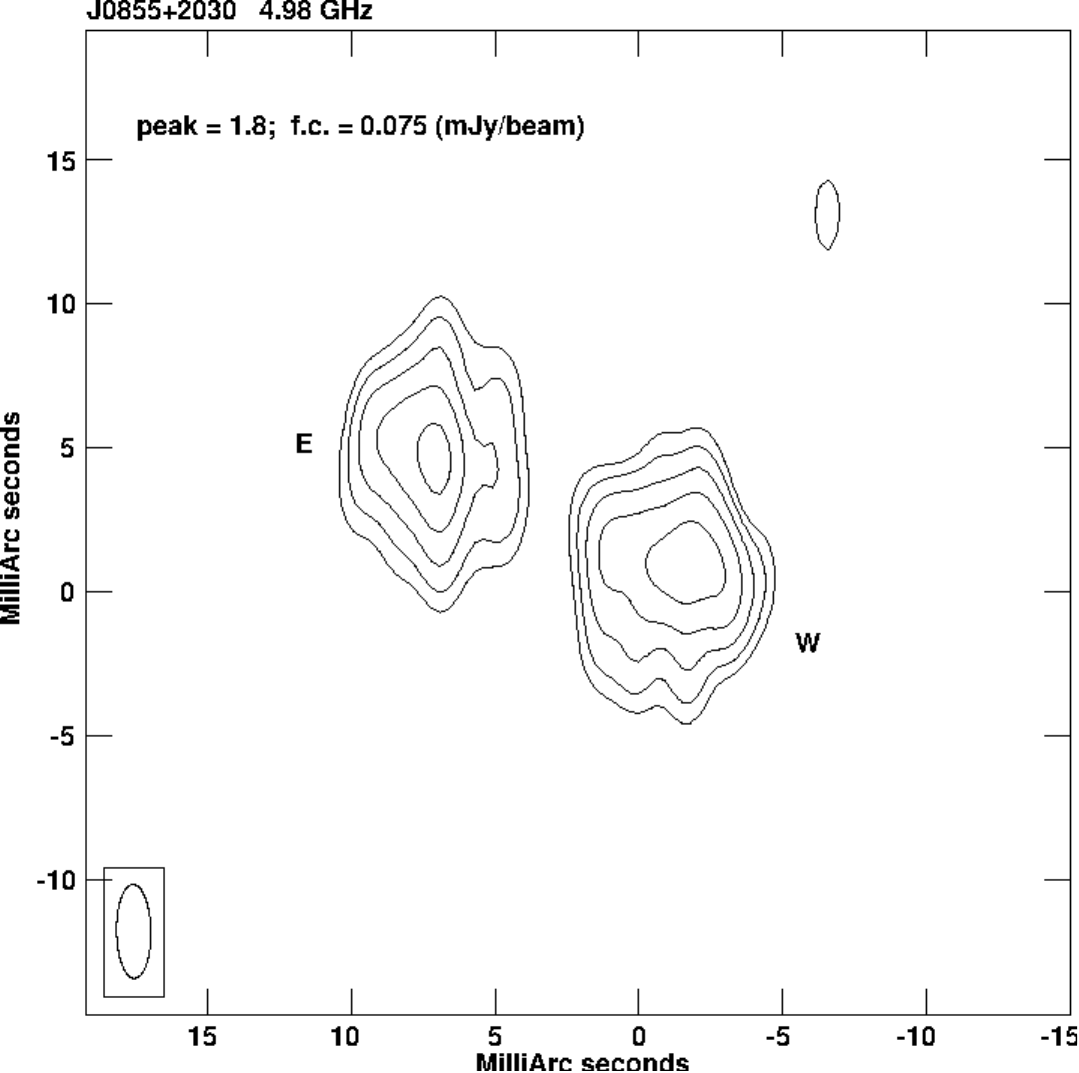}\\
\includegraphics[width=0.60\columnwidth]{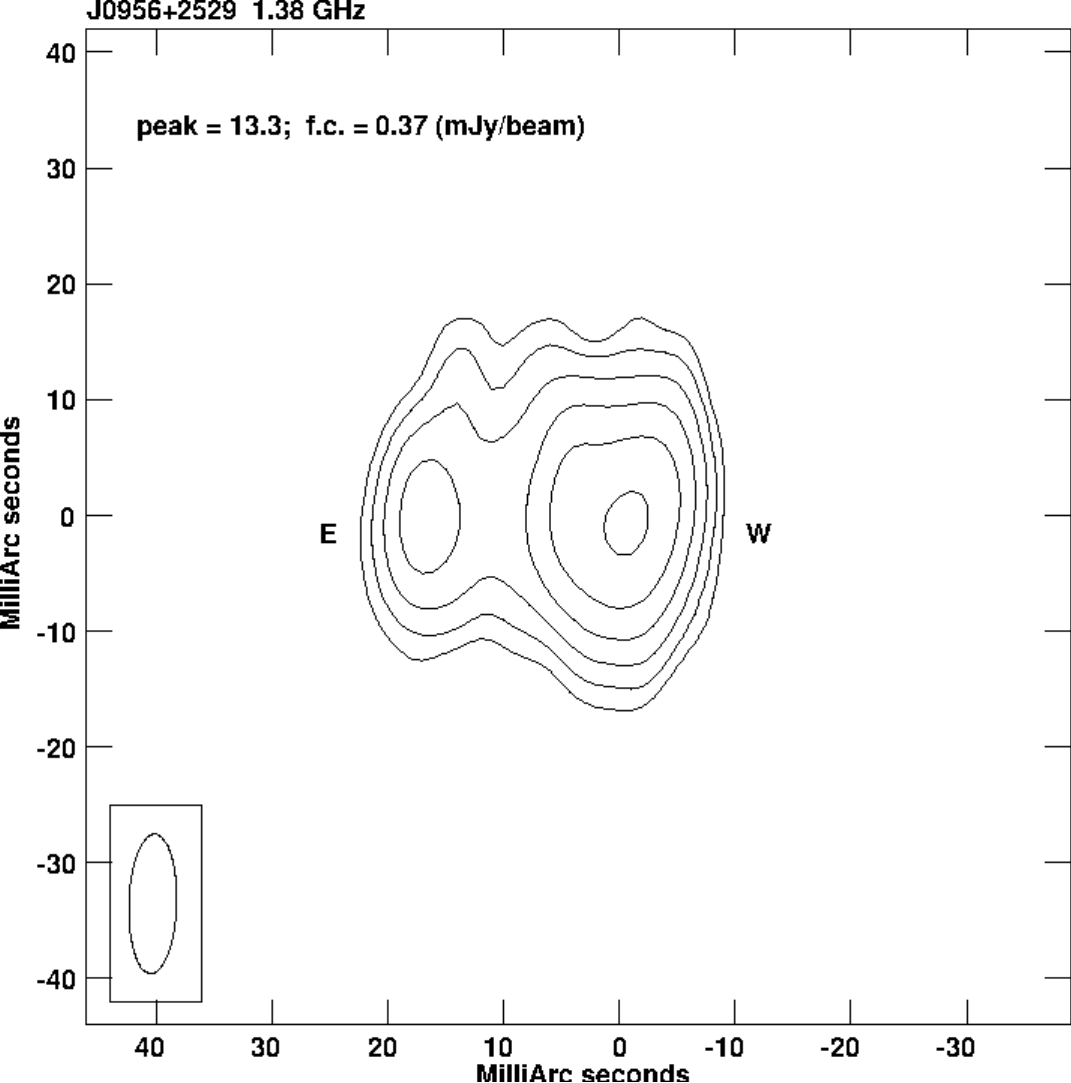}
\includegraphics[width=0.60\columnwidth]{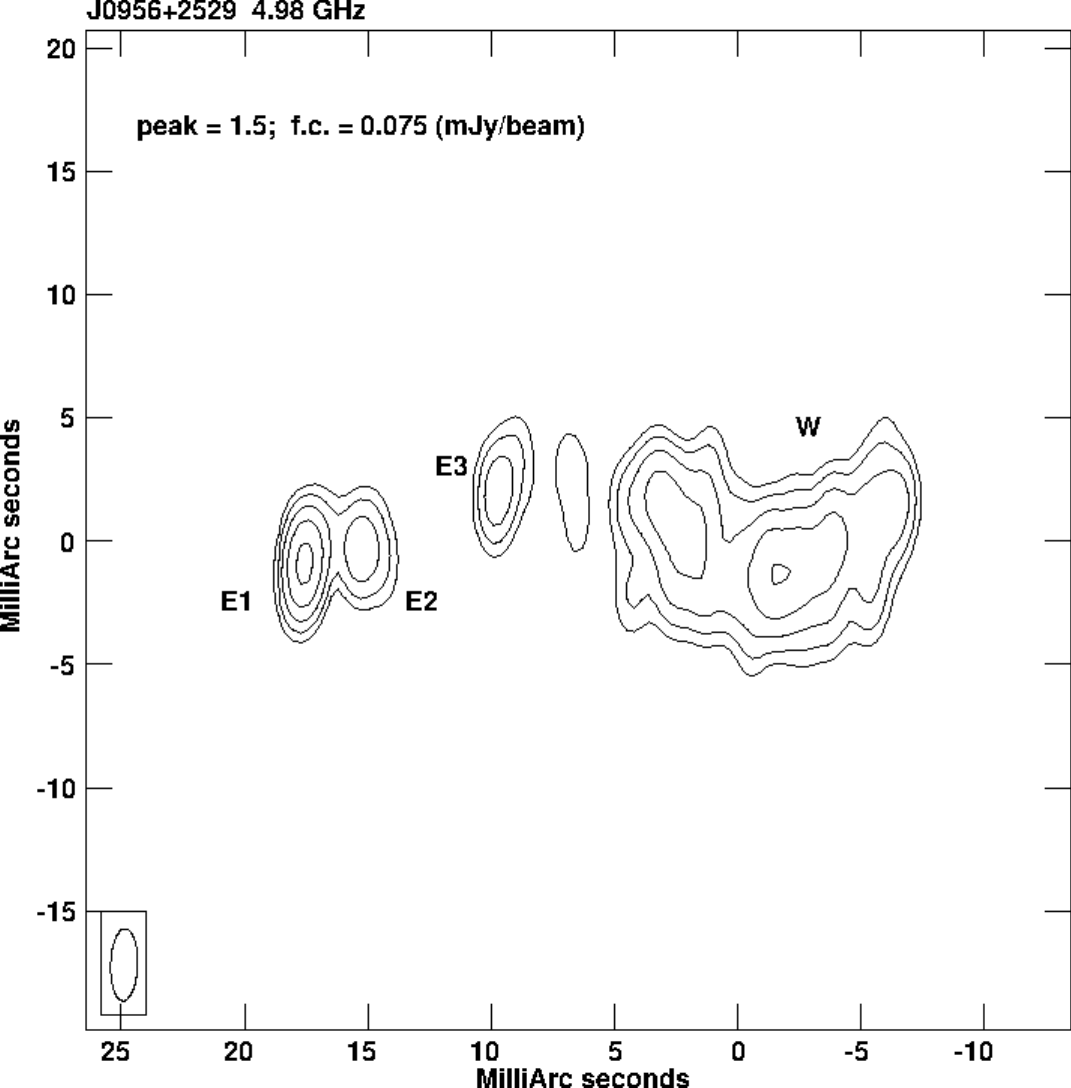}\\
\includegraphics[width=0.60\columnwidth]{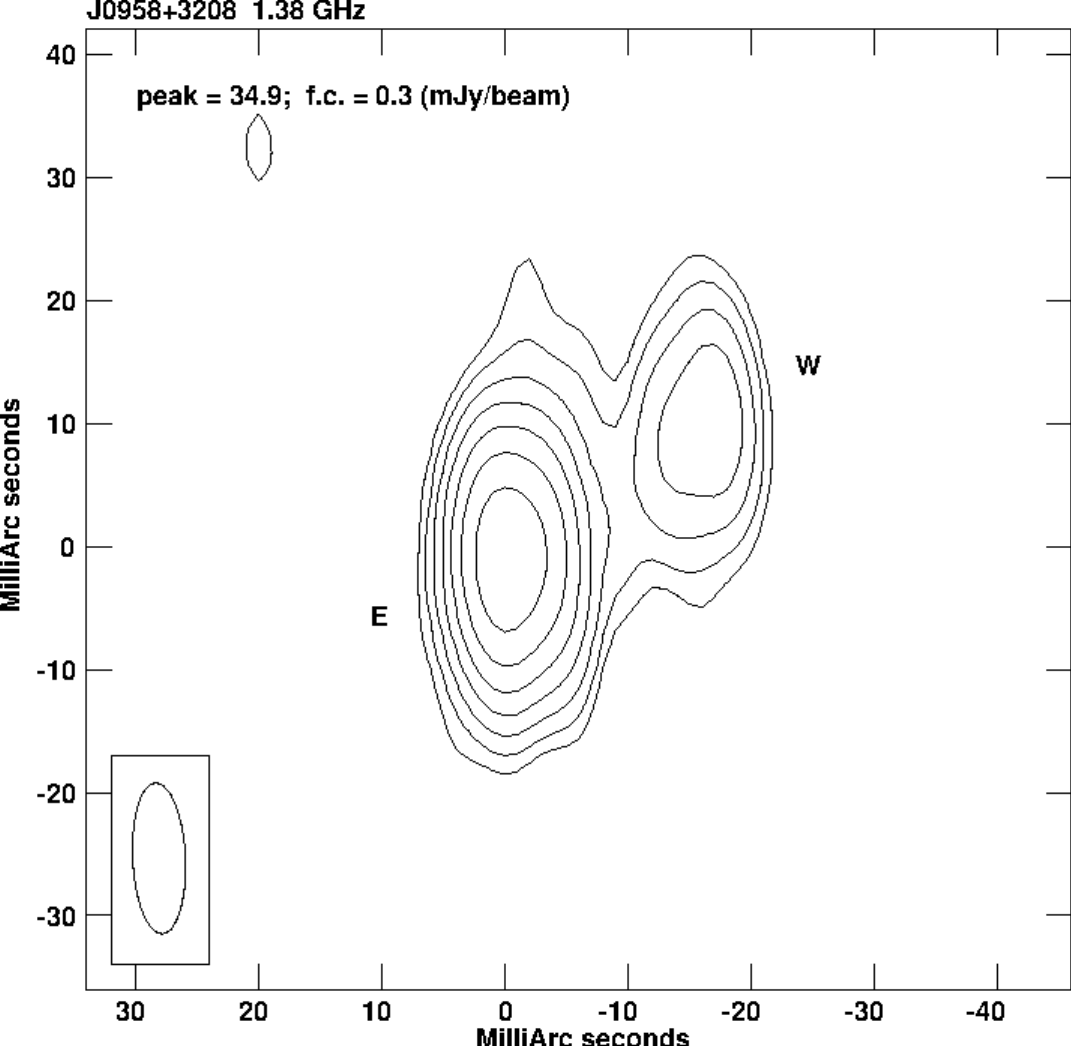}
\includegraphics[width=0.60\columnwidth]{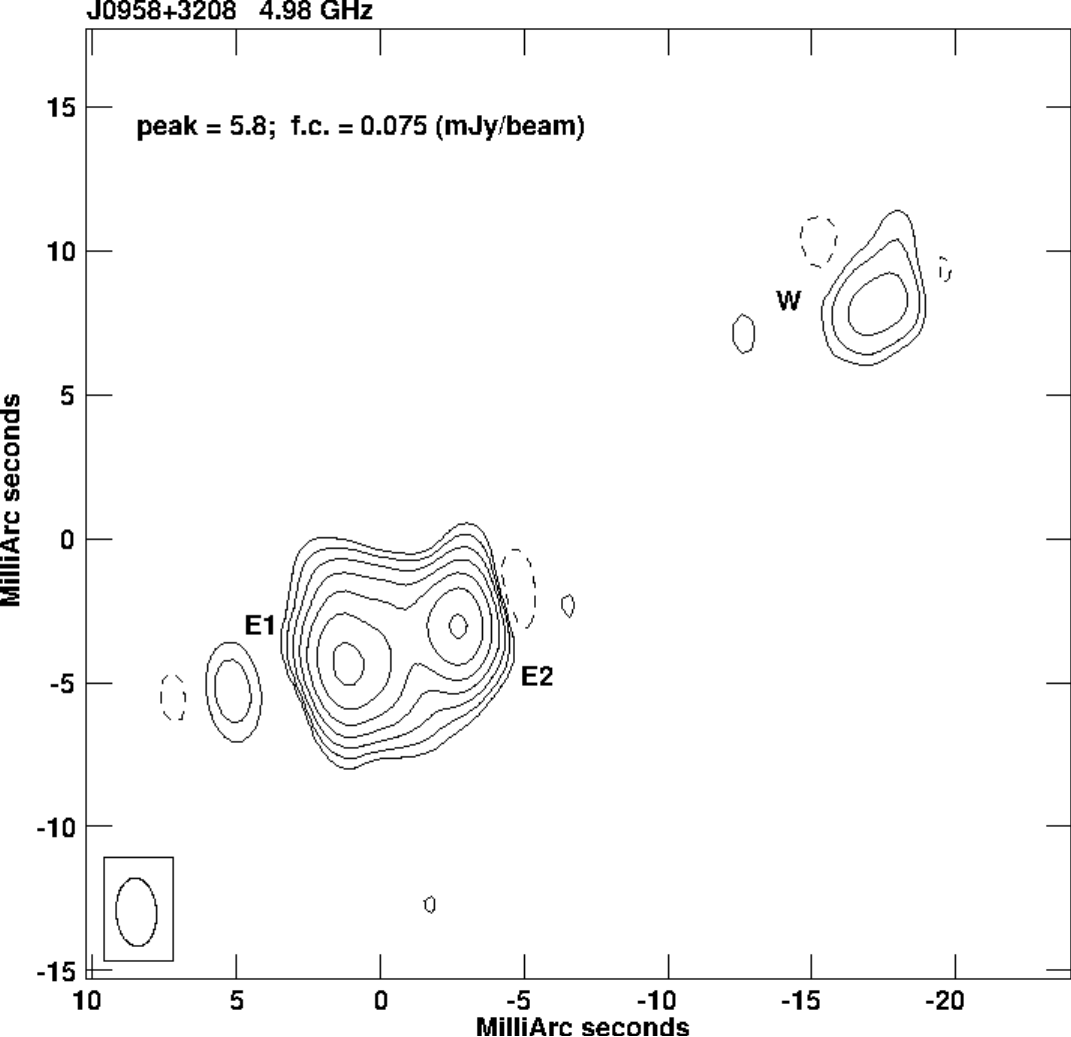}
\caption{VLBA images at 1.38 (left) and at 4.98 GHz (right). In each image, we provide the source name, the observing frequency, the peak brightness, and the first contour (f.c.), which is three times the off-source noise level on the image plane. Contours increase by a factor of 2. 
The restoring beam is plotted in the bottom left-hand corner of each image.}
\label{vlba-fig}
\end{center}
\end{figure*}

\addtocounter{figure}{-1}
\begin{figure*}
\begin{center}
\includegraphics[width=0.60\columnwidth]{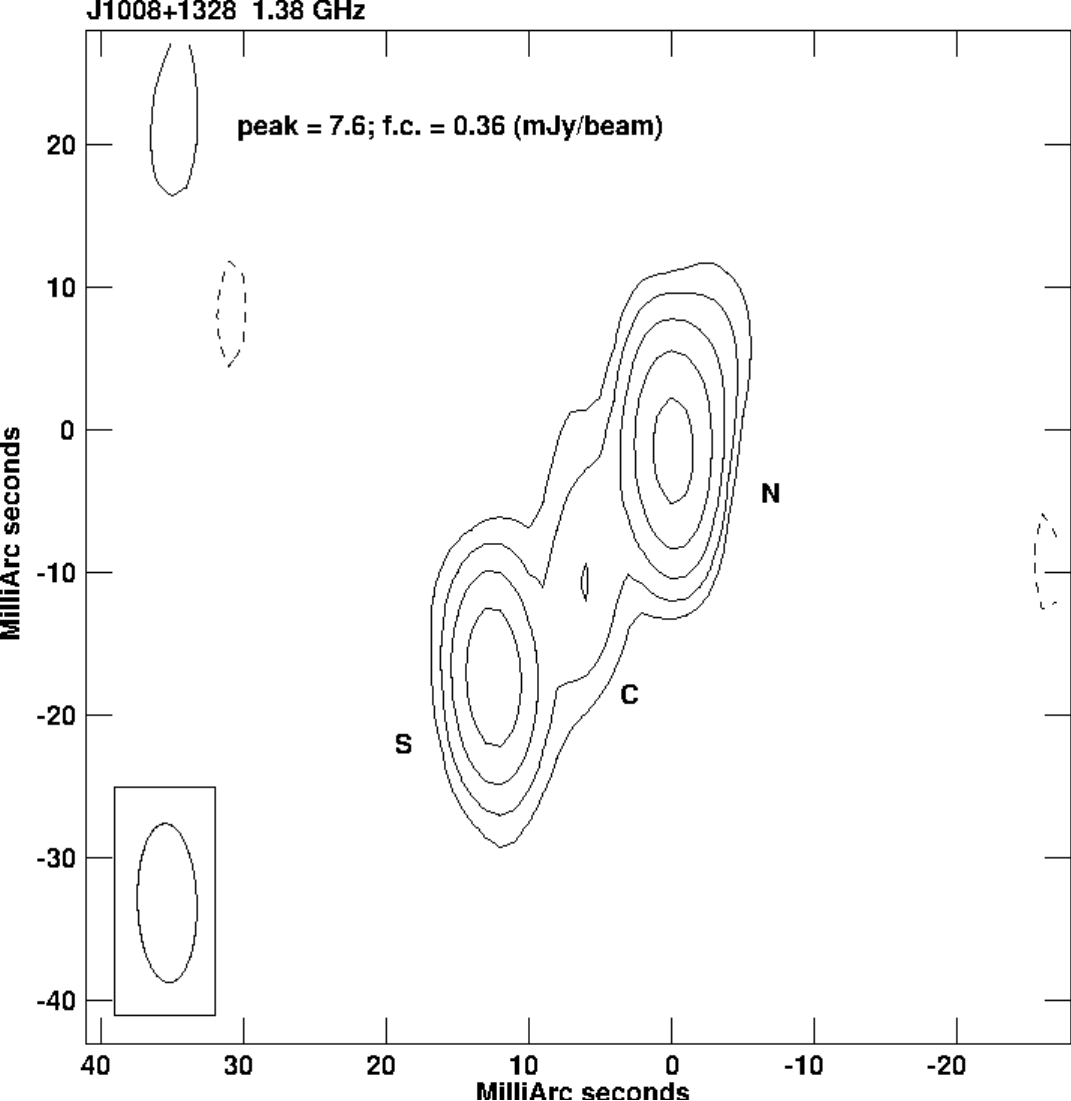}
\includegraphics[width=0.60\columnwidth]{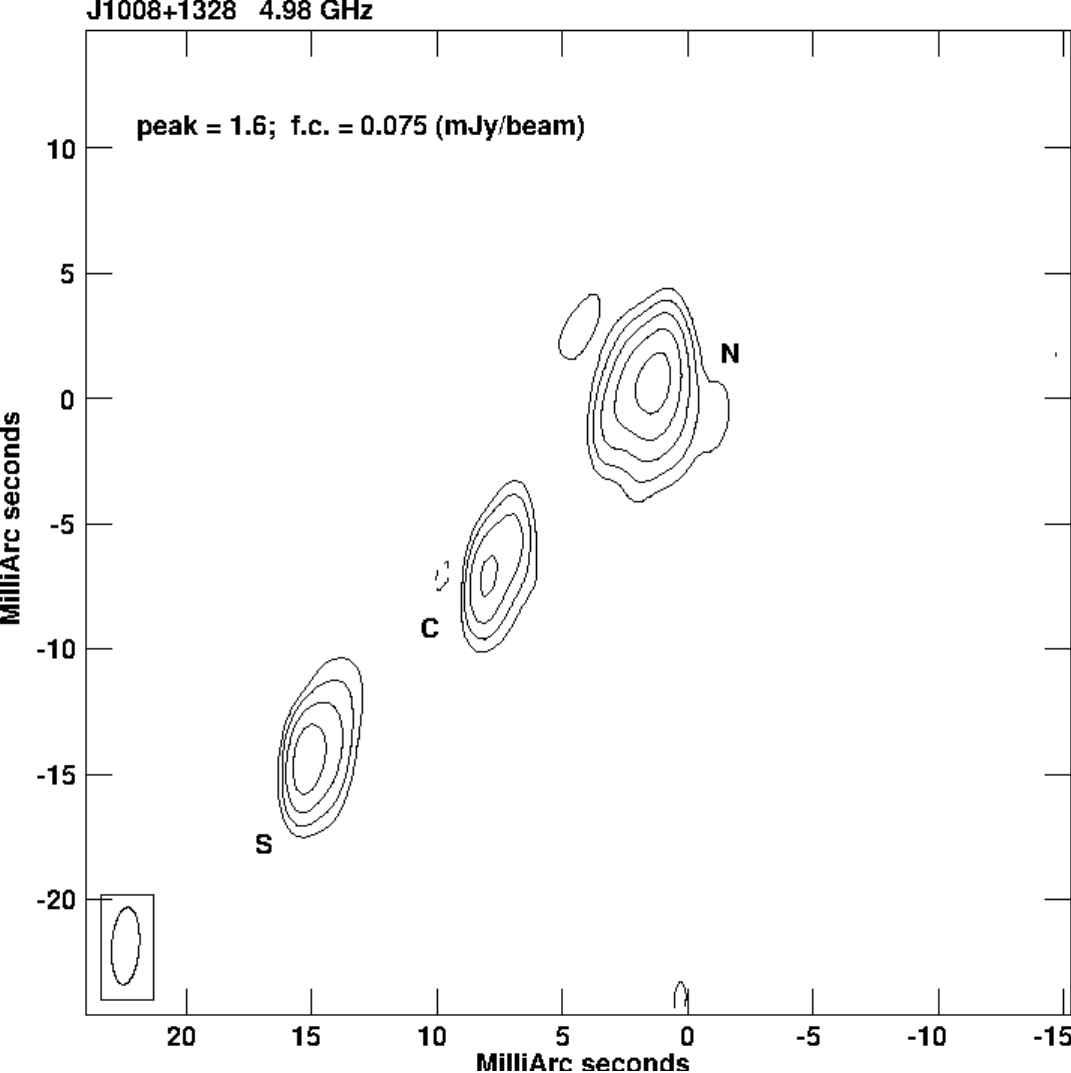}\\
\includegraphics[width=0.60\columnwidth]{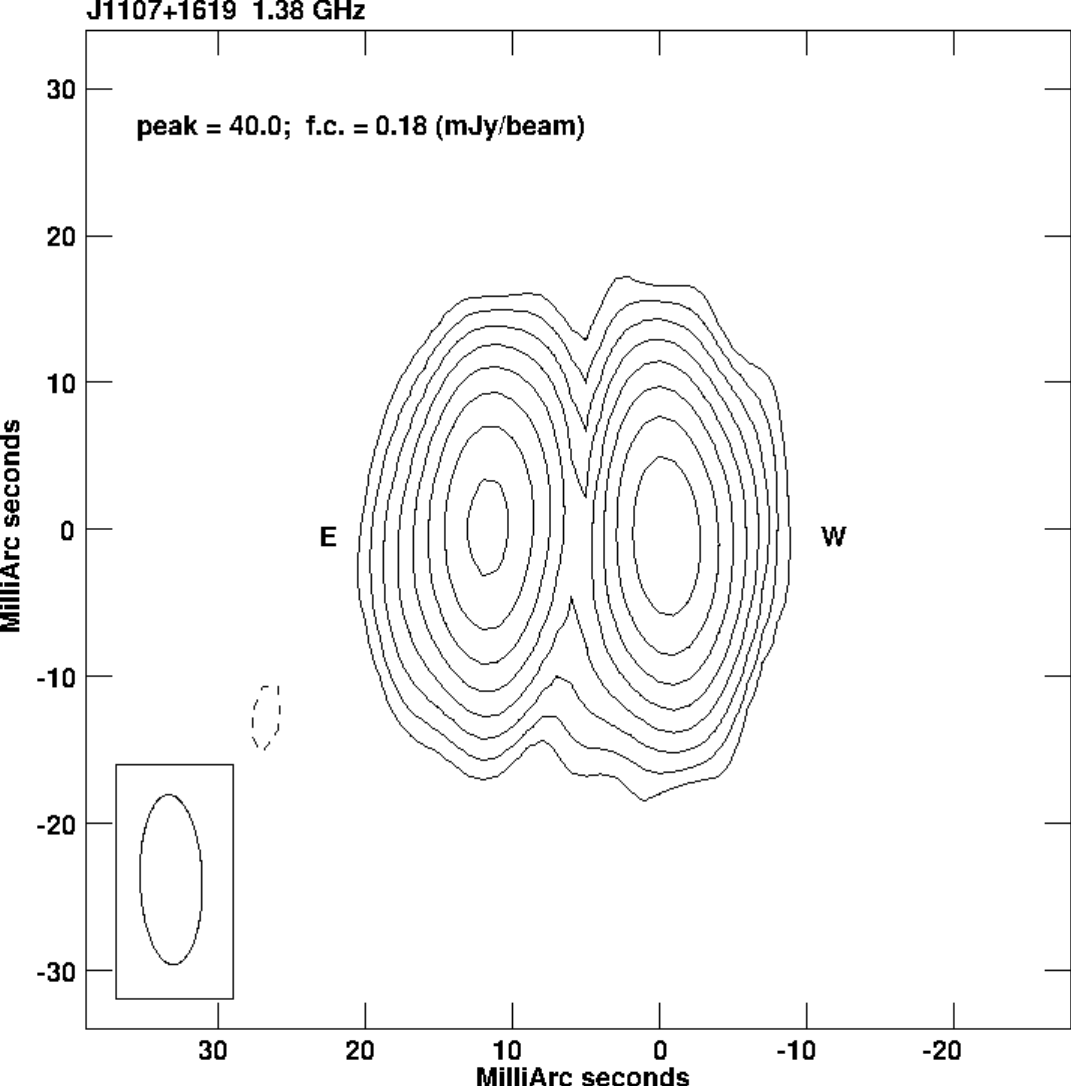}
\includegraphics[width=0.60\columnwidth]{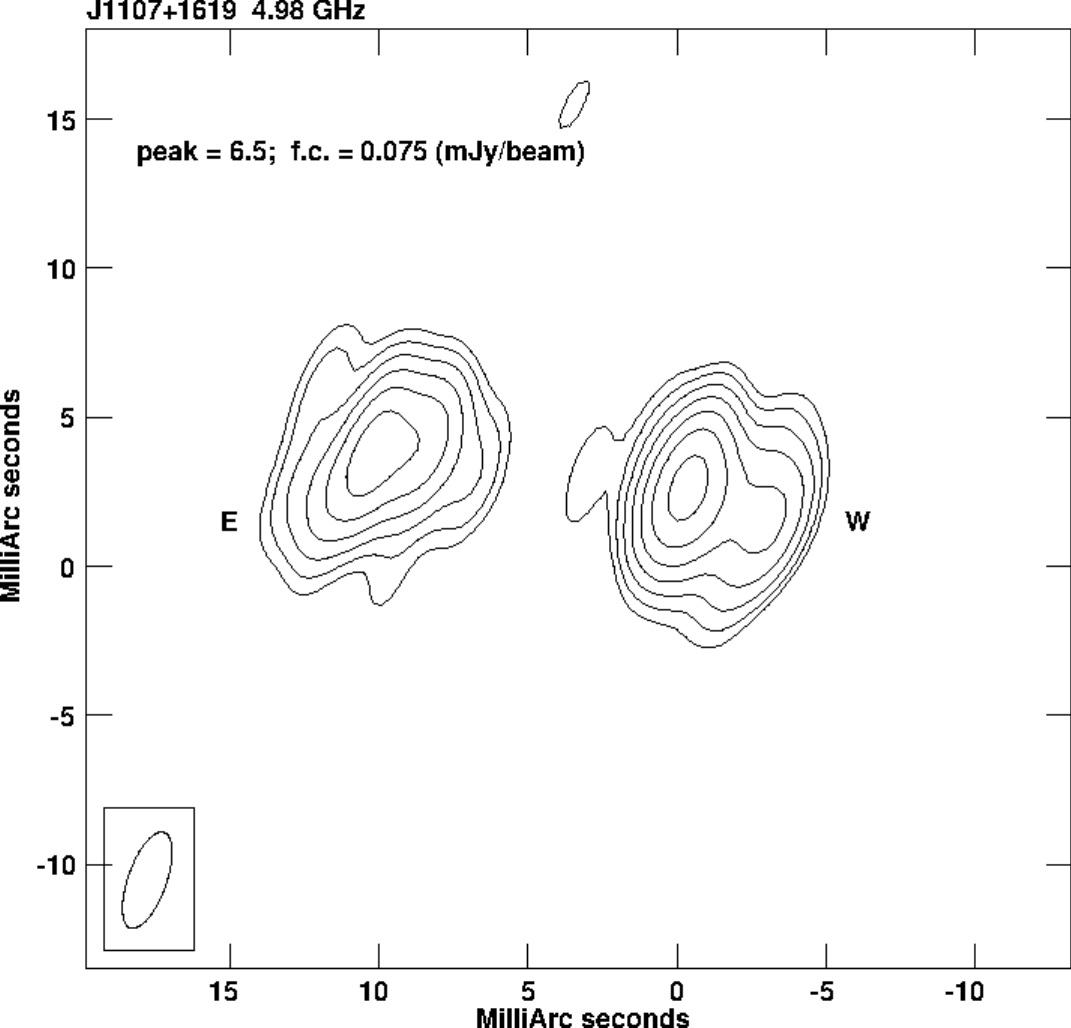}\\
\includegraphics[width=0.60\columnwidth]{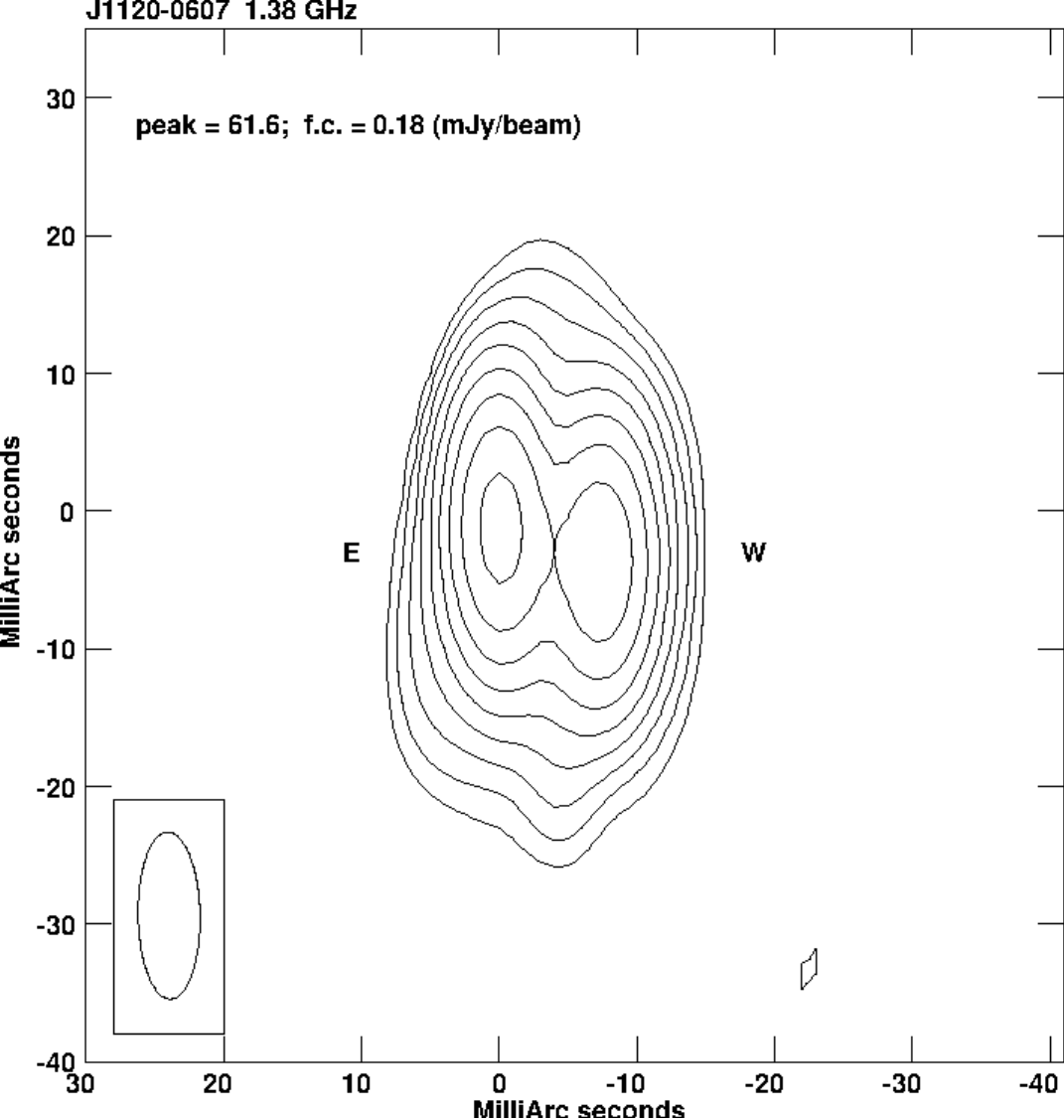}
\includegraphics[width=0.60\columnwidth]{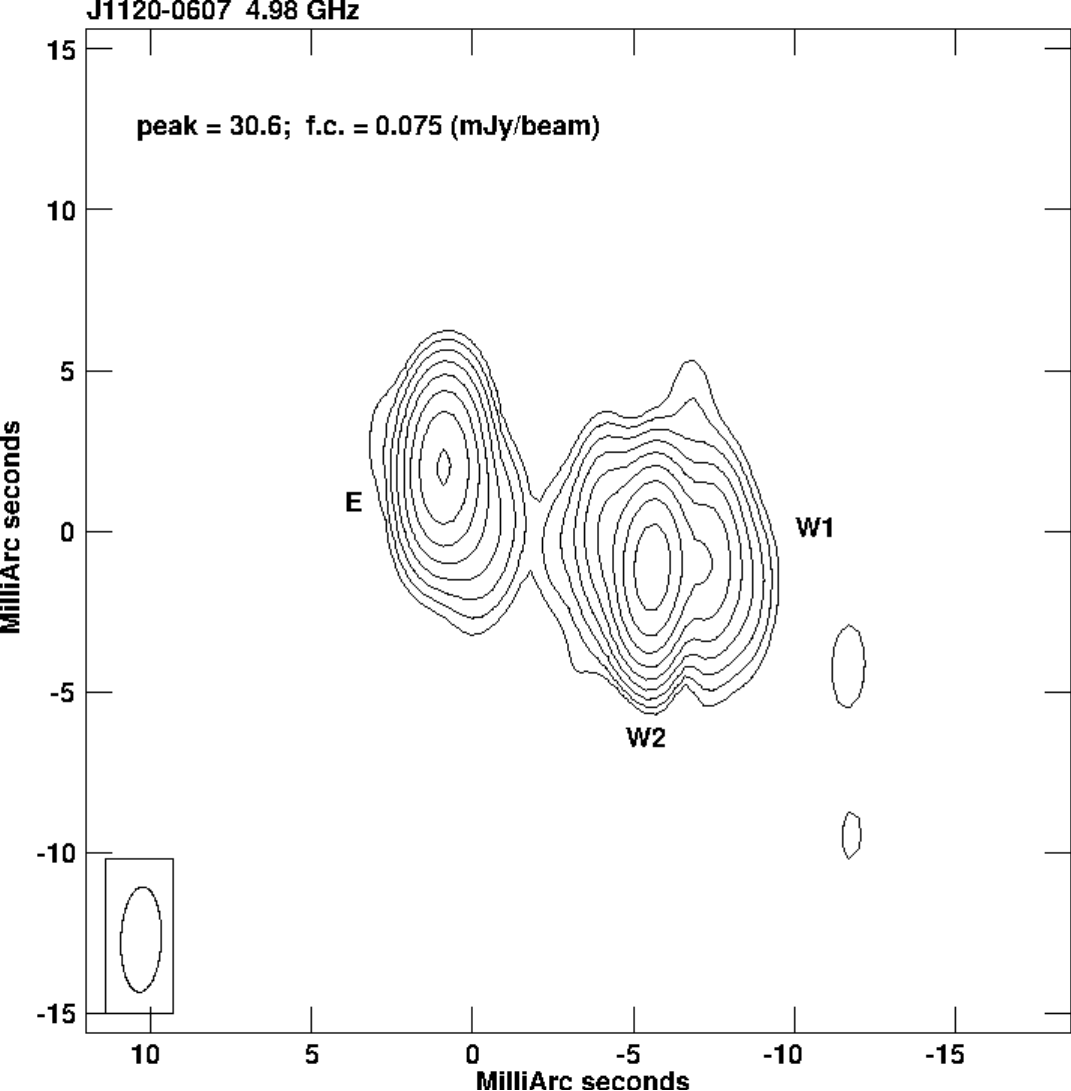}\\
\includegraphics[width=0.60\columnwidth]{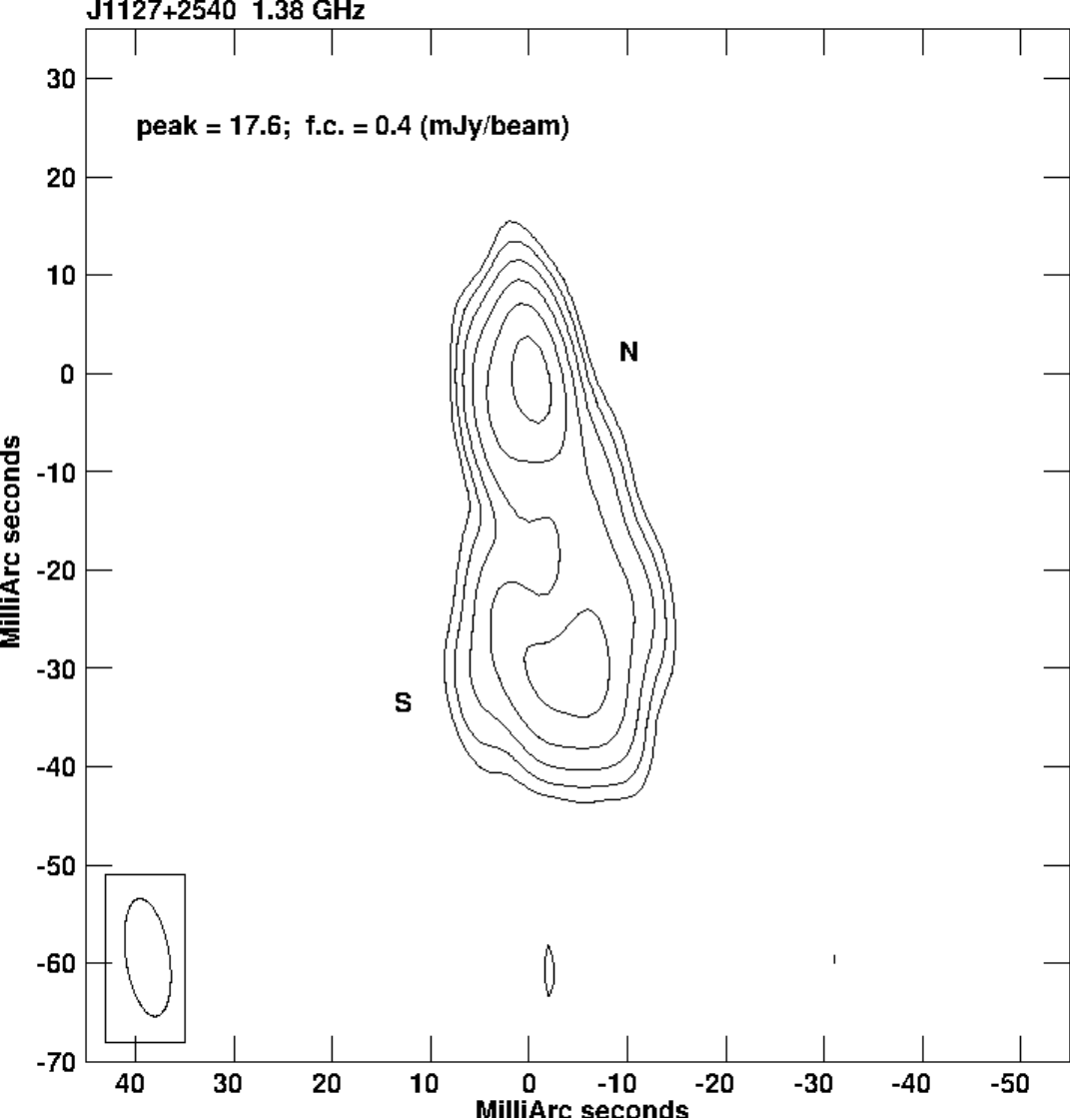}
\includegraphics[width=0.60\columnwidth]{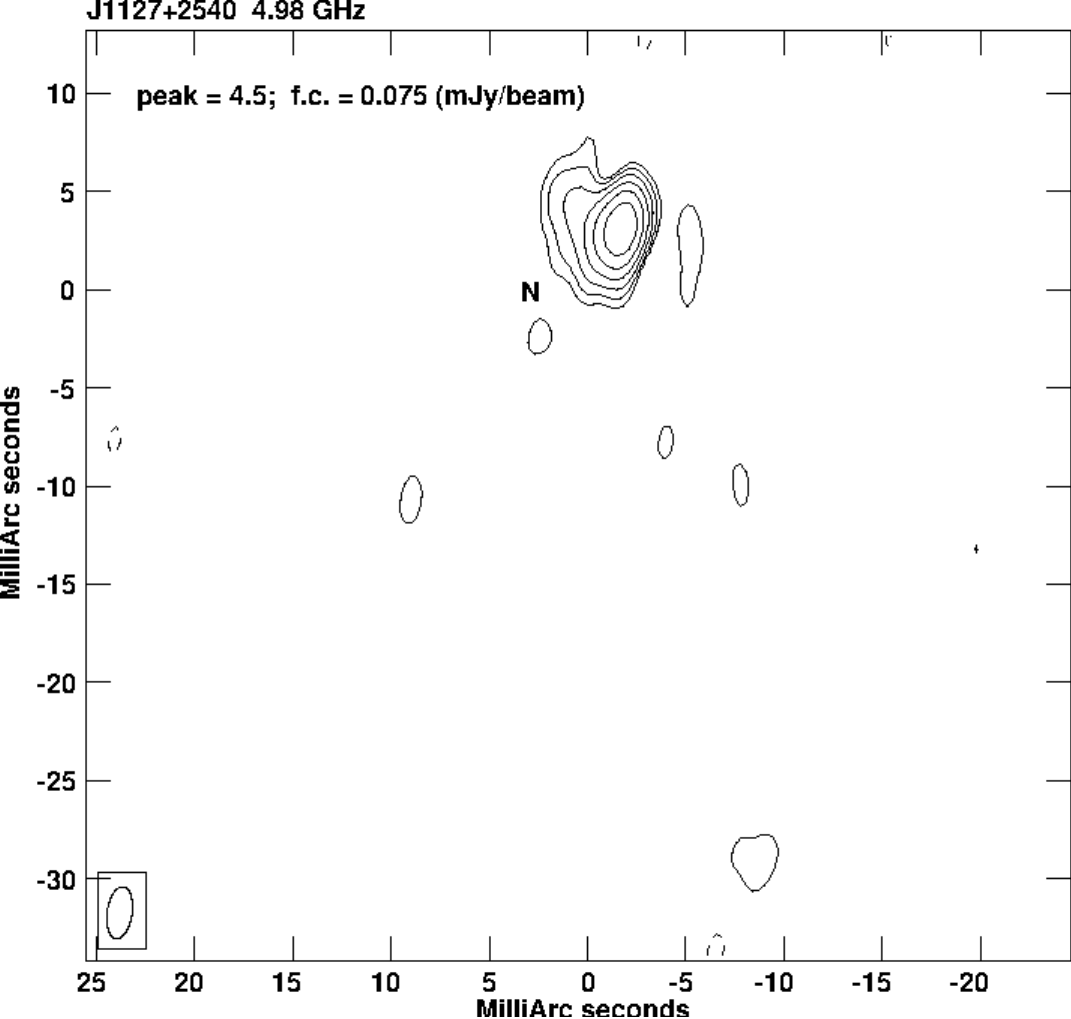}
\caption{continued.}
\end{center}
\end{figure*}

\addtocounter{figure}{-1}
\begin{figure*}
\begin{center}
\includegraphics[width=0.60\columnwidth]{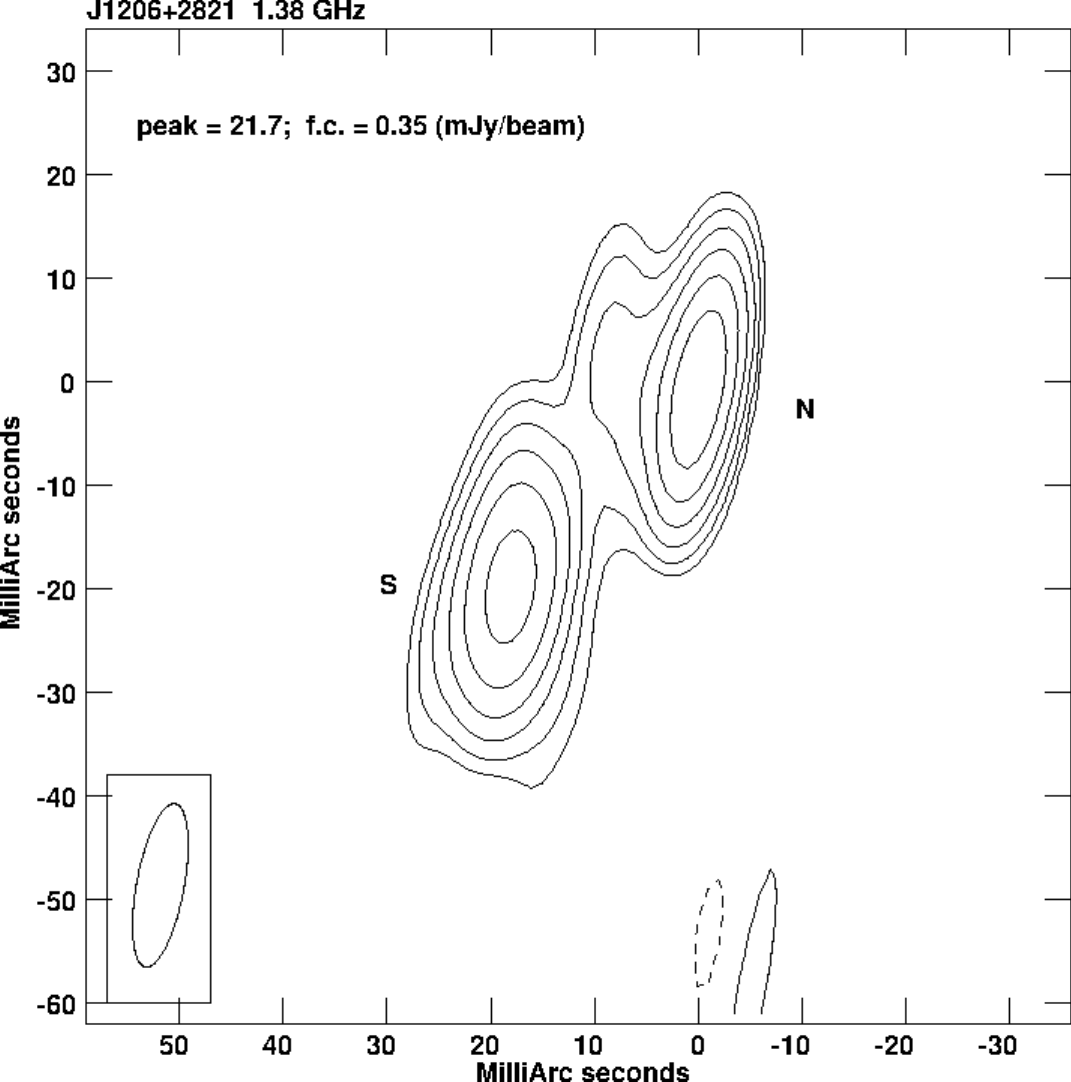}
\includegraphics[width=0.60\columnwidth]{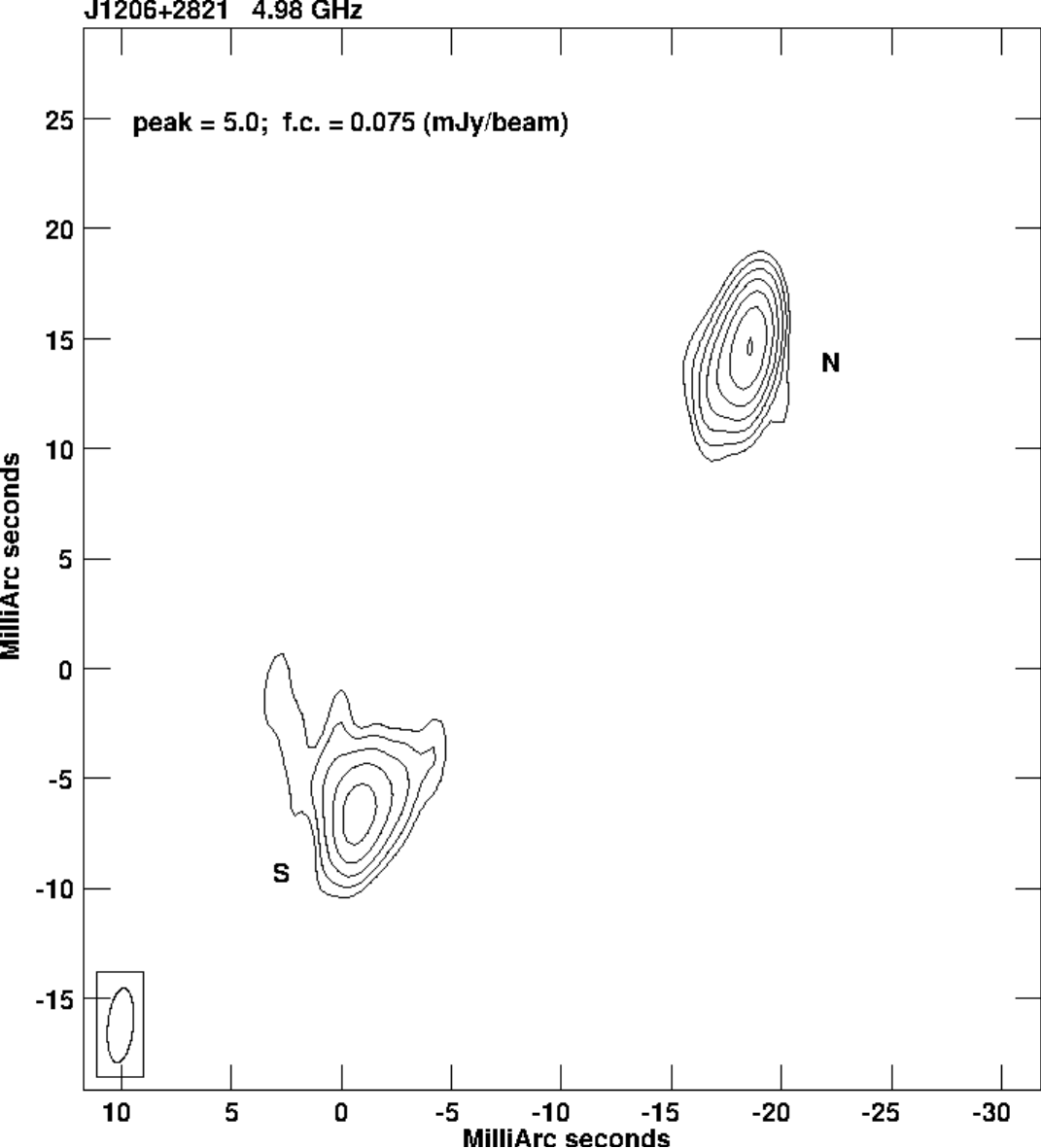}\\
\includegraphics[width=0.60\columnwidth]{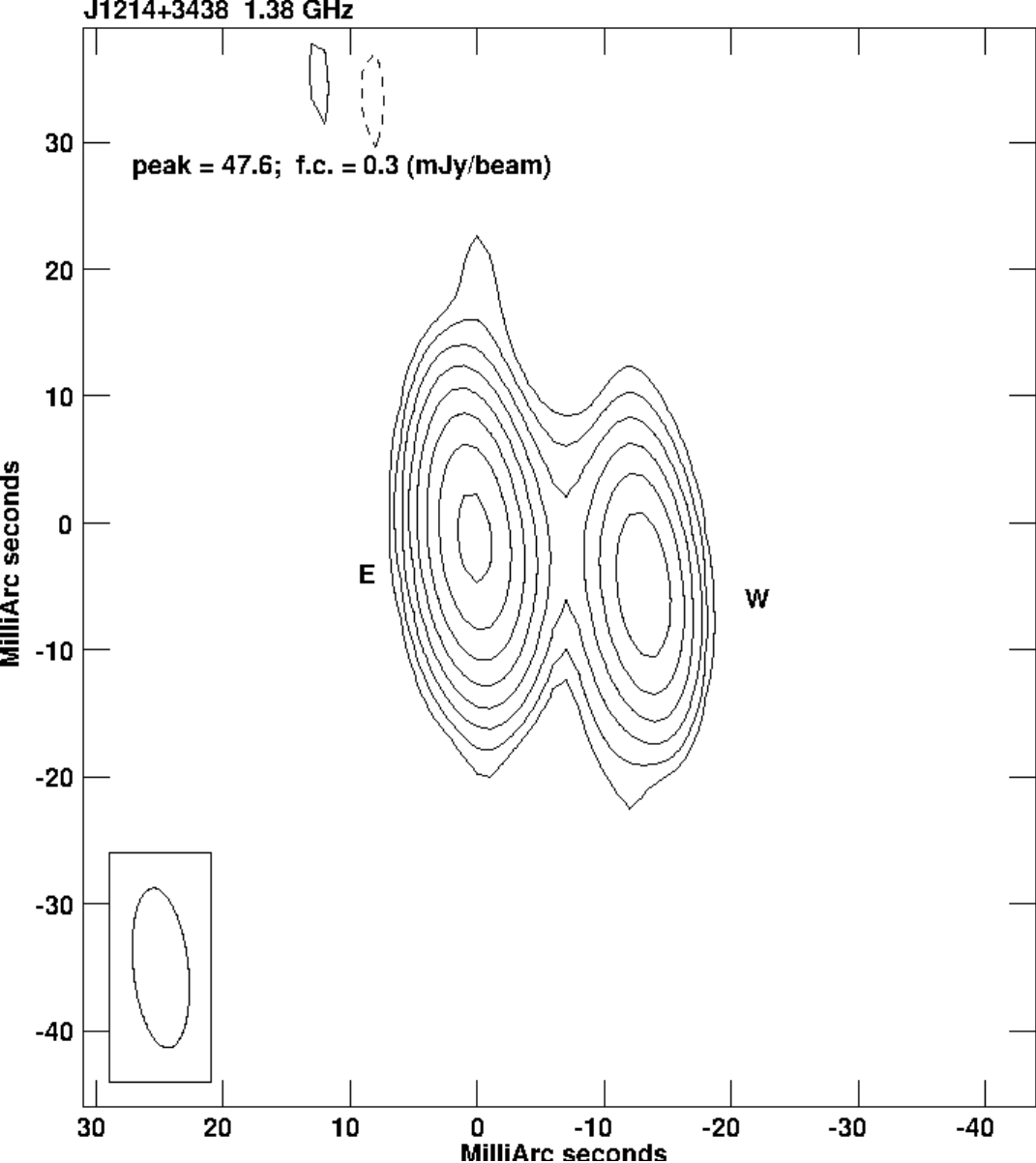}
\includegraphics[width=0.60\columnwidth]{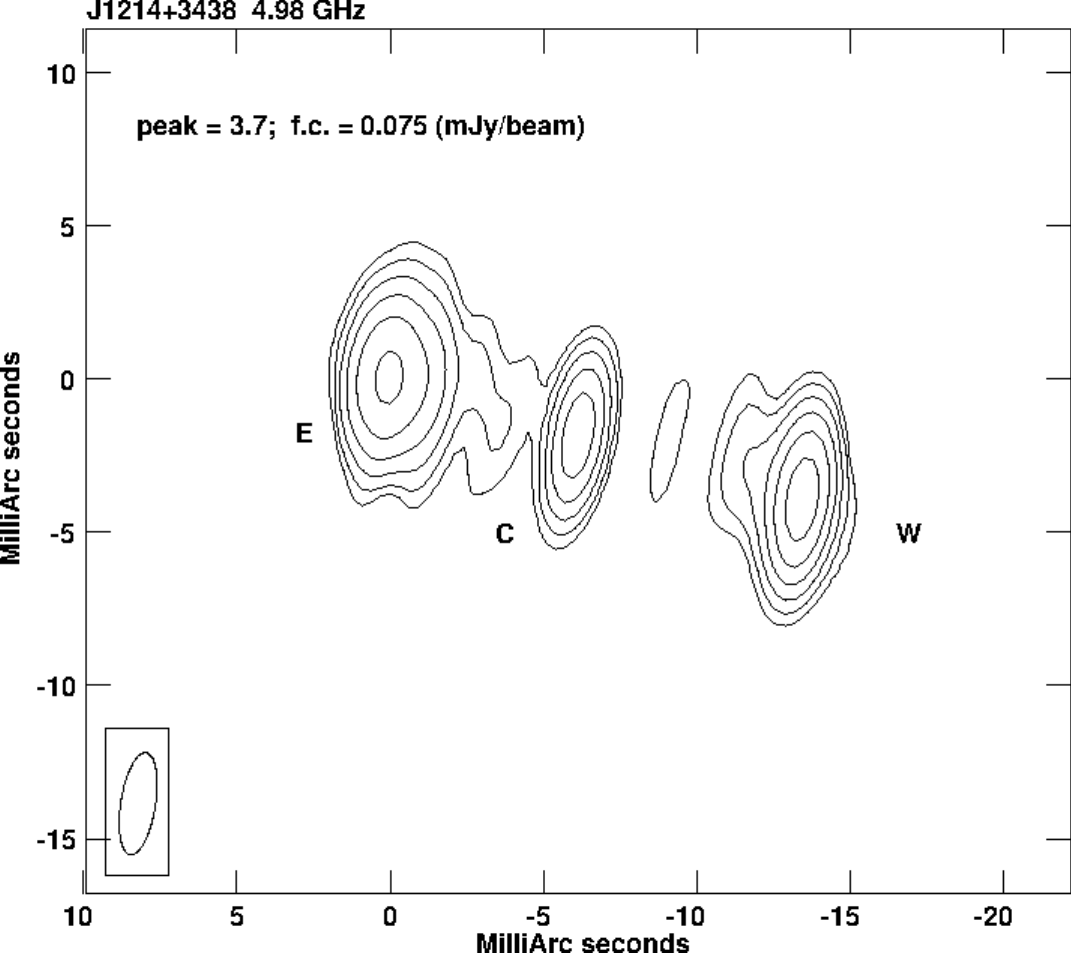}\\
\includegraphics[width=0.60\columnwidth]{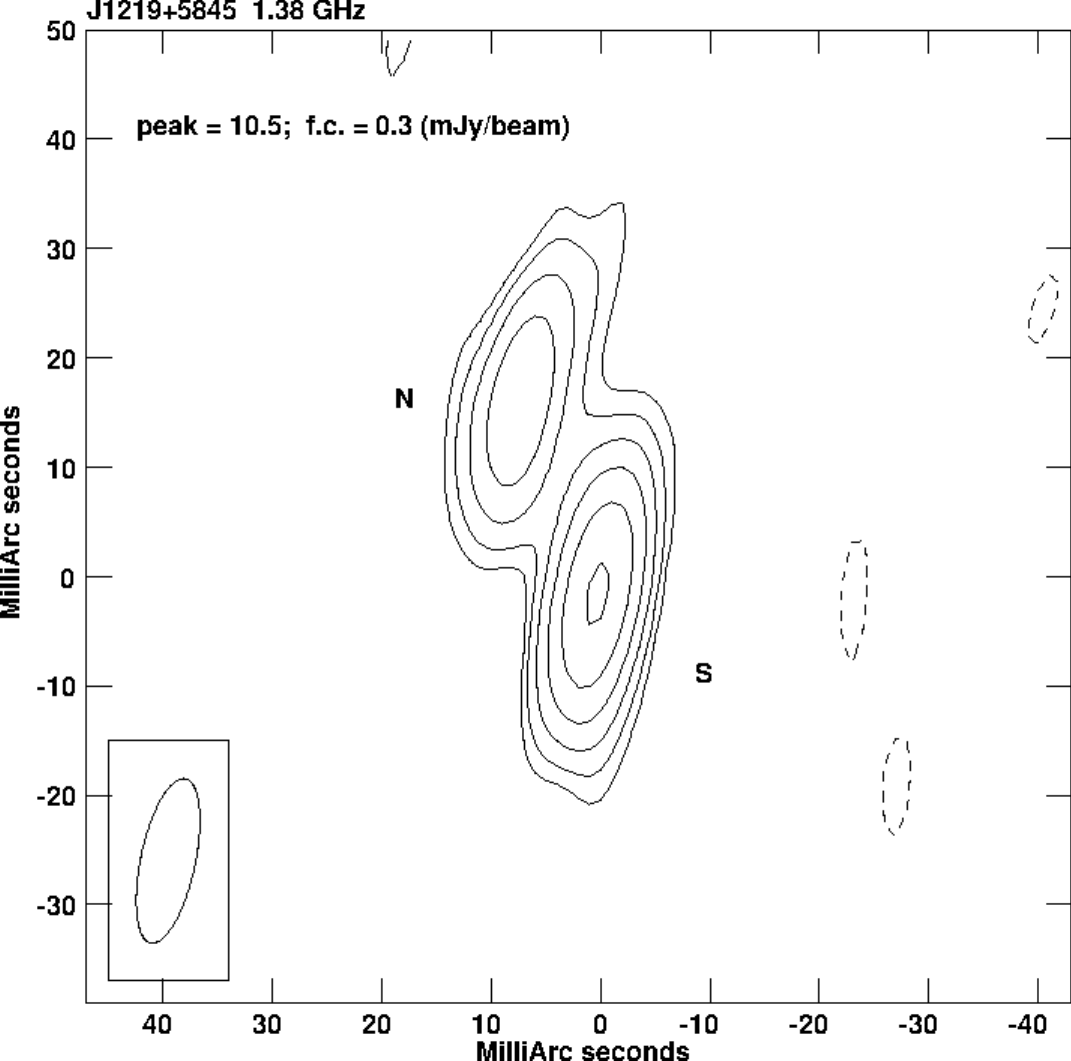}
\includegraphics[width=0.60\columnwidth]{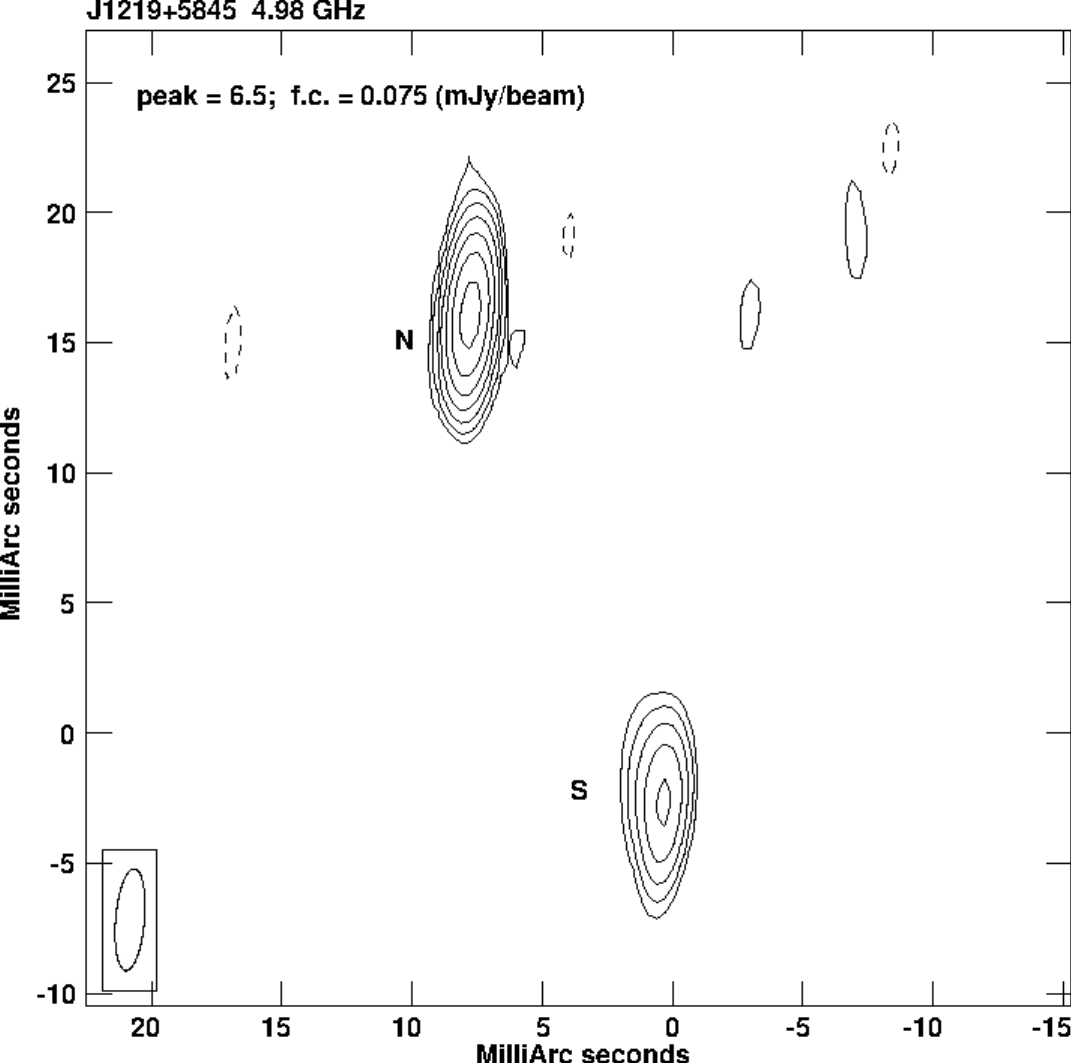}\\
\includegraphics[width=0.60\columnwidth]{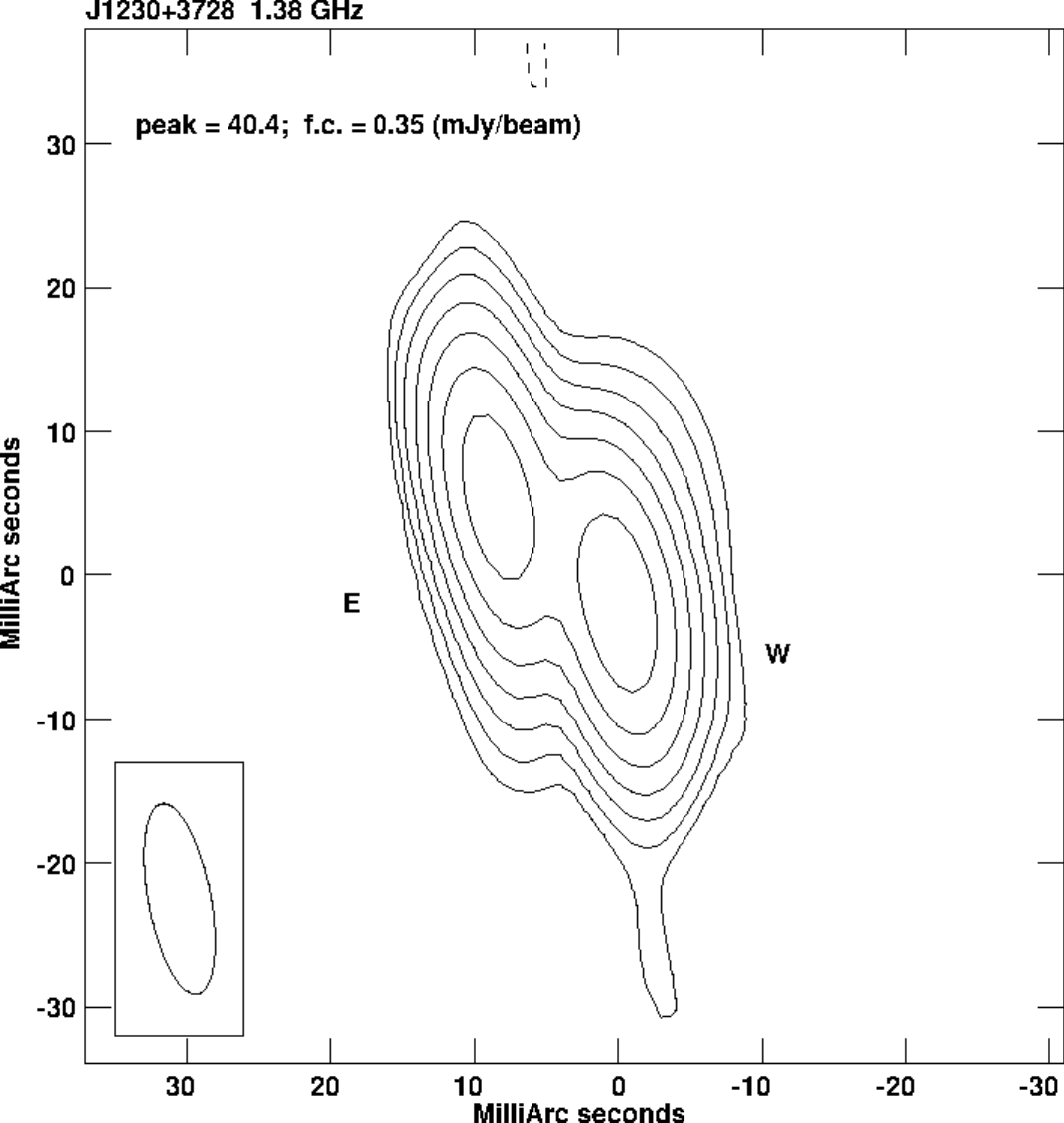}
\includegraphics[width=0.60\columnwidth]{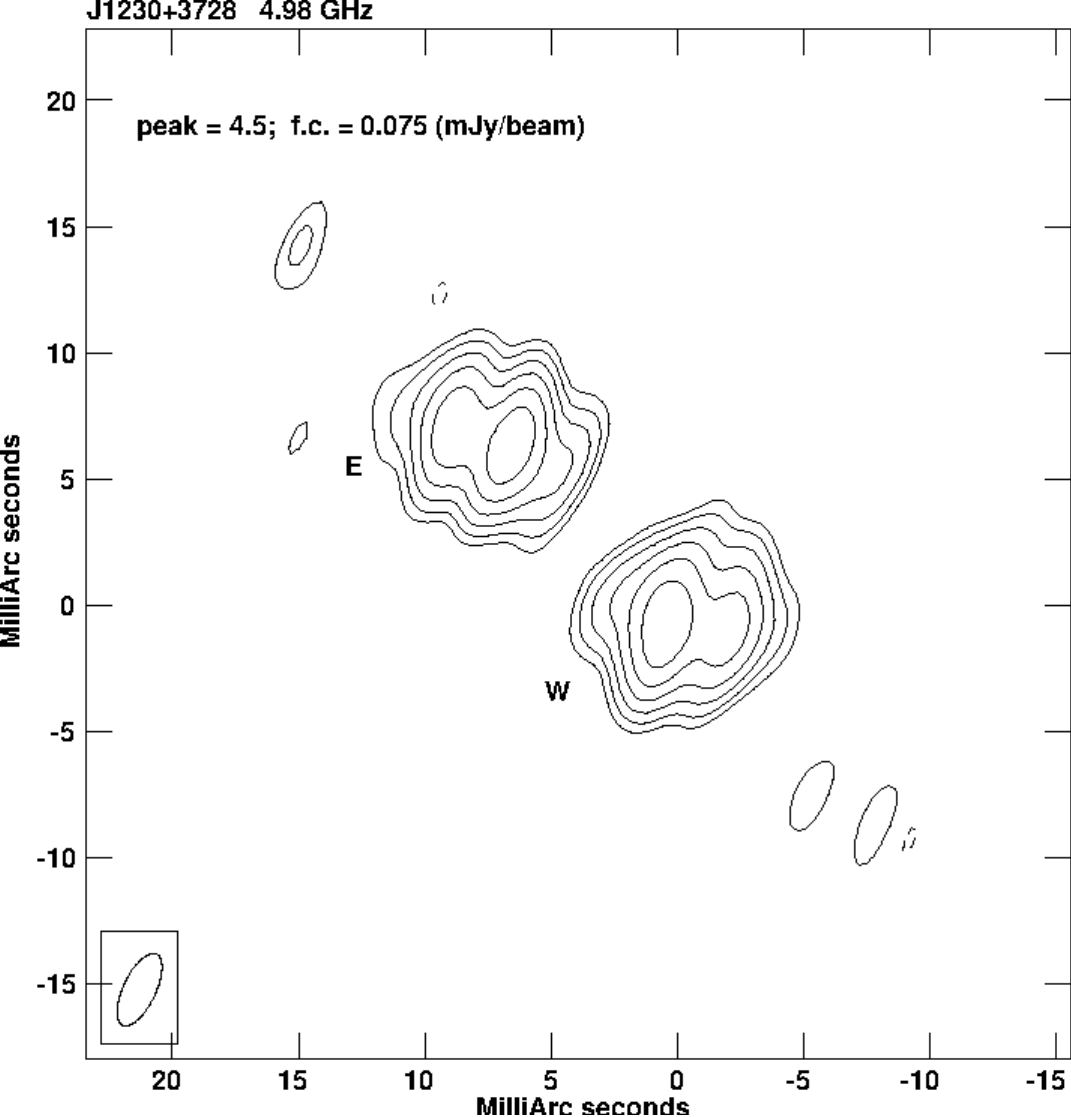}

\caption{continued.}
\end{center}
\end{figure*}

\addtocounter{figure}{-1}
\begin{figure*}
\begin{center}
\includegraphics[width=0.60\columnwidth]{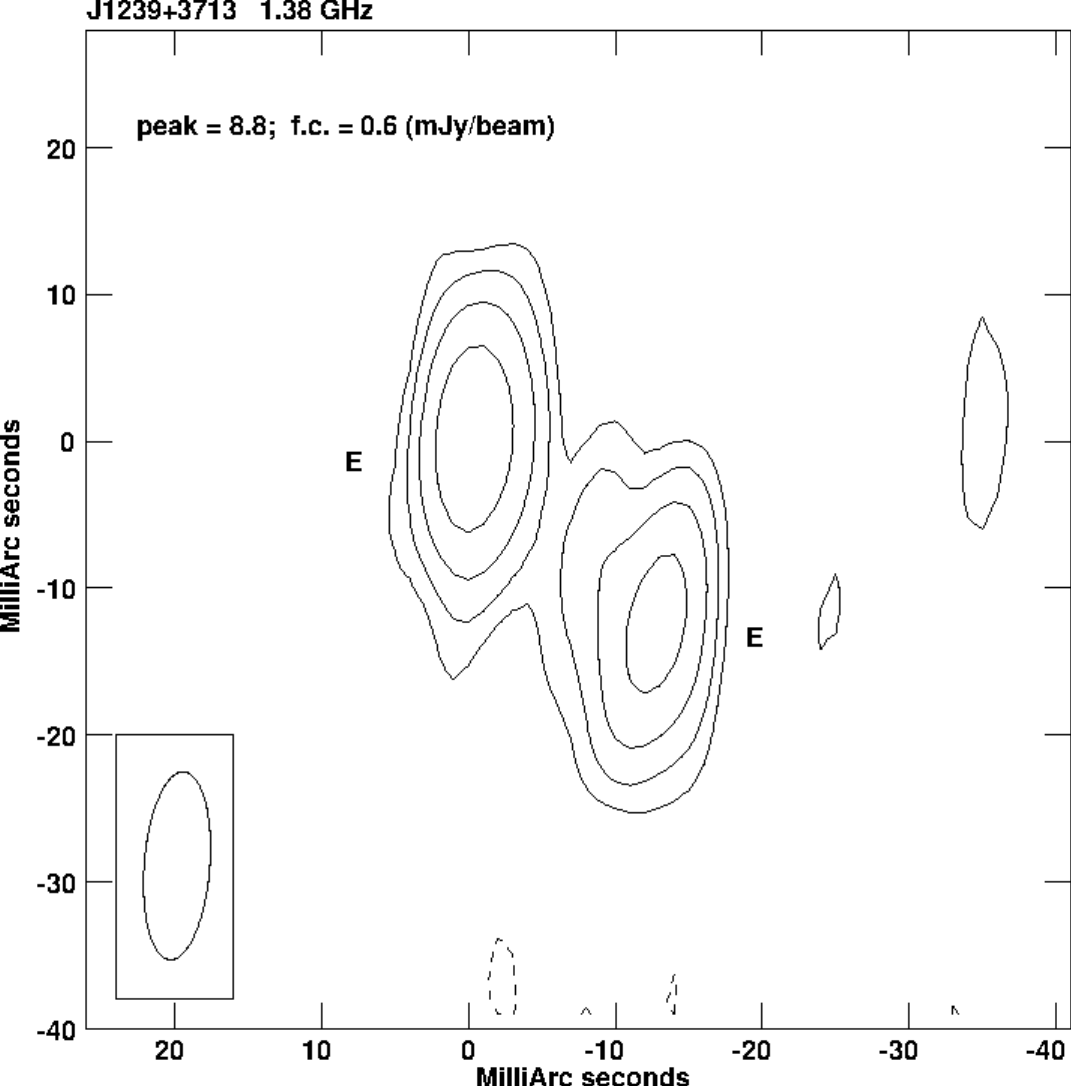}
\includegraphics[width=0.60\columnwidth]{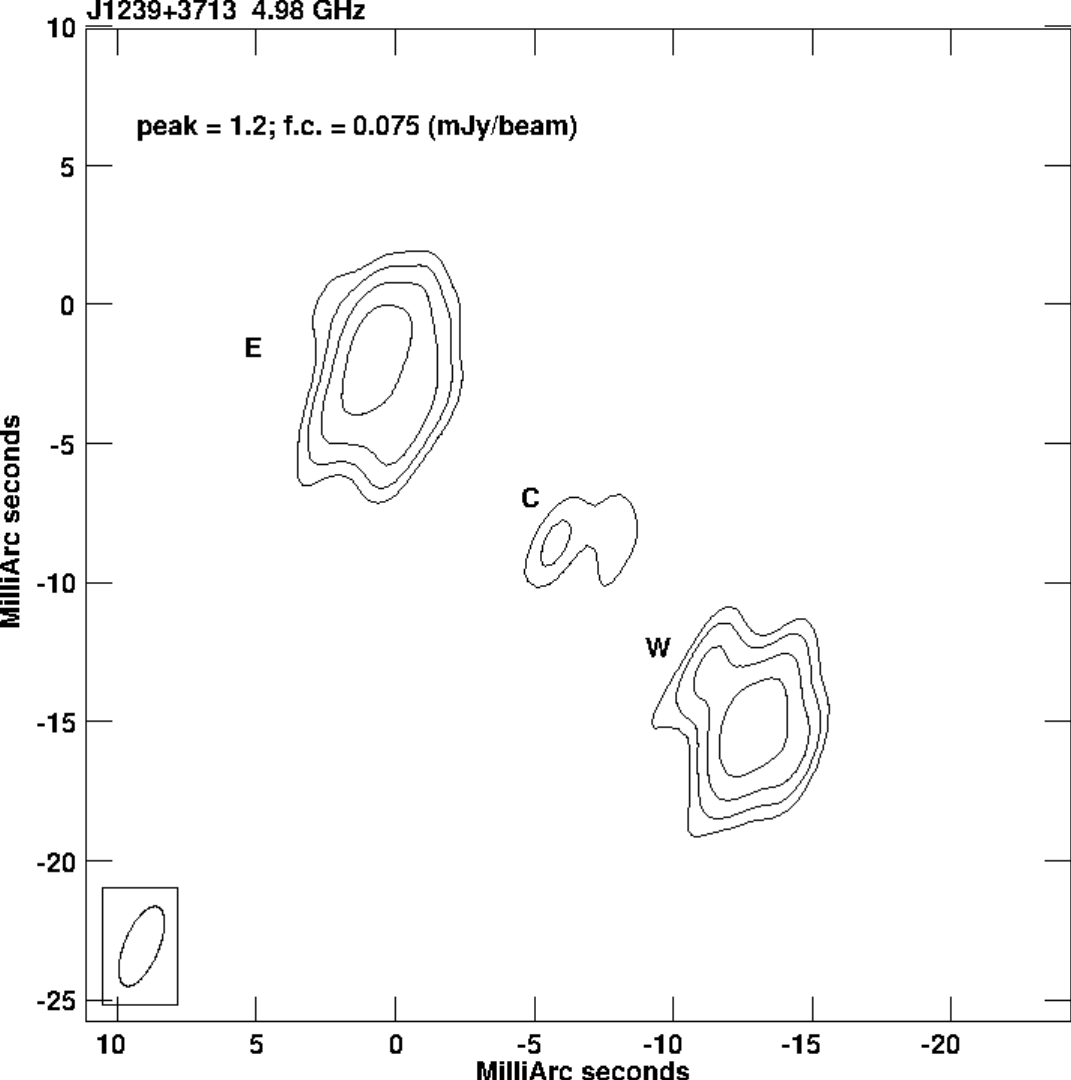}\\
\includegraphics[width=0.60\columnwidth]{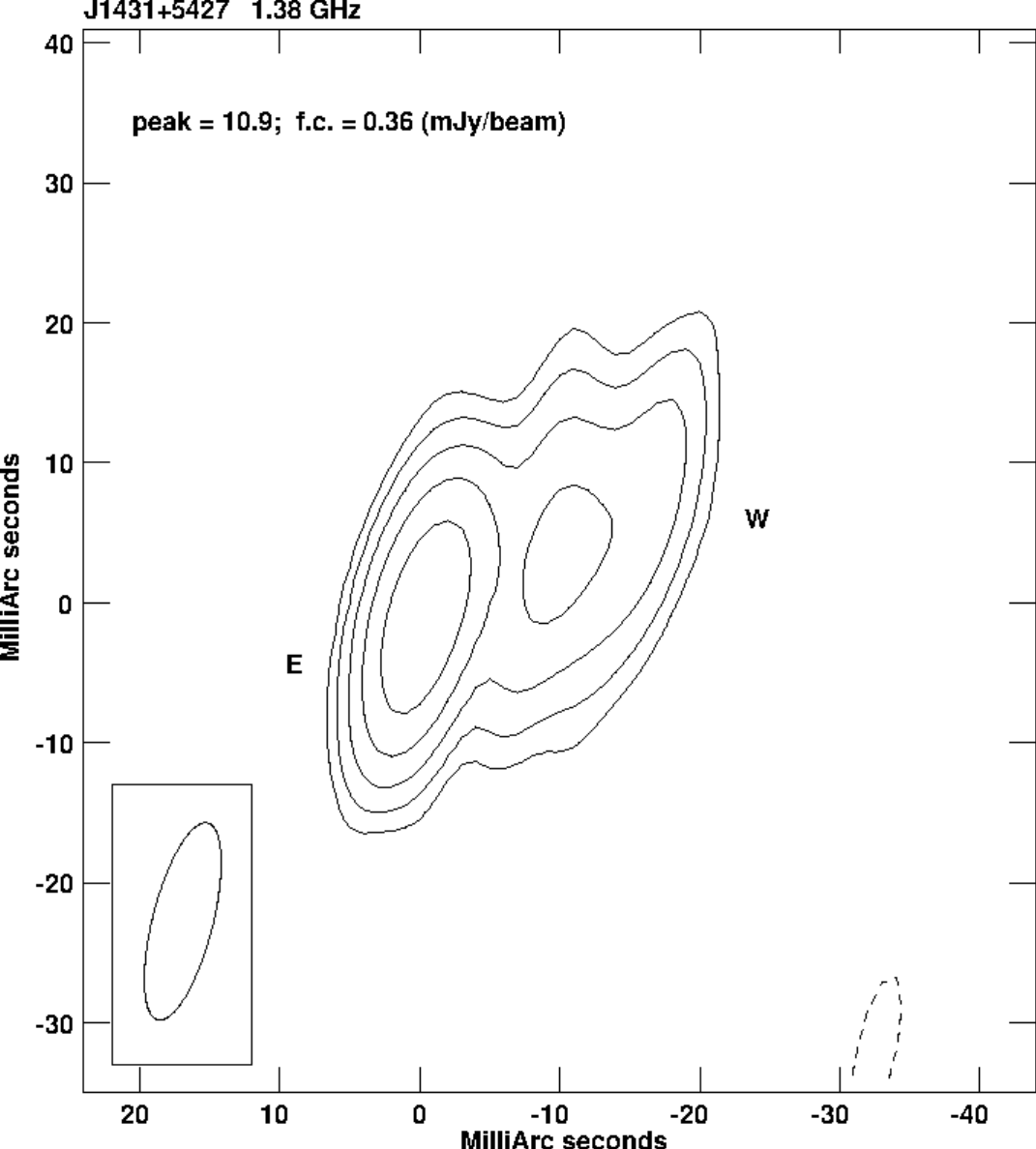}
\includegraphics[width=0.60\columnwidth]{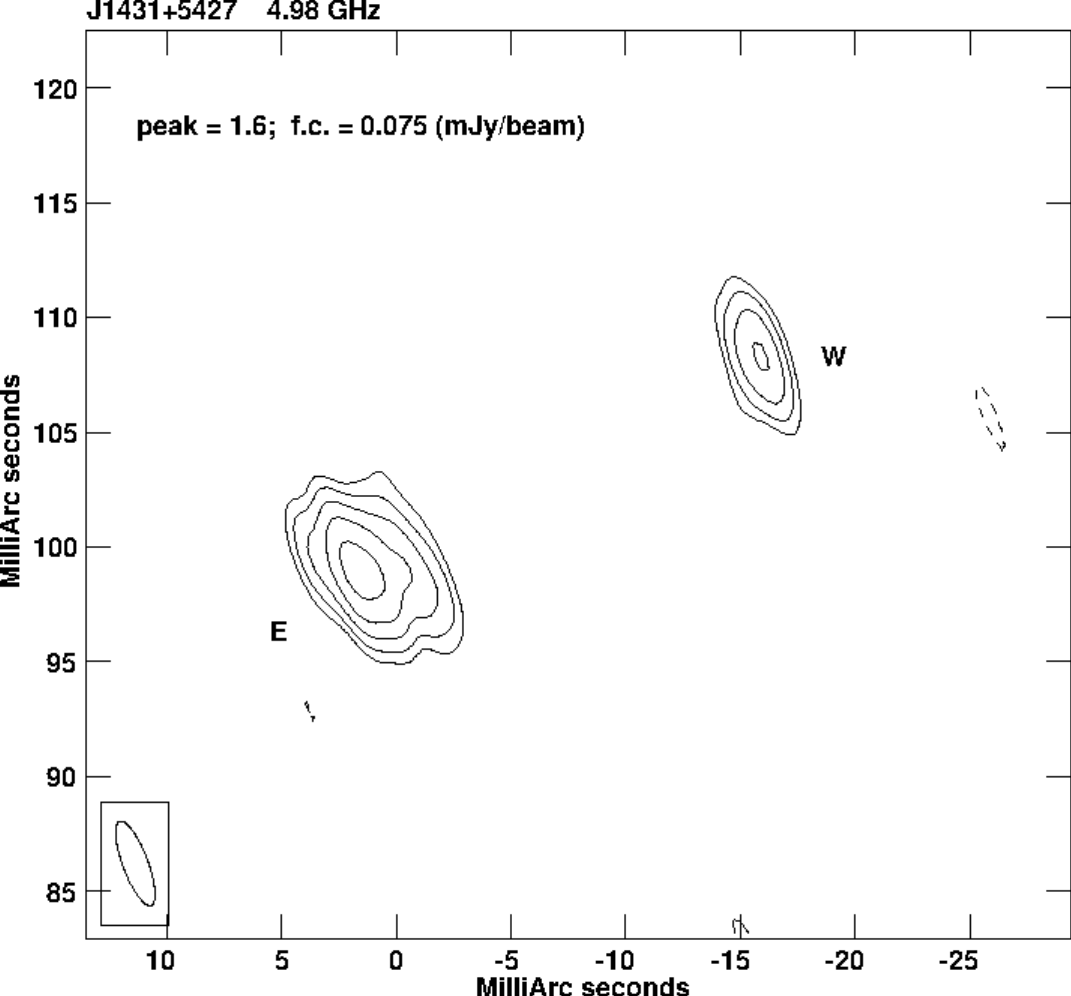}\\
\includegraphics[width=0.60\columnwidth]{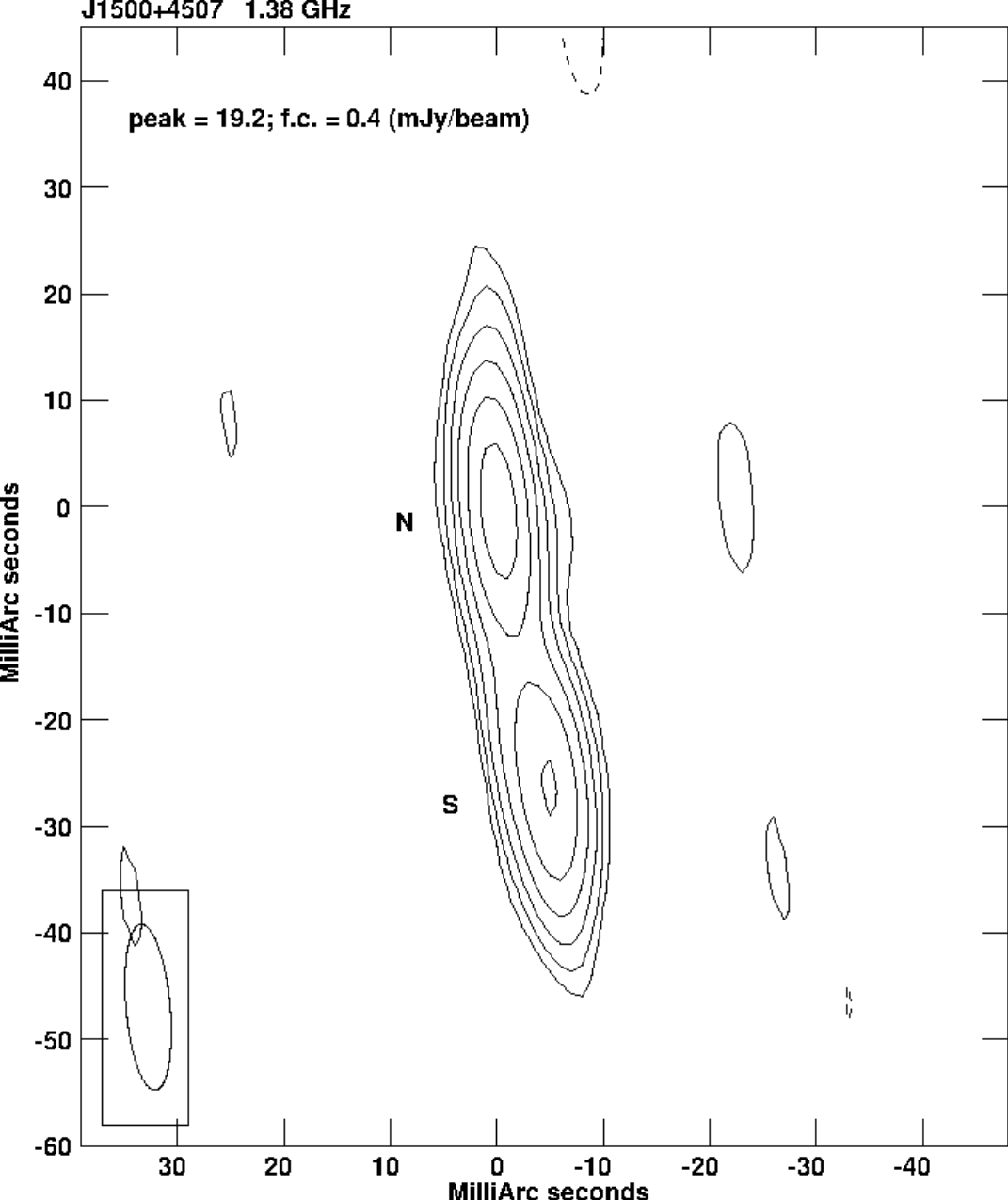}
\includegraphics[width=0.60\columnwidth]{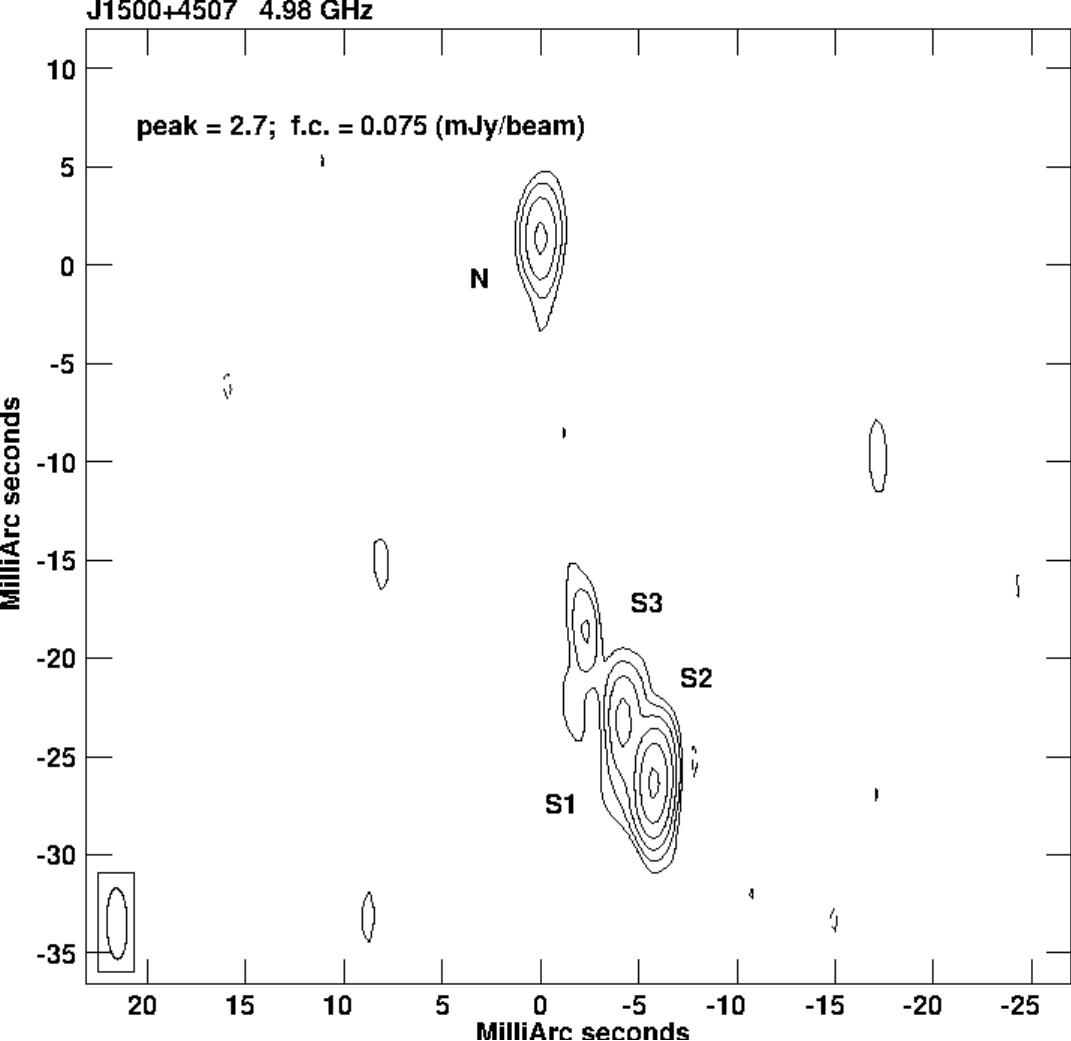}\\
\includegraphics[width=0.60\columnwidth]{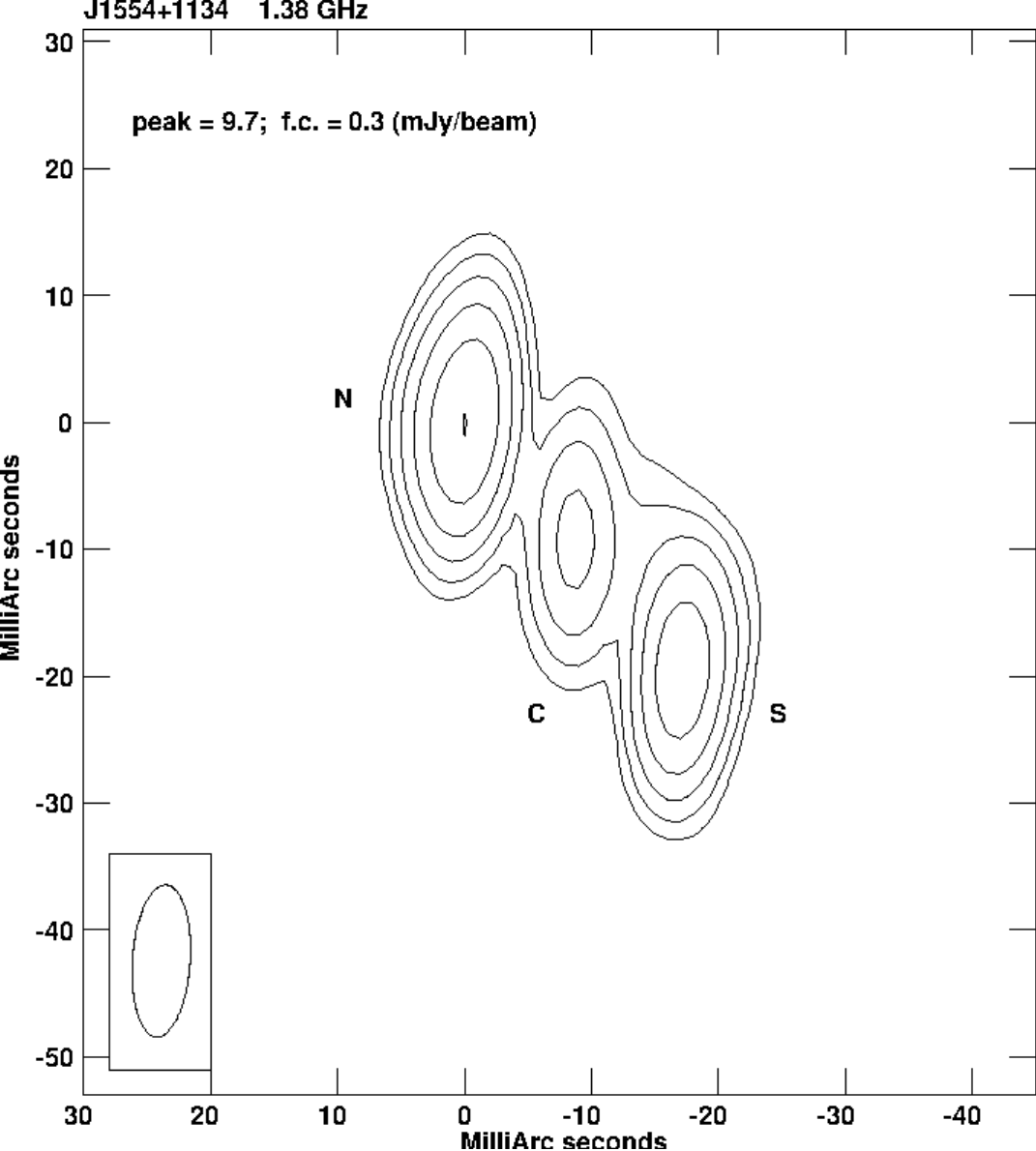}
\includegraphics[width=0.60\columnwidth]{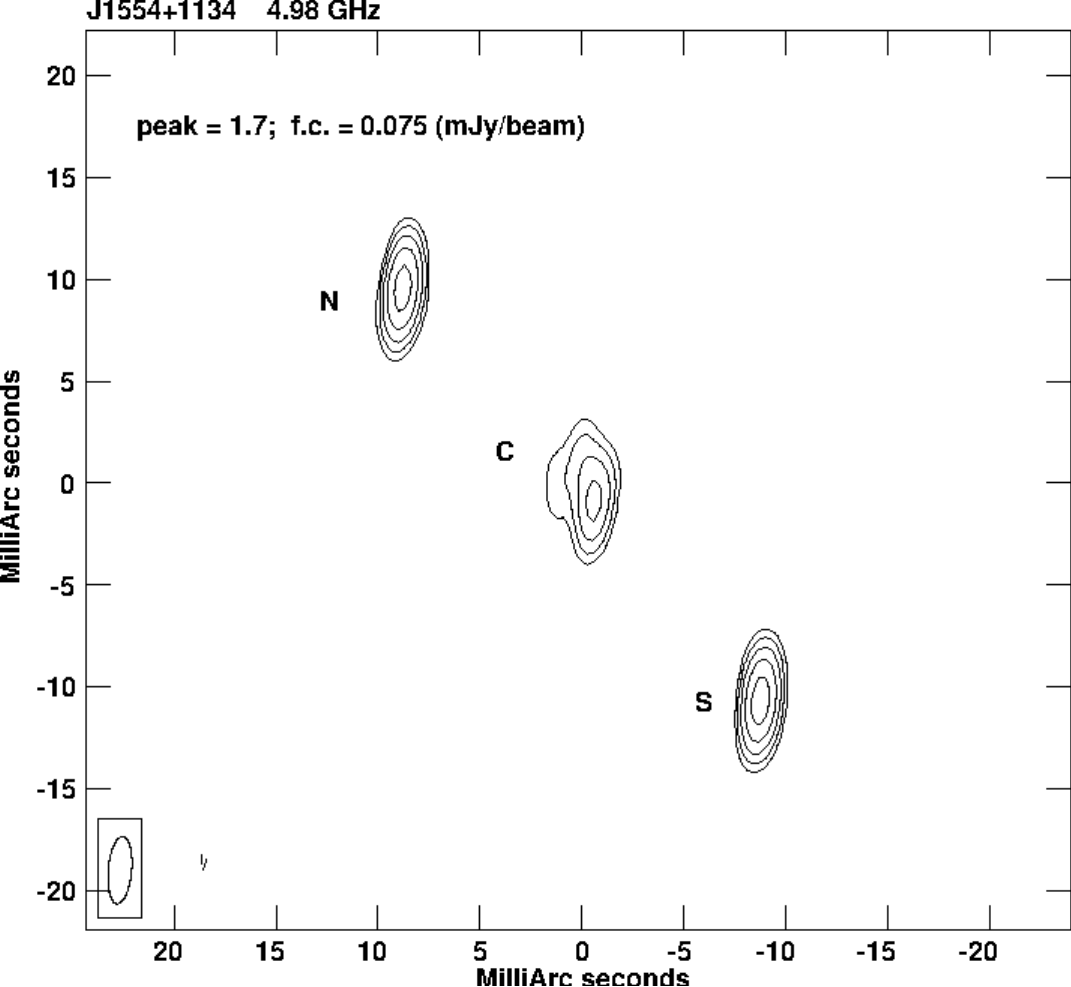}
\caption{continued.}
\end{center}
\end{figure*}

\addtocounter{figure}{-1}
\begin{figure*}
\begin{center}
\includegraphics[width=0.60\columnwidth]{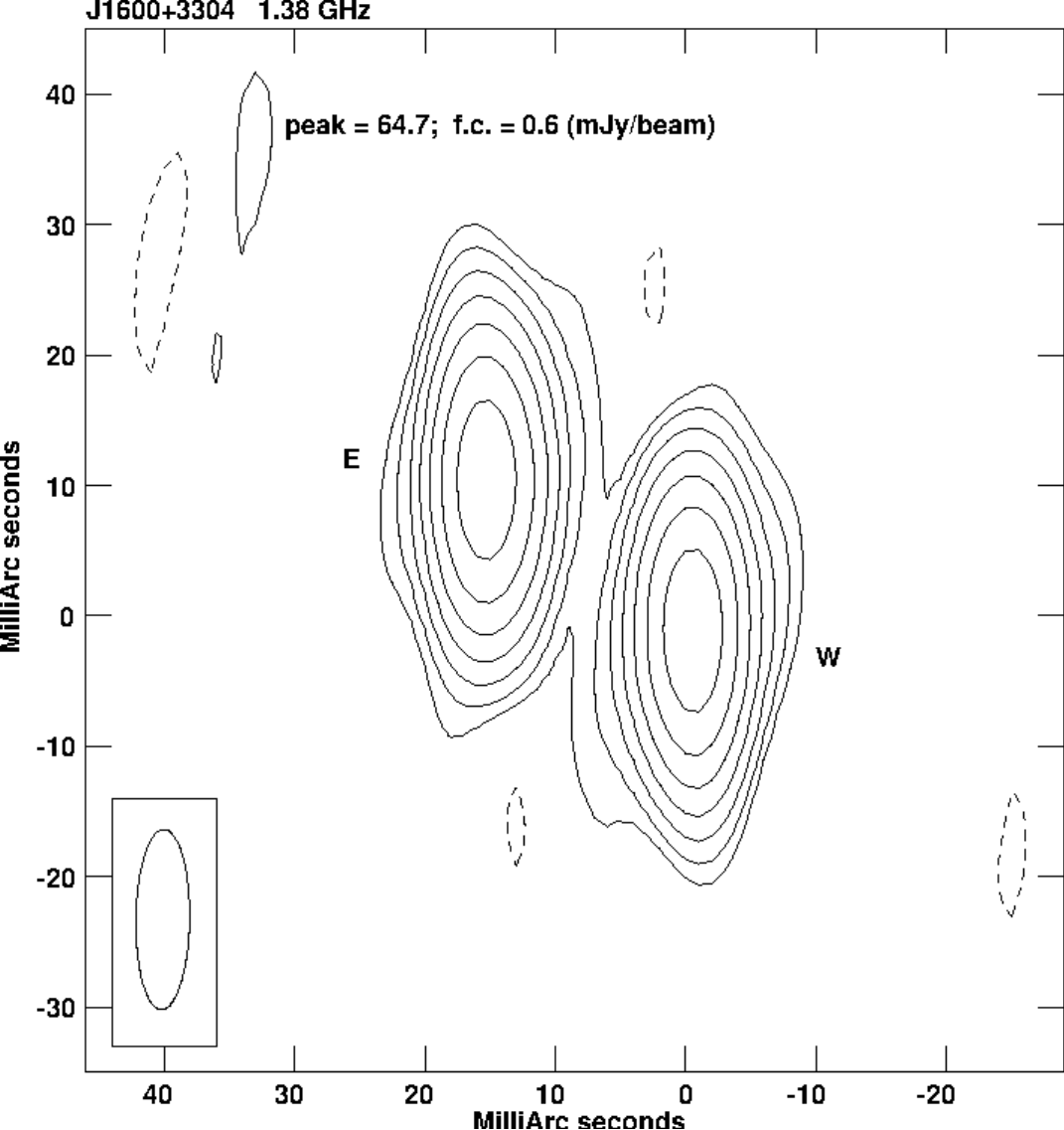}
\includegraphics[width=0.60\columnwidth]{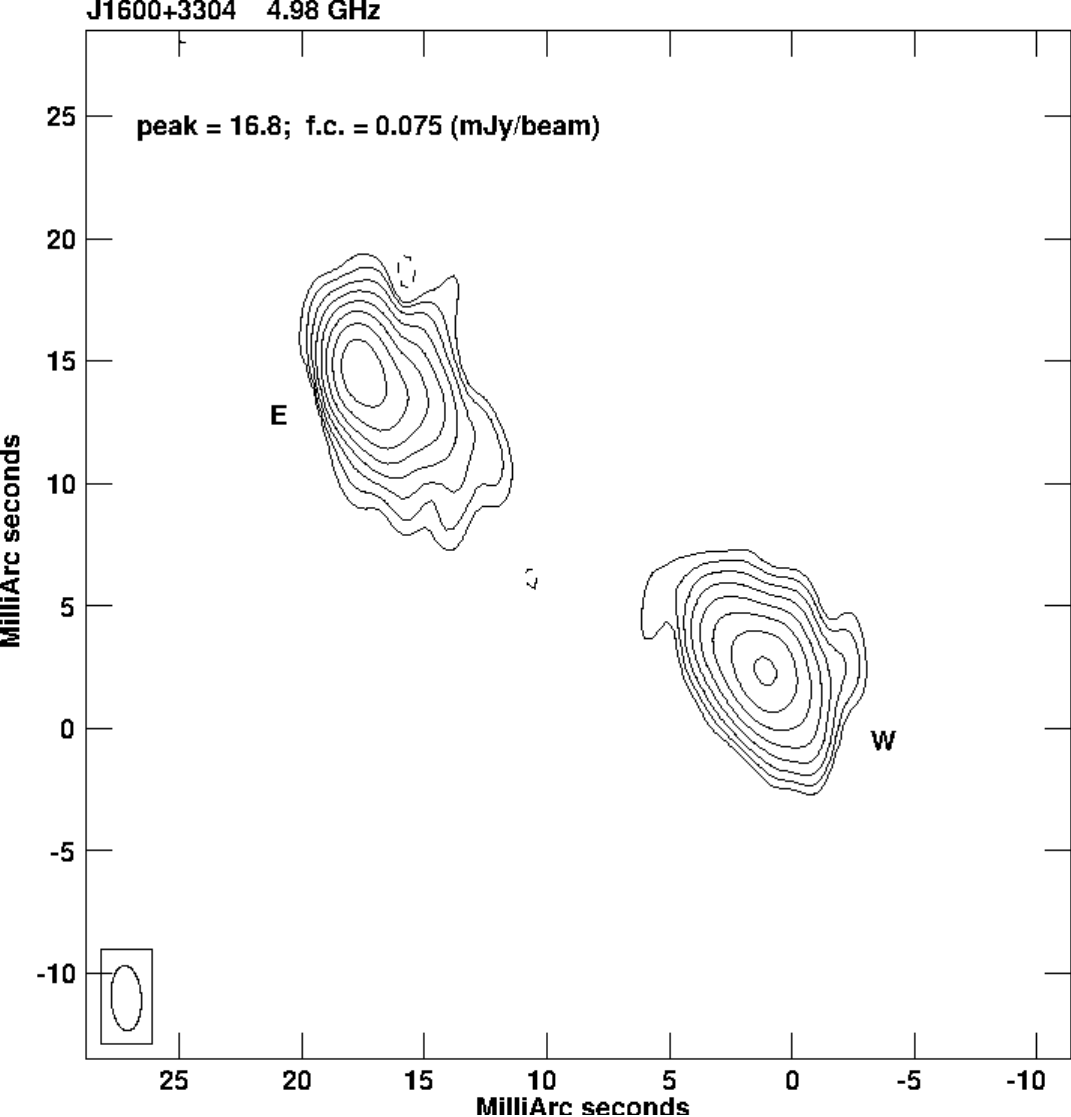}\\
\includegraphics[width=0.60\columnwidth]{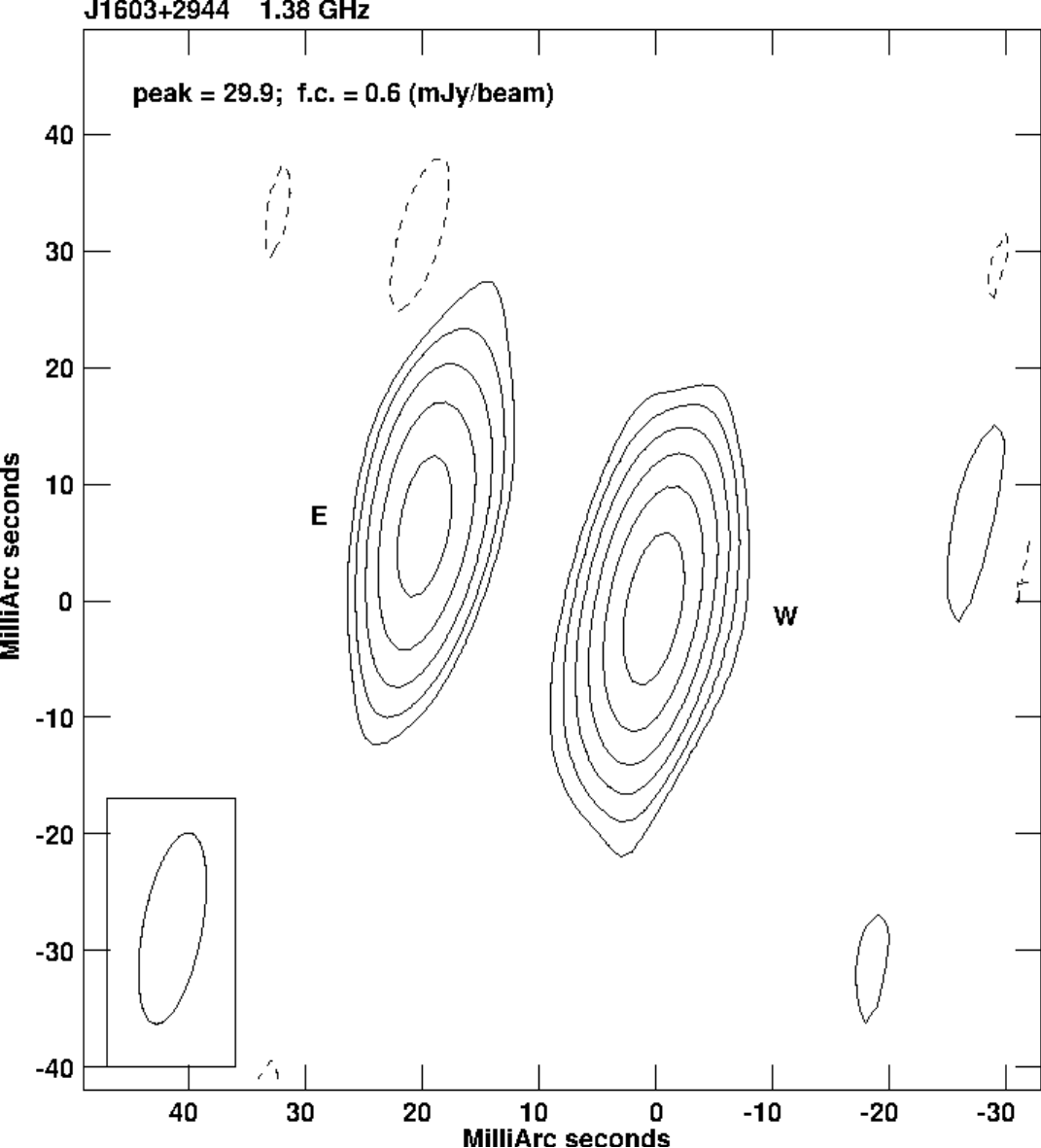}
\includegraphics[width=0.60\columnwidth]{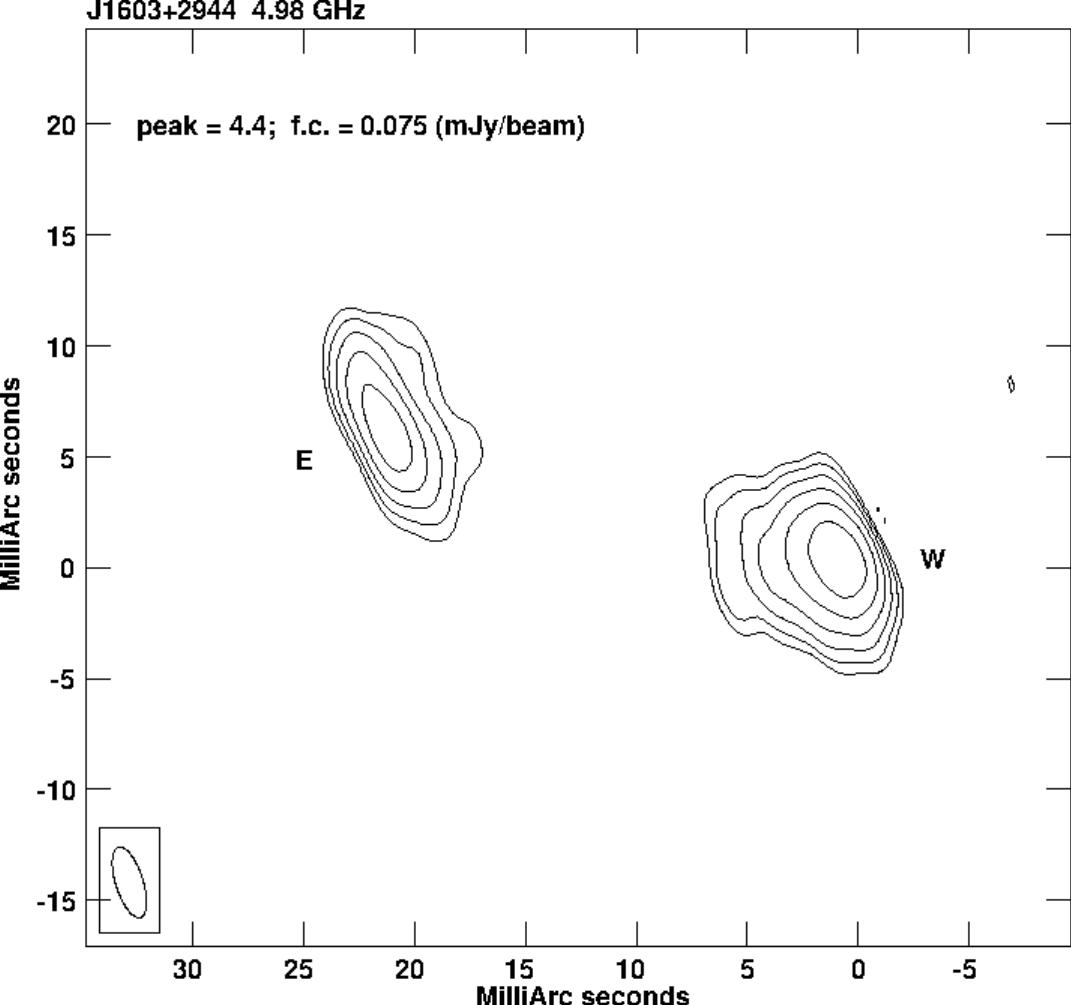}\\
\includegraphics[width=0.60\columnwidth]{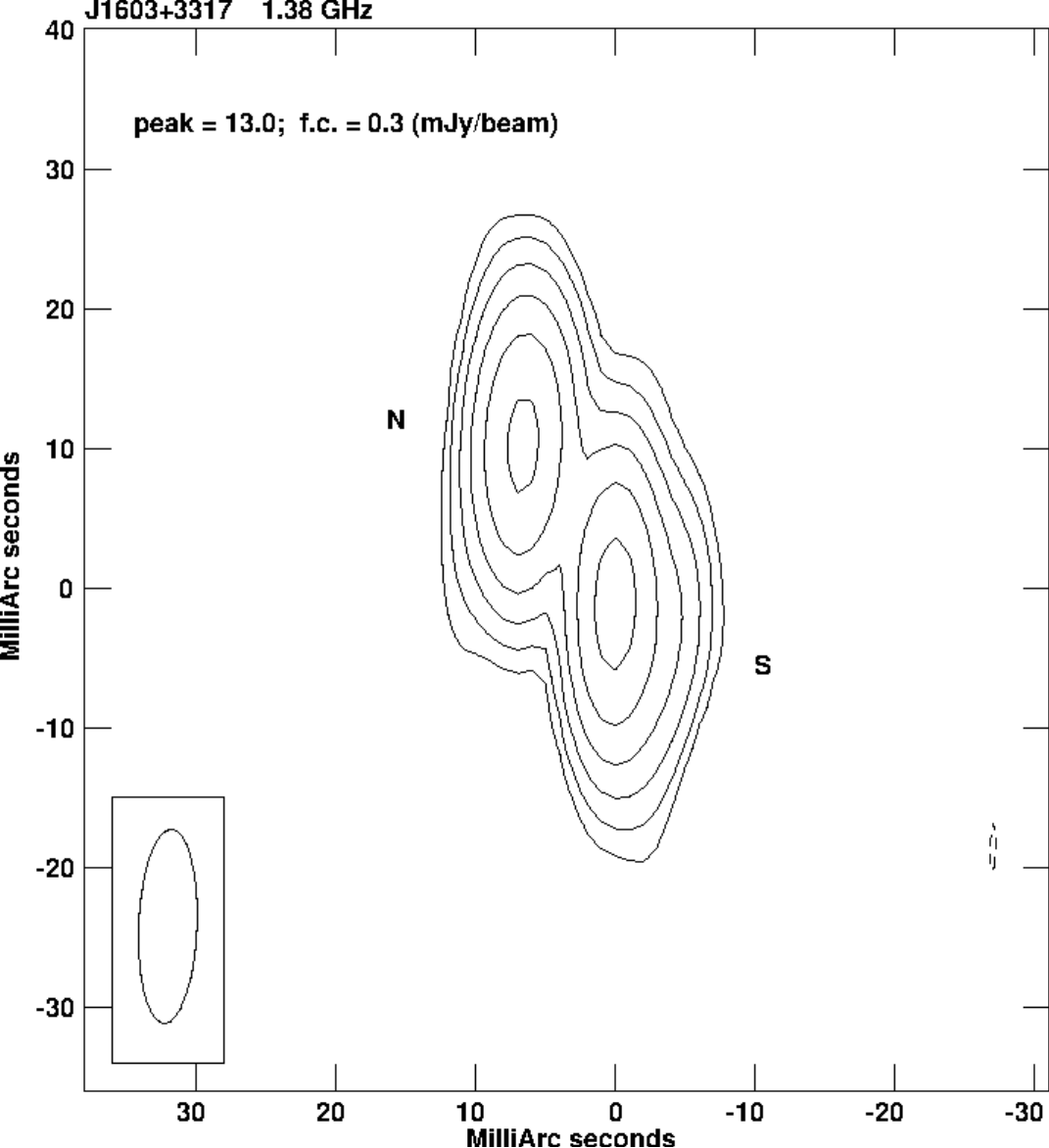}
\includegraphics[width=0.60\columnwidth]{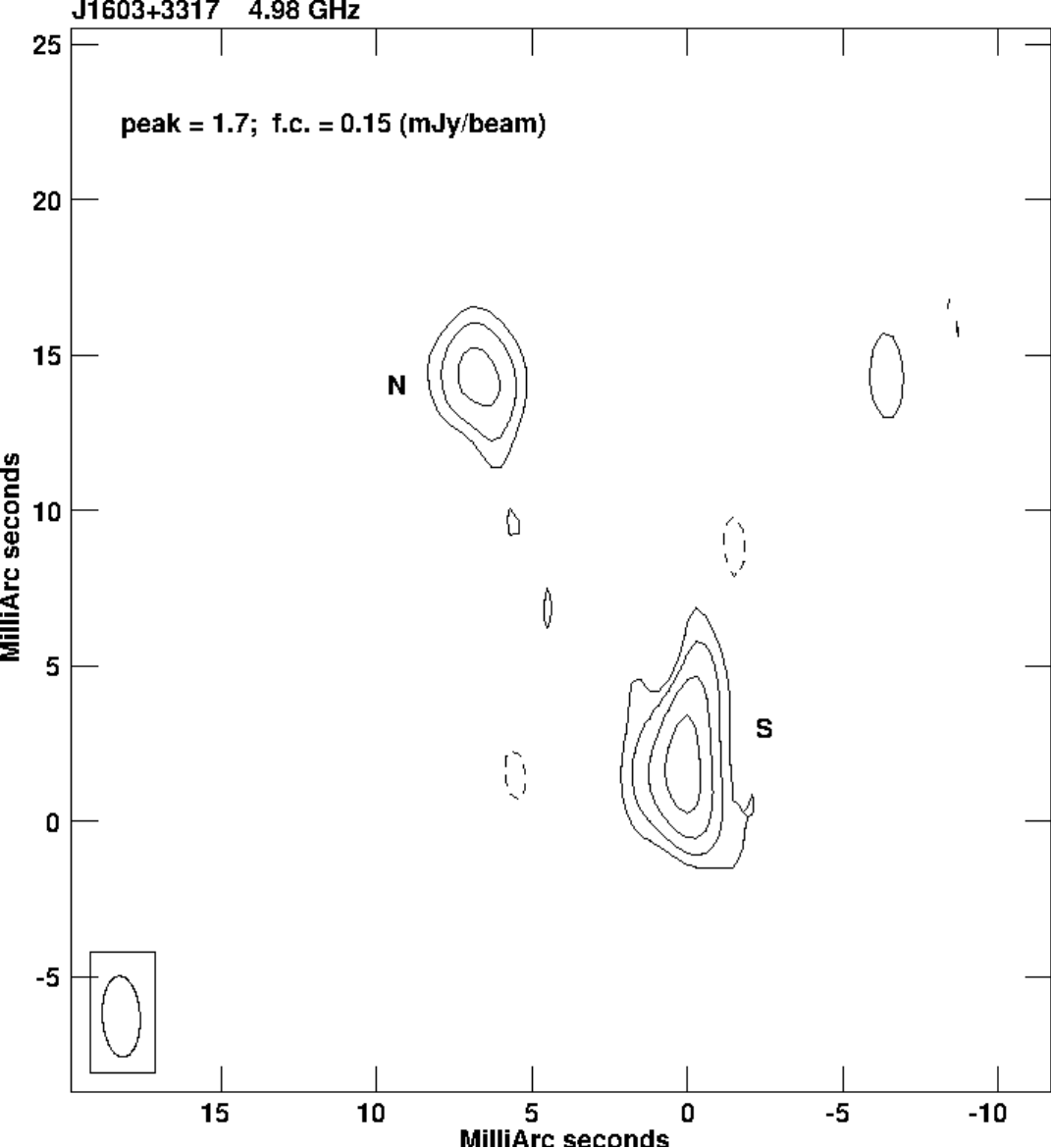}\\
\includegraphics[width=0.60\columnwidth]{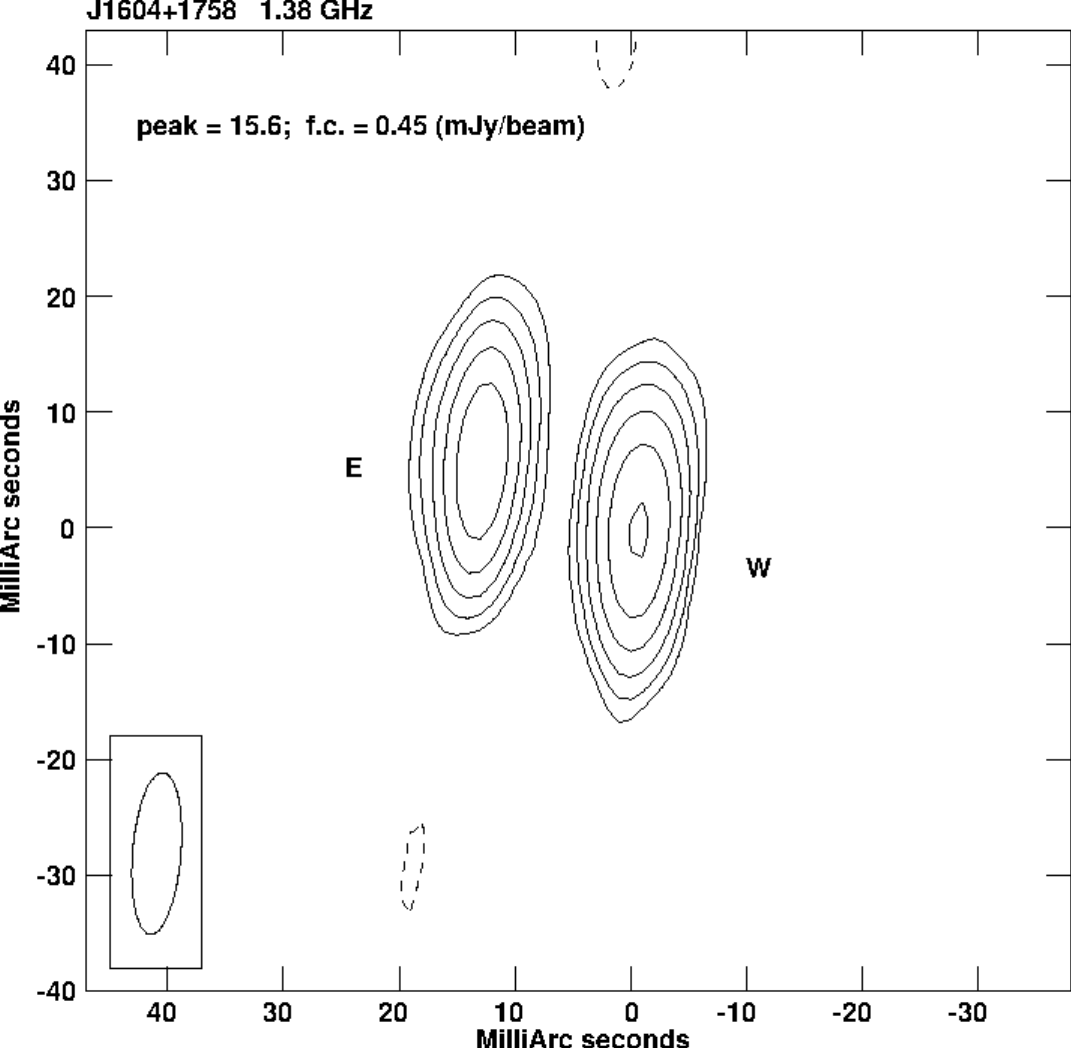}
\includegraphics[width=0.60\columnwidth]{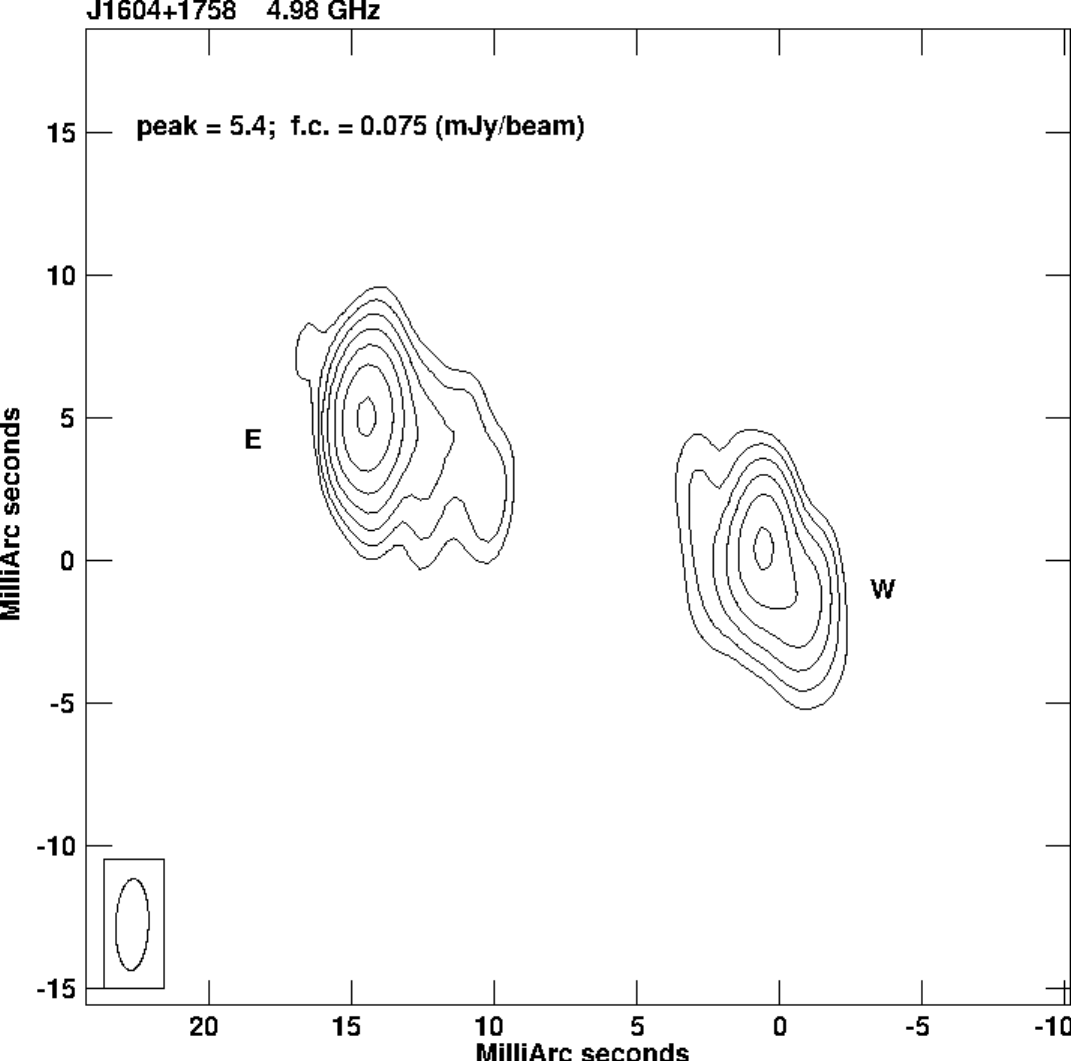}
\caption{continued.}
\end{center}
\end{figure*}


\end{appendix}


\begin{thebibliography}{}

\bibitem[Abdurro'uf et al.(2022)]{abdurrouf22}
Abdurro'uf, A. K., Aerts, C., Silva Aguirre, V., et al. 2022, ApJS, 259, 35

\bibitem[Alexander(2000)]{alexander00}
Alexander, P. 2000, MNRAS, 319, 8

\bibitem[An \& Baan(2012)]{an12a}
An, T., Baan, W.A. 2012, ApJ, 760, 77

\bibitem[An et al.(2012b)]{an12}
An, T., Wu, F., Yank, J., et al. 2012b, ApJS, 198, 5

\bibitem[Ballieux et al.(2024)]{ballieux24}
Ballieux, F.J., Callingham, J.R., R\"ottgering, H.J.A., Slob, M.M. 2024, A\&A, 689, 264

\bibitem[Beasley et al.(2002)]{beasley02}
Beasley, A.J., Gordon, D., Peck., A.B., et al. 2002, ApJS, 141, 3

\bibitem[Becker et al.(1995)]{becker95}
Becker, R.H., White, R.L., Helfand, D.J. 1995, ApJ, 450, 559

\bibitem[Callingham et al.(2017)]{callingham17}
Callingham, J.R., Ekers, R.D., Gaensler, B.M., et al. 2017, ApJ, 836, 174

\bibitem[Carvalho \& O'Dea(2002a)]{carvalho02a}
Carvalho, J.C., O'Dea, C.P. 2002a, ApJS, 141, 337

\bibitem[Carvalho \& O'Dea(2002b)]{carvalho02b}
Carvalho, J.C., O'Dea, C.P. 2002b, ApJS, 141, 371

\bibitem[Dallacasa et al.(2013)]{dd13}
Dallacasa, D., Orienti, M., Fanti, C., Fanti, R., Stanghellini, C. 2013, MNRAS, 433, 147

\bibitem[Dallacasa et al.(2021)]{dd21}
Dallacasa, D., Orienti, M., Fanti, C., Fanti, R. 2021, MNRAS, 504, 2312

\bibitem[Deller \& Middelberg(2014)]{deller14}
Deller, A.T., Middelberg, E. 2014, AJ, 147, 14

\bibitem[de Vries et al.(2009)]{devries09}
de Vries, N., Snellen, I.A.G., Schilizzi, R.T., Mack, K.-H., Kaiser, C.R. 2009, A\&A, 498, 641

\bibitem[Dey et al.(2019)]{dey19}
Dey, A., Schlegel, D.J., Lang, D., et al. 2019, AJ, 157, 168

\bibitem[Fanti et al.(1995)]{fanti95}
Fanti, C., Fanti, R., Dallacasa, D., et al. 1995, A\&A, 302, 317

\bibitem[Fanti et al.(2001)]{fanti01}
Fanti, C., Pozzi, F., Dallacasa, D., et al. 2001, A\&A, 369, 380

\bibitem[Fanti et al.(1990)]{fanti90}
Fanti, R., Fanti, C., Schilizzi, R.T., et al. 1990, A\&A, 231, 333

\bibitem[Helmboldt et al.(2007)]{helmboldt07}
Helmboldt, J.F., Taylor, G.B., Tremblay S., et al. 2007, ApJ, 658, 203

\bibitem[Jeyakumar \& Saikia(2000)]{jeyakumar00}
Jeyakumar, S., Saikia, D.J. 2000, MNRAS, 311, 397

\bibitem[Jorstad et al.(2017)]{jorstad17}
Jorstad, S., Marscher, A.P., Morozova, D.A., et al. 2017, ApJ, 846, 98

\bibitem[Kaiser \& Alexander(1997)]{kaiser97}
Kaiser, C.R., Alexander, P. 1997, MNRAS, 286, 215

\bibitem[Kiehlmann et al.(2024a)]{kiehlmann24a}
Kiehlmann, S., Lister, M.L., Readhead, A.C.S., et al. 2024a, ApJ, 961, 240

\bibitem[Kiehlmann et al.(2024b)]{kiehlmann24b}
Kiehlmann, S., Readhead, A.C.S., O'Neill, S., et al. 2024b, ApJ, 961, 241

\bibitem[Krause(2003)]{krause03}
Krause, M. 2003, A\&A, 398, 113

\bibitem[Kunert-Bajraszewska et al.(2010)]{kb10}
Kunert-Bajraszewska, M., Gawro\'nski, M.P., Labiano, A., Siemiginowska, A. 2010, MNRAS, 408, 2261 

\bibitem[Labiano et al.(2006)]{labiano06}
Labiano, A., Vermeulen, R.C., Barthel, P.D., et al. 2006, A\&A, 447, 481

\bibitem[Lacy et al.(2020)]{lacy20}
Lacy, M., Baum, S.A., Chandler, C.J., et al. 2020, PASP, 132, 5001

\bibitem[Lister et al.(2013)]{lister13}
Lister, M.L., Aller, M.F., Aller, H.D., et al. 2013, ApJ, 146, 120

\bibitem[Massaglia et al.(2016)]{massaglia16}
Massaglia, S., Bodo, G., Rossi, P., Capetti, S., Mignone, A. 2016, A\&A, 596, 12

\bibitem[Massaglia et al.(2019)]{massaglia19}
Massaglia, S., Bodo, G., Rossi, P., Capetti, S., Mignone, A. 2019, A\&A, 621, 132

\bibitem[Massaglia et al.(2022)]{massaglia22}
Massaglia, S., Bodo, G., Rossi, P., Capetti, A., Mignone, A. 2022, A\&A, 659, 139

\bibitem[Mingaliev et al.(2012)]{mingaliev12}
Mingaliev, M.G., Sotnikova, Yu.V., Torniainen, I., Tornikoski, M., Udovitskiy, R.Yu. 2012, A\&A, 544, 25

\bibitem[Mingo et al.(2019)]{mingo19}
Mingo, B., Croston, J.H., Hardcastle, M.J., et al. 2019, MNRAS, 488, 2701

\bibitem[Morganti et al.(2004)]{morganti04}
Morganti, R., Oosterloo, T.A., Tadhunter, C.N., et al. 2004, A\&A, 424, 119

\bibitem[Morganti et al.(2013)]{morganti13}
Morganti, R., Fogasy, J., Paragi, Z., Oosterloo, T., Orienti, M. 2013, Science, 341, 1082

\bibitem[Mukherjee et al.(2018)]{mukherjee18}
Mukherjee, D., Wagner, A.Y., Bicknell, G.V., et al. 2018, MNRAS, 476, 80

\bibitem[Mukherjee et al.(2020)]{mukherjee20}
Mukherjee, D., Bodo, G., Mignone, A., Rossi, P., Vaidya, B. 2020, MNRAS, 499, 681

\bibitem[Murthy et al.(2019)]{murthy19}
Murthy, S., Morganti, R., Oosterloo, T., et al. 2019, A\&A, 629, 58

\bibitem[O'Dea(1998)]{odea98}
O'Dea, C.P. 1998, PASP, 110, 493

\bibitem[Orienti et al.(2007a)]{mo07a}
Orienti, M., Dallacasa, D., Stanghellini, C. 2007a, A\&A, 461, 923

\bibitem[Orienti et al.(2007b)]{mo07}
Orienti, M., Dallacasa, D., Stanghellini, C. 2007b, A\&A, 475, 813

\bibitem[Orienti \& Dallacasa(2008)]{mo08}
Orienti, M., Dallacasa, D. 2008, A\&A, 487, 885

\bibitem[Orienti \& Dallacasa(2012)]{mo12}
Orienti, M., Dallacasa, D. 2012, MNRAS, 424, 532

\bibitem[Orienti \& Dallacasa(2014)]{mo14}
Orienti, M., Dallacasa, D. 2014, MNRAS, 438, 463

\bibitem[Orienti(2016)]{mo16}
Orienti, M. 2016, AN, 337, 9

\bibitem[Orienti \& Dallacasa(2020)]{mo20}
Orienti, M., Dallacasa, D. 2020, MNRAS, 499, 1340

\bibitem[Peacock \& Wall(1981)]{pw81}
Peacock, J.A., Wall, J.V. 1981, MNRAS, 194, 331

\bibitem[Pearson \& Readhead(1988)]{pr88}
Pearson, T.J., Readhead, A.C.S. 1988, ApJ, 328, 114

\bibitem[Peck \& Taylor(2000)]{peck00}
Peck, A.B., Taylor, G.B. 2000, ApJ, 534, 90

\bibitem[Perucho \& Mart\'i(2002)]{perucho02}
Perucho, M., Mart\'i, J.M. 2002, ApJ, 568, 639

\bibitem[Perucho et al.(2014)]{perucho14}
Perucho, M., Mart\'i, J.M., Laing, R.A., Hardee, P.E. 2014, MNRAS, 441, 1488

\bibitem[Perucho(2020)]{perucho20}
Perucho, M. 2020, MNRAS, 494, L22

\bibitem[Polatidis et al.(1995)]{polatidis95}
Polatidis, A.G., Wilkinson, P.N., Xu, W., et al. 1995, ApJS, 98, 1

\bibitem[Readhead et al.(1996)]{readhead96}
Readhead, A.C.S., Taylor, G.B., Pearson, T.J., Wilkinson, P.N. 1996, ApJ, 460, 634

\bibitem[Readhead et al.(2024)]{readhead24}
Readhead, A.C.S., Ravi, V., Blandford, R.D., et al. 2024, ApJ, 961, 242

\bibitem[Rossi et al.(2017)]{rossi17}
Rossi, P., Bodo, G., Capetti, A., Massaglia, S. 2017, A\&A, 606, 57

\bibitem[Rossi et al.(2020)]{rossi20}
Rossi, P., Bodo, G., Massaglia, S., Capetti, A. 2020, A\&A, 642, 69

\bibitem[Rossi et al.(2024)]{rossi24}
Rossi, P., Bodo, G., Massaglia, S., Capetti, A. 2024, A\&A, 685, 4

\bibitem[Scheuer(1974)]{scheuer74}
Scheuer, P.A.G. 1974, MNRAS, 166, 513

\bibitem[Snellen et al.(1996)]{snellen96}
Snellen, I.A.G., Bremer, M.N., Schilizzi, R.T., Miley, G.K., van Ojik, R. 1996, MNRAS, 279, 1294

\bibitem[Snellen et al.(1998)]{snellen98}
Snellen, I.A.G., Schilizzi, R.T., de Bruyn, A.G., et al. 1998, A\&AS, 131, 435

\bibitem[Snellen et al.(2000a)]{snellen00}
Snellen, I.A.G., Schilizzi, R.T., Miley, G.K., et al. 2000a, MNRAS, 319, 445

\bibitem[Snellen et al.(2000b)]{snellen00b}
Snellen, I.A.G., Schilizzi, R.T., van Langevelde, H.J. 2000b, MNRAS, 319, 429

\bibitem[Snellen et al.(2002)]{snellen02}
Snellen, I.A.G., Lehnert, M.D., Bremer, M.N., Schilizzi, R.T. 2002, MNRAS, 337, 981

\bibitem[Snellen et al.(2004)]{snellen04}
Snellen, I.A.G., Mack, K.-H., Schilizzi, R.T., Tschager, W. 2004, MNRAS, 348, 227

\bibitem[Sokolovsky et al.(2011)]{sokolovsky11}
Sokolovsky, K.V., Kovalev, Y.Y., Pushkarev, A.B., Mimica, P., Perucho, M. 2011, A\&A, 535, 24

\bibitem[Stanghellini et al.(2009)]{cstan09}
Stanghellini, C., Dallacasa, D., Orienti, M. 2009, AN, 330, 223

\bibitem[Stanghellini et al.(2025)]{cstan25}
Stanghellini, C., Orienti, M., Spingola, C., et al. 2025, A\&A, 695, 179

\bibitem[Struve \& Conway(2012)]{struve12}
Struve, C., Conway, J.E. 2012, A\&A, 546, 22

\bibitem[Torniainen et al.(2005)]{torniainen05}
Torniainen, I., Tornikoski, M., Ter\"asranta, H., Aller, M.F., Aller, H.D. 2005, A\&A, 435, 839

\bibitem[Tremblay et al.(2016)]{tremblay16}
Tremblay, S.E., Taylor, G.B., Ortiz, A.A., et al. 2016, MNRAS, 459, 820

\bibitem[Webster et al.(2021)]{webster21}
Webster, B., Croston, J.H., Mingo, B., et al. 2021, MNRAS, 500, 4921

\bibitem[Wilkinson et al.(1994)]{wilkinson94}
Wilkinson, P.M., Polatidis, A.G., Readhead, A.C.S., Xu, W., Pearson, T.J. 1994, ApJ, 432, L87

\bibitem[Wright et al.(2010)]{wright10}
Wright, E.L., Eisenhardt, P.R.M., Mainzer, A., et al. 2010, AJ, 140, 1868

\bibitem[Yan et al.(2013)]{yan13}
Yan, L., Donoso, E., Tsai, C.-W., et  al. 2013, ApJ, 145, 55

\bibitem[Zhou et al.(2023)]{zhou23}
Zhou, R., Ferraro, S., White, M., et al. 2023, JCAP, 11, 97

\end{thebibliography}
\end{document}